\documentclass[aip,reprint]{revtex4-1}
\usepackage{subfig}
\usepackage{tikz}
\usetikzlibrary{patterns.meta}
\usetikzlibrary{shapes.geometric, arrows, arrows.meta}
\usetikzlibrary{positioning}
\usetikzlibrary{intersections}
\usetikzlibrary{3d}
\tikzstyle{startstop} = [rectangle, rounded corners, 
minimum width=3cm, 
minimum height=1cm,
text centered, 
draw=black, 
fill=white!30]

\tikzstyle{start} = [ellipse, 
minimum width=3cm, 
minimum height=1cm,
text centered, 
draw=black, 
fill=white!30]

\tikzstyle{ext} = [rectangle, rounded corners, dashed,
minimum width=3cm, 
minimum height=1cm,
text centered, 
draw=black, 
fill=white!30]

\tikzstyle{arrow} = [thick,->,>=stealth]
\tikzstyle{arrowbg} = [thick,->,>=stealth, green!35!black!65]

\tikzstyle{litoBase} = [rectangle, 
minimum width=3cm, 
minimum height=1cm,
text centered, 
draw=black, 
fill=gray!30]

\tikzstyle{insulation} = [rectangle, 
minimum width=3cm, 
minimum height=0.3cm,
text centered, 
draw=black, 
fill=blue!30]

\tikzstyle{insPart} = [rectangle, 
minimum width=1cm, 
minimum height=0.3cm,
text centered, 
draw=black, 
fill=blue!30]

\tikzstyle{resist} = [rectangle, pattern={Lines[angle=+45,distance=4pt]},
minimum width=3cm, 
minimum height=0.2cm,
text centered, 
draw=black]

\tikzstyle{partResist} = [rectangle, pattern={Lines[angle=+45,distance=4pt]},
minimum width=1cm, 
minimum height=0.2cm,
text centered, 
draw=black]

\tikzstyle{mask} = [rectangle,
minimum width=3cm, 
minimum height=0.2cm,
text centered, 
draw=black,
fill=pink!60]

\tikzstyle{mask2} = [rectangle,
minimum width=1cm, 
minimum height=0.2cm,
text centered, 
draw=black,
fill=black]

\begin{document}


\title{The influence of implantation conditions on dopant activation in Al-implanted 4H-SiC:
A MD study applying an Al potential fitted to DFT barriers.} 
\author{Sabine Leroch}

\affiliation{CDL for Multi-Scale Process Modeling of Semiconductor Devices and Sensors, at the 
 Institute for Microelectronics, TU Wien, 1040 Vienna, Austria}
 \email[]{leroch@iue.tuwien.ac.at}
 \author{Robert Stella}
 \affiliation{CDL for Multi-Scale Process Modeling of Semiconductor Devices and Sensors, at the 
 Institute for Microelectronics, TU Wien, 1040 Vienna, Austria}
 \author{Andreas Hössinger}
\affiliation{Silvaco Europe Ltd., St Ives, Cambridgeshire, PE27 5JL, United Kingdom}
\author{Lado Filipovic}
\affiliation{CDL for Multi-Scale Process Modeling of Semiconductor Devices and Sensors, at the 
 Institute for Microelectronics, TU Wien, 1040 Vienna, Austria}
\date{\today}
\bibliographystyle{rsc}
\begin{abstract}
The non-monotonic dependence of Al dopant activation on implantation temperature in 4H-SiC has been experimentally observed but its atomistic origin remains unclear. We present a molecular dynamics (MD) study of Al implantation in 4H-SiC at implantation temperatures of $500$\,K and $900$\,K, spanning seven doses from $1\times10^{13}$ to $7.5\times10^{14}$\,cm$^{-2}$, to describe how implantation temperature and dose govern defect formation, defect evolution, and dopant activation during early-stage annealing (here defined as the first $100$\,ns of annealing at temperatures between $1500$ and $2500$\,K as accessible to molecular dynamics). Using the Gao-Weber potential combined with a reparameterized Morse potential for Al-SiC interactions fitted to DFT migration and kick-in/out barriers, we demonstrate that implantation at both $500$\,K and $900$\,K substantially reduces Frenkel pair creation and extended amorphous pocket formation compared to room-temperature implantation, consistent with experimental observations. However, subsequent annealing reveals a non-monotonic temperature dependence at concentrations above the Al solubility limit ($>10^{20}$\,cm$^{-3}$). Systems implanted at $900$\,K develop substantially larger interstitial clusters that are kinetically stable, persisting throughout the $100$\,ns annealing window. These clusters act as sinks and trapping centers for Al, thereby reducing substitutional incorporation. Despite the lower as-implanted crystallinity for the $500$\,K implantation, a significantly larger fraction of Al occupies lattice sites after annealing for samples implanted at $500$\,K compared to those implanted at $900$\,K.
The simulations identify two distinct regimes, separated by the Al solubility limit: a low-dose regime dominated by isolated point defects and small complexes, and a high-dose regime characterized by defect clustering and planar defect formation with strong dependence on implantation temperature. These results provide an atomistic explanation for the experimentally observed activation window for optimal implantation temperatures between $500$ and $900$\,K, and suggest a kinetic processing regime in which controlled nanoscale amorphization at $500\,K$ enhances chemical activation via regrowth-assisted incorporation while suppressing extended defect formation. In addition, MD simulations identified a new basal-plane diffusion path for Al and revealed an activation mechanism involving the kick-out of the carbon antisite; both findings were confirmed by DFT-NEB calculations.
\end{abstract}
\maketitle
\section{Introduction}
Due to its wide bandgap of $3.26$\,eV, high breakdown voltage, excellent thermal stability, and low dopant diffusivity, 4H-SiC is particularly well suited for high-power and high-temperature semiconductor applications. 
This work focuses on p-type doping of 4H-SiC using aluminum. Al ion implantation is the standard method for forming p-type regions in 4H-SiC power devices, including MOSFET body regions, junction termination extensions, and ohmic contact regions, because conventional dopant diffusion is ineffective in SiC.
Substitutional Al on a Si lattice site introduces acceptor levels with thermal ionization energies of $198$\,meV and $210$\,meV above the valence band for the hexagonal and cubic lattice sites, respectively.

Selective ion implantation is employed to form well-defined doped regions with precisely controlled depth and lateral extent in 4H-SiC power devices.
Implantation is a non-equilibrium process thus, dopant concentrations in the range $10^{17}$\,cm$^{-3}$ - $10^{21}$\,cm$^{-3}$  can be achieved.
Although implantation enables dopant concentrations far beyond the thermal solubility limit achievable by epitaxy, it simultaneously introduces substantial lattice damage that limits electrical activation.
Implanted Al atoms may occupy interstitial positions or form defect complexes instead of occupying substitutional sites. In addition, ion bombardment generates interstitial-vacancy pairs (Frenkel pairs, FPs), which can recombine, migrate, or agglomerate into extended defect structures or amorphous pockets.

Post-implantation annealing at temperatures between $1600$ and $2400$\,K is therefore required to at least partially restore crystallinity and promote dopant activation. Nevertheless, defects that survive annealing above $1600$\,K frequently remain, included in small point-defect complexes (vacancy-interstitial and impurity-vacancy pairs, Al-interstitial configurations), dislocation loops, stacking faults in the basal plane, voids, and so-called black spot defects (BSDs), which correspond to nanometer-sized ($1-5$\,nm) interstitial clusters\,\cite{Ko2017}. 
These defects degrade electrical performance and act as sinks or traps for dopants, thereby reducing activation efficiency. A recent comprehensive review of electrically active defects in 3C-, 4H-, and 6H-SiC polytypes assessed by deep-level transient spectroscopy\,\cite{Capan2025} confirms that Al-related complexes and radiation-induced intrinsic point defects constitute the dominant compensating levels in p-type 4H-SiC material.

Experimental studies have demonstrated that both implantation temperature and dopant concentration critically influence defect evolution. Transmission electron microscopy studies by Nipoti \emph{et al.}\,\cite{Nipoti2018-2} revealed various kinds of defect structures in the implanted layer over a wide Al concentration range ($1\times 10^{19}$\,cm$^{-3}$ to $1\times 10^{21}$\,cm$^{-3}$) for implantation at $700$\,K.  For Al concentrations in the range of $2\times 10^{18}$\,cm$^{-3}$ - $2\times 10^{20}$\,cm$^{-3}$ only point defects, interstitial clusters and loops that could form network like structures were visible, while
at concentrations exceeding $2\times10^{20}$\,cm$^{-3}$, additional defect structures were observed, including Al precipitates and extended stacking faults free of Al. These stacking faults formed near the center of the implantation peak via insertion of an additional Si-C bilayer in the $\langle 0001\rangle$ plane, were framed by partial dislocations \cite{kimoto2014}, and increased in size with increasing Al concentration.
Similarly, Persson \emph{et al.}\,\cite{Persson2003} reported dose-dependent formation of small dislocation loops after hot implantation at $900$\,K and subsequent annealing. At medium Al concentrations in the range $0.1-1\times10^{20}$\,cm$^{-3}$, both the density and size of these defects visible in TEM images increased with increasing implantation dose. 

Because stacking faults and dislocation loops degrade device performance, processing strategies aim to avoid exceeding a critical Al concentration while employing high-temperature annealing above $2000$\,K \cite{Nipoti2018}. 
The critical concentration for extended defect formation has been reported to vary between $5\times10^{19}$ and $2\times10^{20}$\,cm$^{-3}$ depending on implantation temperature\,\cite{Nipoti2018-2}.
Although elevated implantation temperatures are commonly used to suppress complete amorphization through dynamic defect recovery, several studies indicate a more complex and non-monotonic behavior at high Al concentrations. 
Wendler \emph{et al.}\,\cite{Wendler1998} suggested that hot implantation promotes the formation of more stable defect agglomerates at medium and high Al concentrations that require prolonged annealing for recovery.
Michaud \emph{et al.}\,\cite{Michaud2013} reported that implantation at intermediate temperatures (around $500$\,K) results in the lowest electrical resistance for Al concentrations of $1\times10^{20}$\,cm$^{-3}$ after annealing compared to implantation at room temperature or elevated temperature of $900$\,K.
Similarly, Negoro \emph{et al.}\,\cite{Negoro2004} observed higher carrier concentrations after annealing at $2100$\,K for room-temperature implantation compared to implantation at $900$\,K at a very high Al concentration of $1.5\times10^{21}$\,cm$^{-3}$. 

More recently, Wang \emph{et al.}\,\cite{WANG2023} investigated supersaturated Al implantation at temperatures between $800$ and $1200$\,K. They reported progressive agglomeration of defects into larger clusters with increasing temperature, eventually leading to dislocation loop formation. The density of loops increased with implantation temperature, while the critical Al concentration required for planar defect formation shifted to higher values at lower temperatures. 
In a related study, Zang \emph{et al.}\,\cite{ZANG2025} demonstrated that implantation at $700$\,K followed by annealing at $1000$\,K significantly reduces planar defect density compared to prolonged implantation at $1000$\,K. Based on these observations, a temperature window between $500$ and $900$\,K was proposed \cite{WANG2023} in which complete amorphization is suppressed while extended planar defect formation remains limited.  Collectively, these results point to a temperature-dependent transition between amorphous damage, defect clustering, and planar defect nucleation.

Despite extensive experimental work, the atomistic mechanisms governing this transition remain unclear. In particular, it is not yet understood why moderate implantation temperatures can enhance dopant activation at high doses, whereas higher temperatures promote defect clustering and dislocation loop formation even though amorphization is reduced. The interplay between Frenkel-pair separation, interstitial mobility, cluster nucleation, defect agglomeration, and substitutional Al incorporation has not yet been resolved on the atomic scale for 4H-SiC.

To address these questions, we perform large-scale molecular dynamics (MD) simulations of shallow Al implantation in 4H-SiC over a broad range of implantation doses and temperatures. Ion cascades are simulated explicitly, followed by extended annealing up to $100$\,ns to capture defect recombination, clustering, and dopant incorporation. Recent MD studies of Al$^+$ implantation in 4H-SiC using the GW potential have characterized dose-dependent damage accumulation and interface stress evolution\,\cite{Yang2025,Xuanyuan2025}; however, these studies employ short annealing windows of $1$-$3$\,ns, which is insufficient to capture extended defect coarsening and dopant incorporation occurring on longer time scales. In contrast to previous MD studies on Al-implanted 3C-SiC that employed considerably shorter annealing times by using interatomic potentials with limited recovery accuracy\,\cite{wu2021,samolyuk2015}, the present work combines the Gao--Weber (GW) potential with a DFT-parameterized Morse interaction for Al, enabling a more reliable description of defect formation and migration processes in 4H-SiC.

By directly correlating implantation temperature, defect clustering, and Al incorporation, we identify two distinct regimes separated by the Al saturation limit: a low-concentration regime dominated by isolated point defects and small complexes, and a high-concentration regime characterized by defect clustering and the emergence of planar defect precursors. The simulations provide an atomistic explanation for the experimentally observed activation window and establish a kinetic framework for optimizing implantation temperature at high Al doses.

This manuscript is structured as follows.
Section~\ref{sec:Methods} describes the simulation methods and setup.
In Section~\ref{sec:Results}, defect structures formed at different implantation temperatures and doses are analyzed before and after annealing and compared with available experimental and density functional theory (DFT) data.

\section{Methods}
\label{sec:Methods}

Ion implantation in 4H-SiC was simulated using the Gao-Weber (GW) interatomic potential\,\cite{gao2002}, combined with a Ziegler-Biersack-Littmark (ZBL) potential to accurately describe strong short-range repulsive interactions during high-energy atomic collisions. 

The GW potential was developed by fitting a Brenner-type potential to DFT data, including defect formation and migration energies for silicon and carbon defects in 3C-SiC in the neutral state\,\cite{mattausch2005,matsushima2019,bockstedte2003,kobayashi2019}. It has been extensively used to investigate irradiation damage in SiC, including displacement 
cascade-induced cluster formation\,\cite{gao_weber2004,gao2003,weber2004} and recrystallization of amorphous 4H-SiC\,\cite{Gao_zhang_2006,leroch2024}.
Recent benchmark studies confirm that the GW potential remains the most reliable empirical potential currently available for large-scale irradiation simulations in SiC\,\cite{samolyuk2015,Yu2024}.

Interactions between Al and host atoms (Si or C) were modeled using a Morse potential originally proposed by Dandekar and Shin\,\cite{dandekar2011molecular}. In this work, the Morse parameters were reparameterized to DFT results to better reproduce migration and kick-in/-out barriers of Al in the neutral state.
The optimized Morse parameters are listed in Table~\ref{tab:morse} using an interaction cutoff of 5\,\AA{}. Pure Al-Al interactions were described using an embedded atom method (EAM) potential distributed with the LAMMPS package\,\cite{thompson2022}.

\begin{table}[ht!]
\caption{Optimized Morse potential parameters used for Al-Si and Al-C interactions. The interaction cutoff distance was set to $5$\,$\AA{}$.}
    \centering
    \setlength{\tabcolsep}{9pt}   
    \begin{tabular}{lcc}
   \textbf{System} & \textbf{Parameters} & \textbf{Morse-Potential} \\  
    \hline
    Al-Si & $D_0$ &  0.4 \\
    & $\alpha$ & 1.6\\
    & $r_0$ & 2.8 \\ 
    Al-C & $D_0$ & 0.4 \\
    & $\alpha$ & 1.4\\
    & $r_0$ & 2.05\\ 
    \hline
    \end{tabular}
\label{tab:morse}
\end{table}

To validate the reparameterized Al interactions, migration pathways and energy barriers of neutral Al defects in 4H-SiC were recalculated using DFT. These calculations serve as reference data for assessing the accuracy of the GW-Morse description of Al diffusion and activation mechanisms.

DFT simulations were performed using the CP2K package\,\cite{kuhne2020cp2k} within the Gaussian plane-wave (GPW) formalism\,\cite{vandevondele2007gaussian} and Goedecker-Teter-Hutter (GTH) pseudopotentials\,\cite{goedecker1996separable}. Double-$\zeta$ Gaussian basis sets were employed together with a plane-wave energy cutoff of 800\,Ry. Exchange-correlation effects were treated using the semilocal PBE functional\,\cite{perdew1996generalized}. Electronic minimization was carried out using classical Broyden mixing, which showed improved convergence compared to the orbital-transformation (OT) method. Migration barrier heights were calculated using the climbing-image nudged elastic band (CI-NEB) method\,\cite{henkelman2000climbing} in an orthogonal supercell containing 480 atoms with dimensions $15.44$\,\AA\,$\times$\, $16.03$\,\AA\,$\times$\, $20.22$\,\AA. Initial and final configurations were optimized using the conjugate-gradient algorithm, and a force convergence criterion of 20\,meV/\AA\ was applied in all calculations.

For a given Fermi level, the thermodynamically stable charge state of a defect is determined by its formation energy. Since binding energies and migration barriers depend on the defect charge state, these quantities usually vary with Fermi level. In MD simulations, a neutral charge state is conventionally assumed to describe the defect kinetics, and the Gao-Weber potential was also developed entirely for neutral intrinsic defects. This simplification is delicate but can be partially justified if the simulation is considered valid for certain regions of Fermi level. 
Immediately after implantation, the semiconductor contains a high concentration of implantation-induced defects, many of which introduce deep donor- and acceptor-like states in the band gap. Because the electrical activation of dopants is initially low, these defect states dominate the charge balance and can pin the Fermi level near mid-gap.
Moreover, MD simulations can only cover a short period of several $100$ \,ns of the entire annealing process. During MD simulations at high temperatures the semiconductor is in the so-called transient enhanced diffusion (TED) regime, which is characterized by increased dopant diffusion, low electrical and chemical activation and by the Ostwald ripening of defect clusters. The system is therefore not in a steady state at the end of the annealing cycle, which in turn supports the assumption of a Fermi level around mid-gap. Most of the point defects and defect complexes treated in this work are stable in the neutral state from the intrinsic up to n-type regime. To show this more clearly, in the supplementary information formation plots with charge transition levels for all point defects and defect complexes discussed in this work are summarized in Tables S1 -- S3. and Figs. S1 -- S4.
\subsection{Validation of GW-Morse potential}
\subsubsection{Al kick-out and migration barriers}
\begin{figure*}[ht!]
\centering
    \subfloat[]{\includegraphics[width=0.5\linewidth]{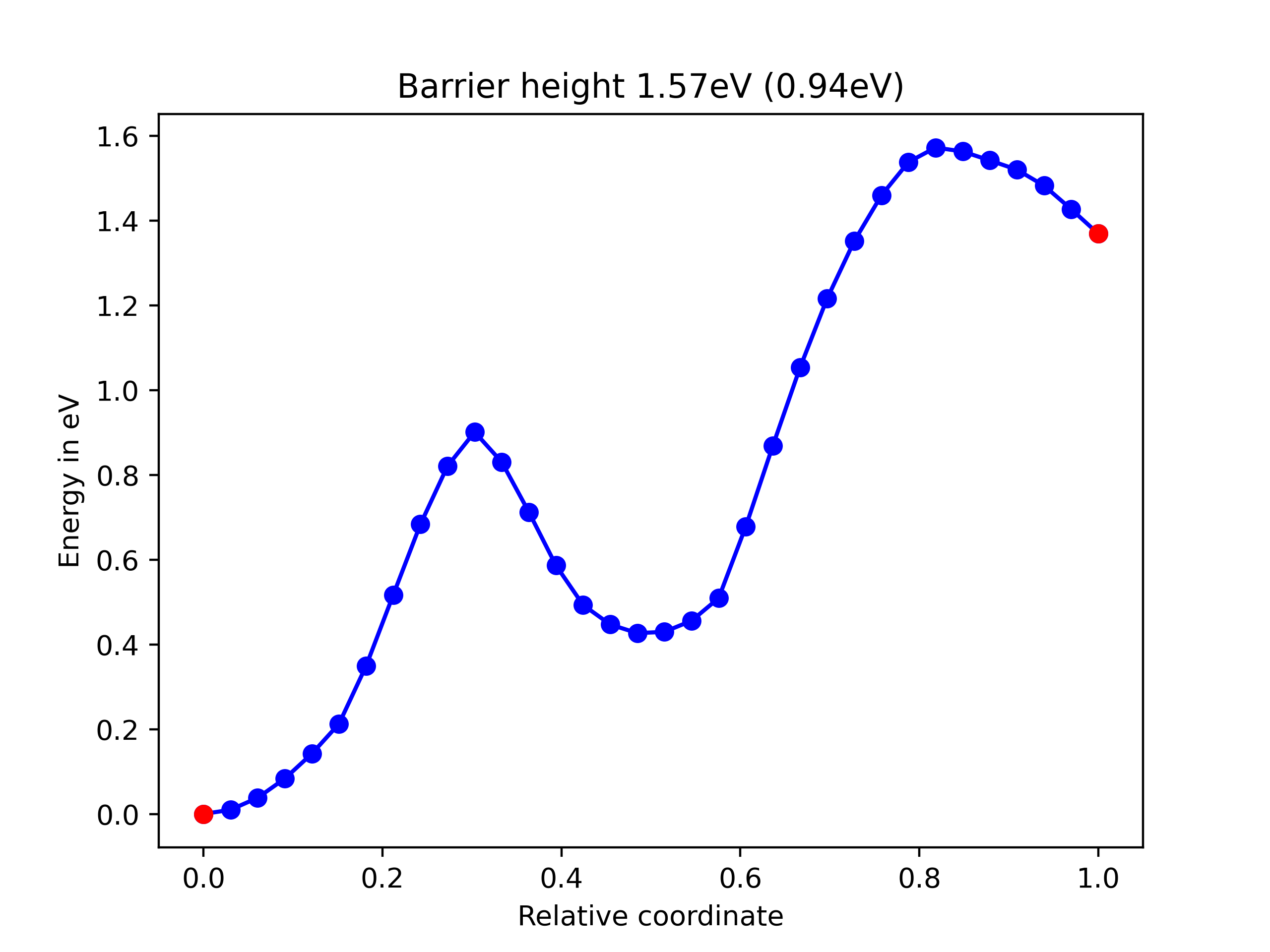}}\\
    \subfloat[]{\includegraphics[width=0.3\linewidth]{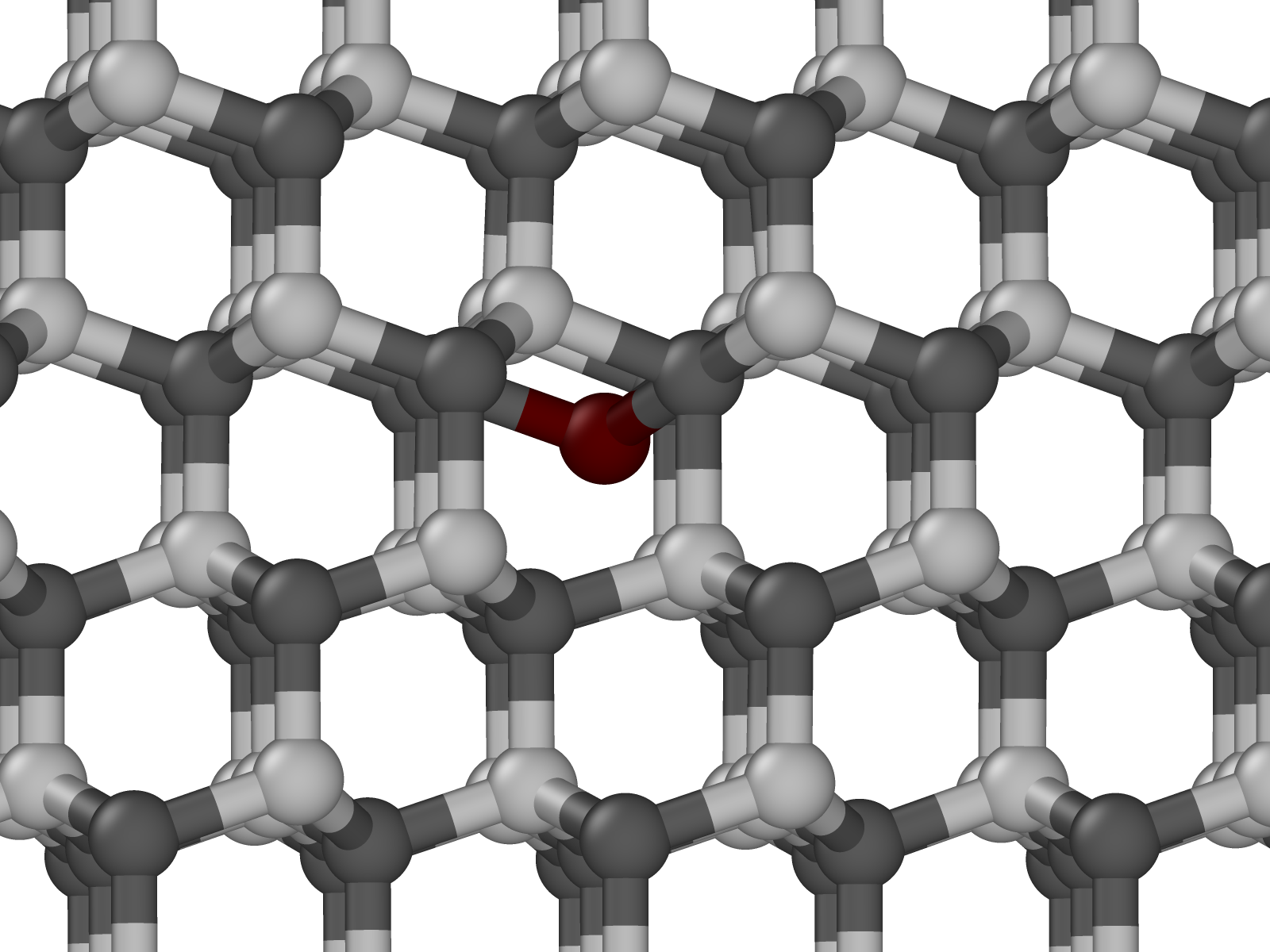}}\hspace{5pt}
    \subfloat[]{\includegraphics[width=0.31\linewidth]{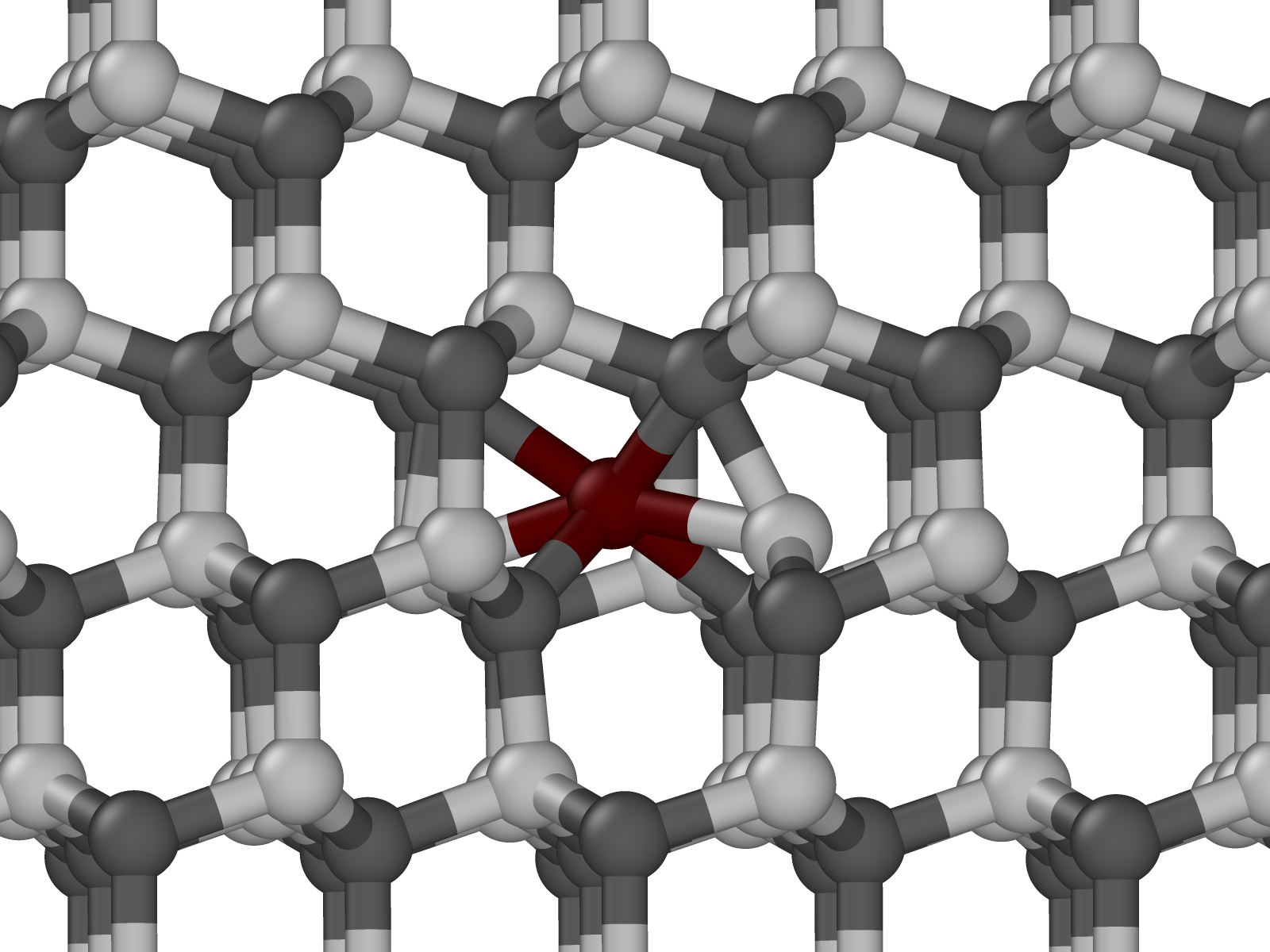}}\hspace{5pt}
    \subfloat[]{\includegraphics[width=0.31\linewidth]{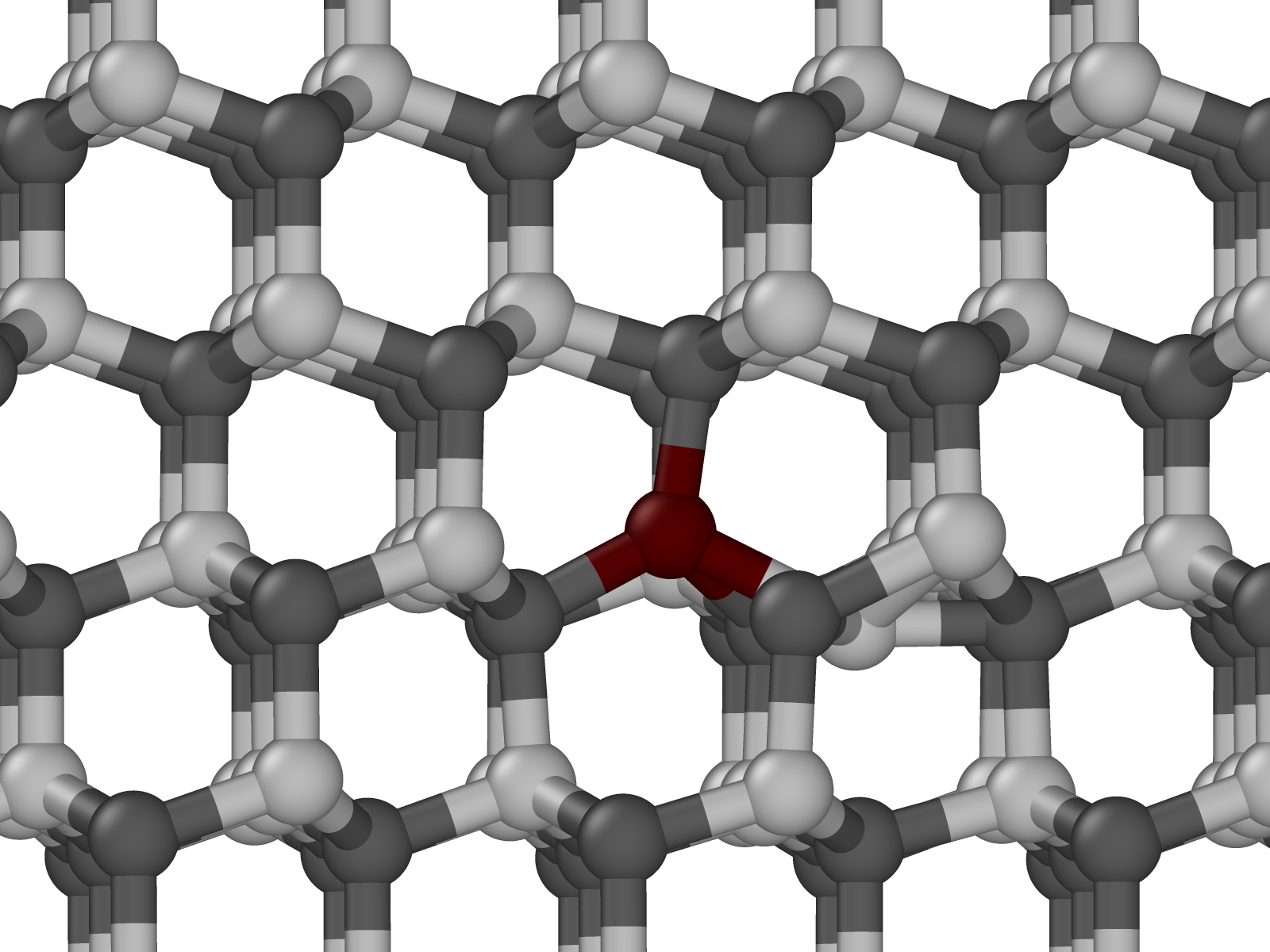}}
\caption{(a) CI-NEB calculation of the kick-in process of the neutral Al interstitial in the hexagonal plane. (b) Al$_{I,h}$ interstitial on the hexagonal site. 
(c) Formation of the metastable basal split interstitial in $\langle\overline{1}2\overline10\rangle$ direction at the minimum. 
(d) After further rotation of the split and breakage of Al bonds, the Al atom occupies the lattice site while Si is expelled to the position on the cubic site, where its final equilibrium position has not yet been reached by the DFT calculation. Al is shown in brown and host atoms are C (dark gray) and Si (light gray).}
\label{fig:kick-out}
\end{figure*}

The migration and kick-in/-out diffusion mechanisms of Al in 4H-SiC, where Al preferentially substitutes for Si, have been previously investigated for various charge states\,\cite{mattausch2005Al,huang2022}. However, the barriers for the neutral and positive charge state deduced by our CI-NEB calculations show significantly lower values than in \cite{huang2022}.
Figure~\ref{fig:kick-out} depicts the kick-in path for the neutral aluminum starting from the Al interstitial configuration at the hexagonal site (Al$_\mathrm{I,h}$). The kick-in (activation) process proceeds via formation of an in-basal-plane split interstitial with a lattice Si atom, providing a metastable intermediate state (minimum) that rotates into an out-of-plane split configuration at the maximum barrier height, before 
Al occupies the lattice site, displacing Si to the interstitial position (Si$_\mathrm{I,c}$) in the cubic plane where finally an Al$_{Si}$Si$_I$ complex is formed.  
Deactivation enabling diffusion of Al$_I$ requires the reverse kick-out process. The kick-in barrier in the neutral state is significantly higher than the kick-out barrier shown in Table~\ref{tab:barrier}. Moreover, activation requires the dissociation of the (Al$_\mathrm{Si}$Si$_\mathrm{I}$) complex formed at the end of kick-in, combined with Si interstitial diffusion, which introduces an additional kinetic barrier. 
The formation and dissolution of the Al--Si complex starting from the Al interstitial in the cubic or hexagonal layer, in dependence of charge state, is discussed in more detail in the supplementary information in Figs. S15 to S16 and summarized in Table S7. The (Al$_\mathrm{Si}$Si$_\mathrm{I}$) complex forms a trapping state that decays via Al kick-out rather than via Si interstitial diffusion in the entire band gap. Consequently, direct kick-in via the Si lattice site is unlikely to be the dominant activation pathway during annealing.

For the Al interstitial in charge state $+3$, the lowest-energy diffusion pathway reported in the literature proceeds via cage-to-cage transitions within the basal plane\,\cite{huang2022}, connecting Al$_\mathrm{TC}$ and Al$_\mathrm{hex}$ configurations via Al$_\mathrm{TSi}$ or Al$_\mathrm{CSi}$ transition states, with a migration barrier of approximately $2.9$\,eV. Using the same pathway for the neutral Al interstitial in CI-NEB calculations resulted in a considerably lower barrier of $2.0$\,eV. The Al-SiC interactions have been fitted to the cage and the kick-in/out diffusion barriers just discussed.
The GW-Morse description reproduces the neutral DFT migration and activation barriers  within $\sim0.4$\,eV in Table~\ref{tab:barrier}.
\subsection{New diffusion path of the neutral Al interstitial}
\begin{figure*}[ht!]
\centering
     \subfloat[]{\includegraphics[width=0.5\linewidth]{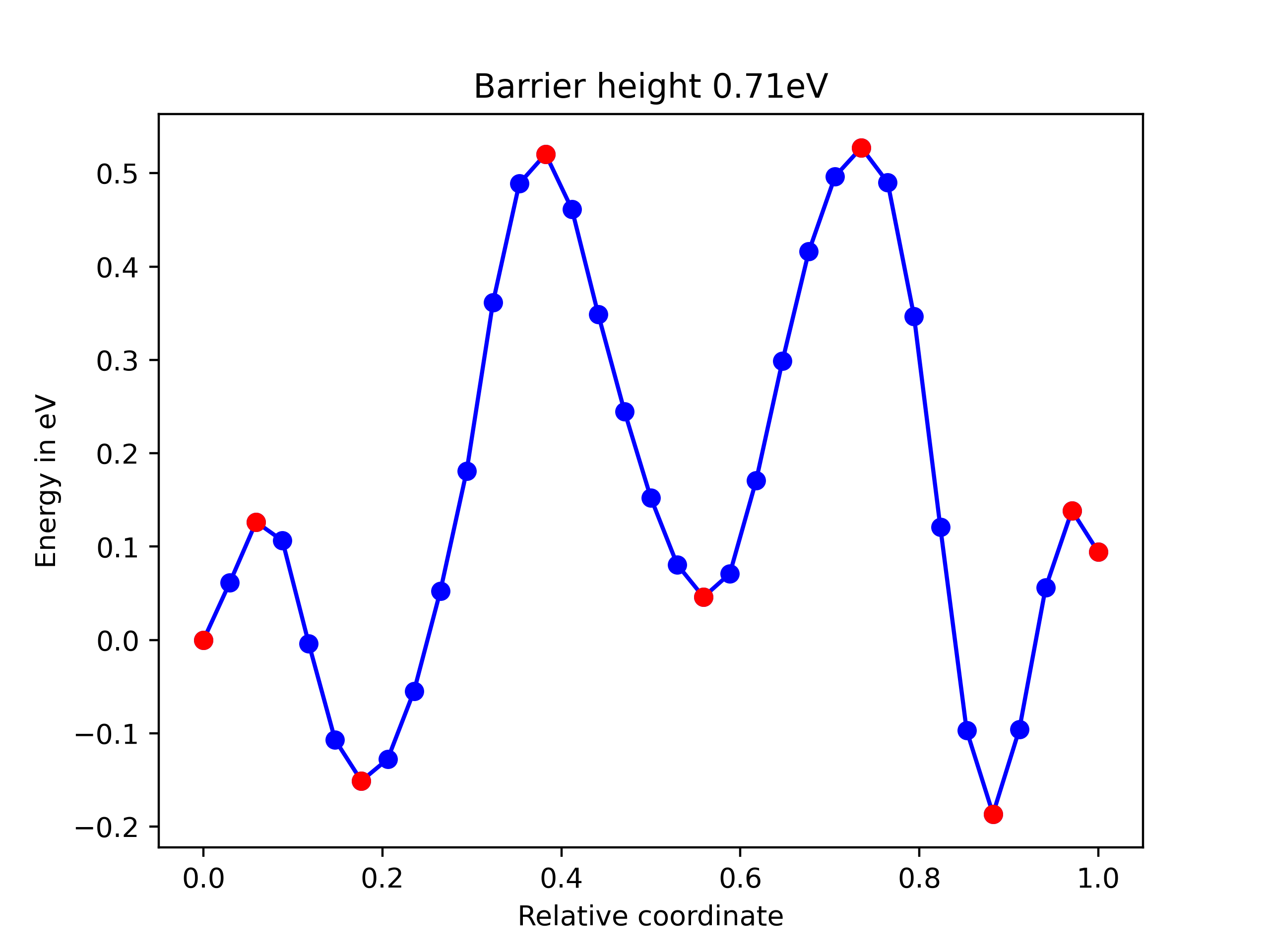}}\\
     \subfloat[]{\includegraphics[width=0.29\linewidth]{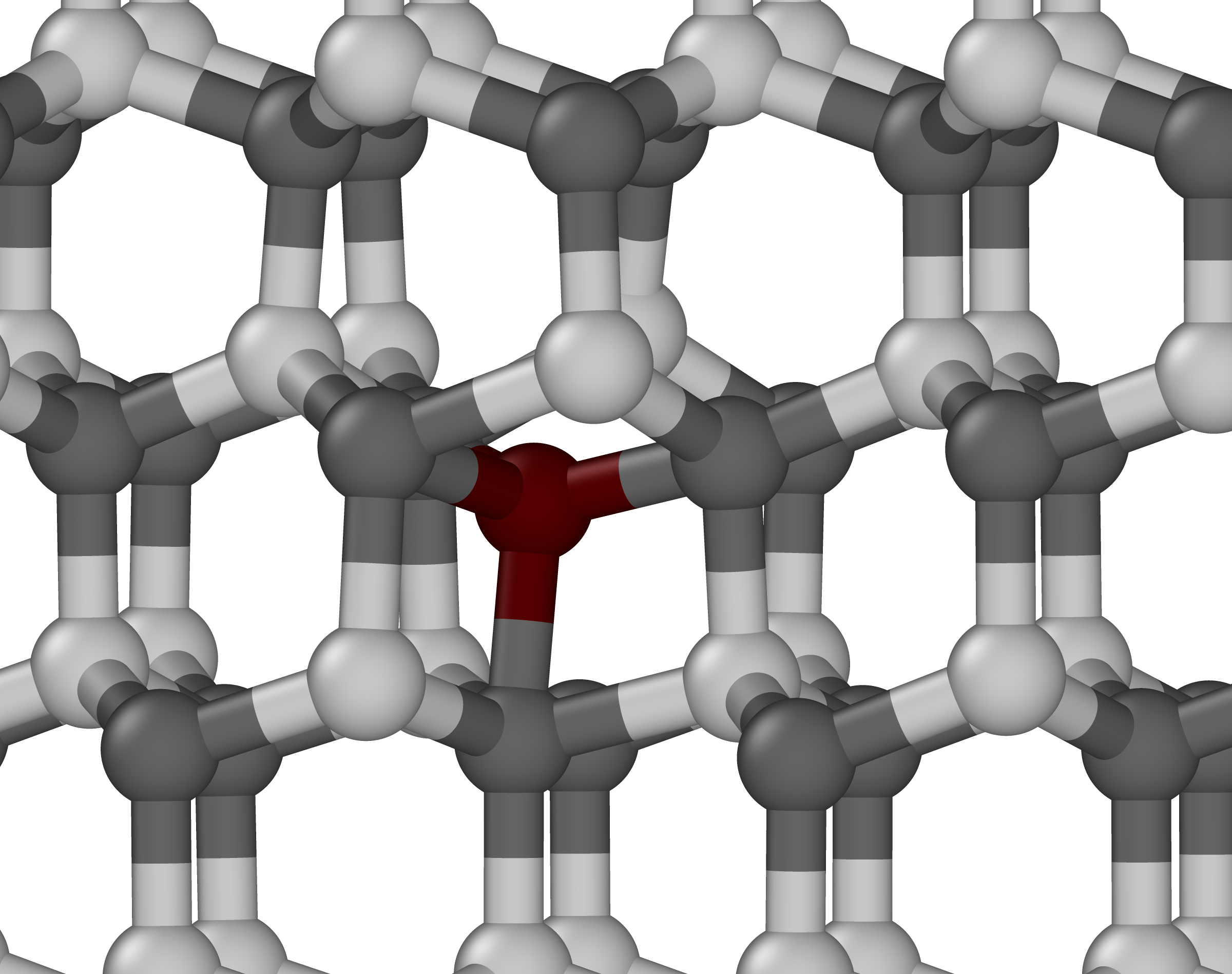}}\hspace{5pt}
     \subfloat[]{\includegraphics[width=0.29\linewidth]{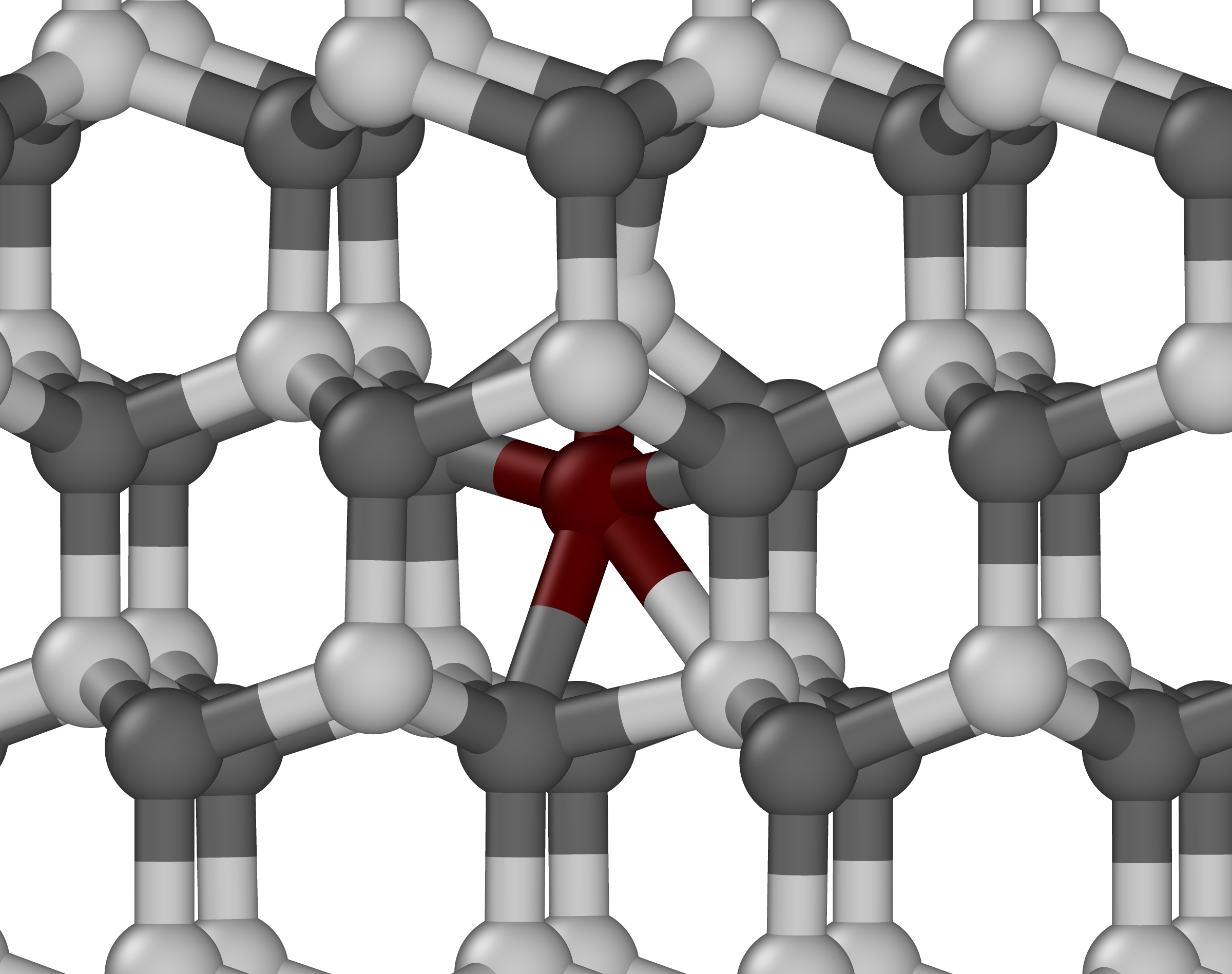}}\hspace{5pt}
     \subfloat[]{\includegraphics[width=0.29\linewidth]{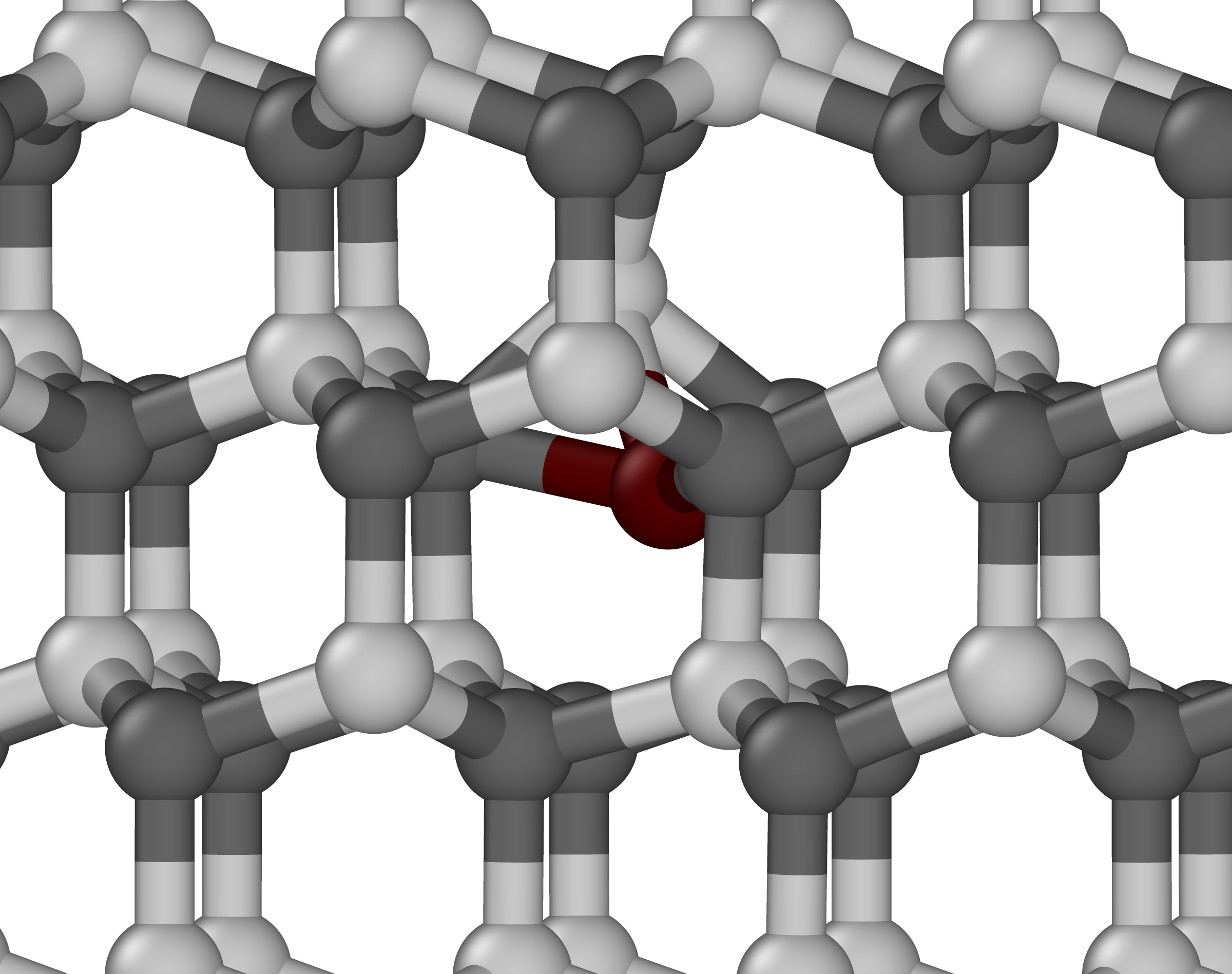}}\hspace{5pt}
\caption{(a) Basal plane diffusion path of the neutral Al interstitial via an (Al-Si) split configuration calculated applying CI-NEB DFT. Diffusion starts with a distorted Al$_\mathrm{TC}$ in its lowest energy configuration (b) (first minimum in (a) marked by a red dot). From there, the Al interstitial moves into a tilted split with a lattice Si atom, corresponding to the first maximum in the diffusion barrier (c). The split rotates until it lies in the basal plane and coincides with the $\langle 110 \rangle$ orientation (d), corresponding to the center minimum in (a). The Al interstitial rotates around the split and moves further in the basal plane to the neighbor cage on the right where it adopts a tilted split like in (c) again, until it finally returns to the Al$_\mathrm{TC}$ configuration corresponding to (b). Al shown in brown and host atoms: C (dark gray) and Si (light gray).}
\label{fig:barrier}
\end{figure*}

\begin{table*}[t]
\caption{Migration and kick-in/-out barriers of Al in 4H-SiC in dependence of charge state. DFT calculations from this work.}
\label{tab:barrier}
\centering
\begin{tabular}{lcccc}
\hline
\textbf{Process} & \textbf{GW/Morse (eV)} & \textbf{DFT (0) (eV)} & \textbf{DFT (+3) (eV)} & \textbf{DFT (+1) (eV)}\\
\hline
basal-plane diffusion (cage) & 1.6 & 2.0 & 2.9 & 2.4\\
diffusion (via split)        & 1.1 & 0.7 & 2.5   & 0.8\\
Al kick-in                   & 1.9 & 1.6 & 2.9 & 1.7\\
Al kick-out                  & 1.2 & 0.7 & 2.1 & 0.7\\
\hline
\end{tabular}
\end{table*}
After running our MD simulations we identified an alternative diffusion mechanism for the neutral Al interstitial involving transient split-interstitial formation with a lattice Si atom, followed by rotation of the split configuration as shown in Fig.~\ref{fig:barrier}. Unlike the kick-in process, this mechanism does not expel Si from its lattice site. Instead, Al is transferred to a neighboring cage, enabling both in-plane and out-of-plane diffusion with significantly lower barriers. The recalculated diffusion barrier using DFT-NEB yielded a value of $0.7$\,eV for the neutral Al shown in Table~\ref{tab:barrier}.
The two global minima in the diffusion barrier in Fig.~\ref{fig:barrier} can be attributed to a distorted configuration of the Al interstitial on the cubic site Al$_{TC}$. This effect arises from the three valence electrons of Al, which prefer to form bonds with three surrounding carbon atoms rather than four, as would occur at the highest-symmetry position. The local minimum in the center of the NEB curve is caused by the formation of an in-plane split interstitial ${\langle110\rangle}$(Al-Si)$_{Si}$. Therefore, diffusion in the cubic plane proceeds by alternating hopping between distorted Al$_{TC}$ configurations and the meta-stable (Al-Si) split \cite{matsushima2019}, similar to the diffusion path of boron in 3C-SiC \cite{Rurali2003}. 

For a Fermi level pinned above mid-gap, the carbon split interstitial in the positive and neutral charge states, as well as the Si split interstitial ${\langle110\rangle}$Si in the neutral state, are the most stable intrinsic point defects \cite{bockstedte2004} contributing to diffusion necessary for the recombination of FP during annealing. The Al interstitial exists in both positively and neutrally charged states with migration barriers in the $1+$ and $0$ charge state differing only slightly from each other, as shown in 
Table~\ref{tab:barrier}. Moreover, diffusion paths and barriers as a function of charge state for Al interstitials in the hexagonal and cubic planes are analyzed in the supplementary materials in Fig. S12 and summarized in Table S6. The results show that diffusion in the hexagonal plane in the intrinsic regime is significantly faster than in the cubic plane, where the migration barrier amounts to $1.0$\,eV for Al$_{I,h}^{2+}$. Moreover, as shown in Table~S6 and S7 of the supplementary information the reparameterized interactions of the GW-Morse potential provide a reliable representation of Al diffusion and recombination processes for Fermi levels around and above mid-gap for large-scale MD simulations.
\newpage
\subsection{Molecular Dynamics Simulation}
All MD simulations were performed using the LAMMPS software package\,\cite{thompson2022}, while OVITO was employed for subsequent visualization\,\cite{stukowski2009visualization}.
\subsubsection{Domain}
\begin{figure}[ht!]
\centering
     \subfloat[]{\includegraphics[width=0.6\linewidth]{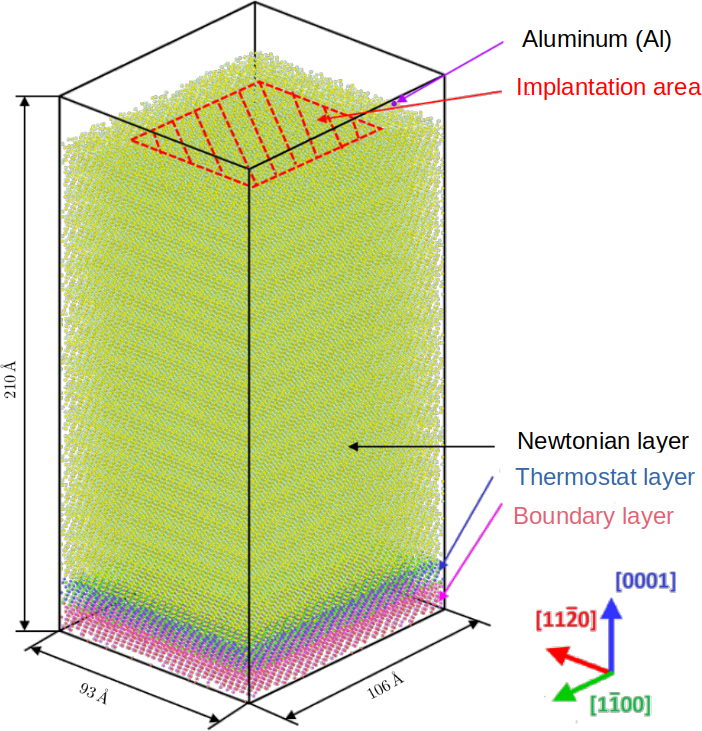}}
     \subfloat[]{\includegraphics[width=0.4\linewidth]{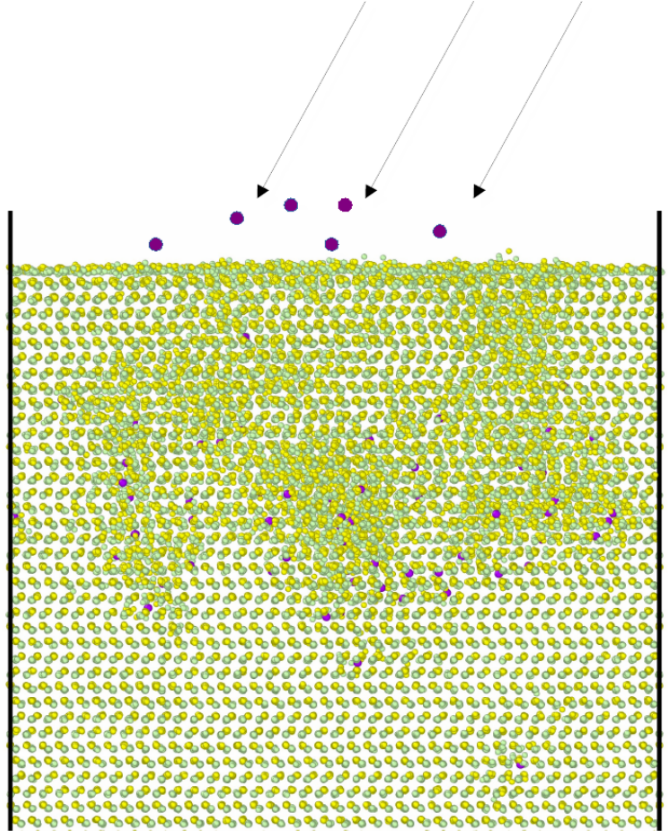}}
\caption{Molecular-dynamics (MD) implantation model setup.
(a) Simulation domain partitioned into a Newtonian (NVE) region, a thermostat region (heat sink), and a fixed boundary layer used during implantation.
(b) Snapshot of a $2$\,keV Al ion cascade in 4H-SiC for an incidence angle of $7^\circ$ relative to the surface normal ($\langle0001\rangle$ direction).}
\label{fig:setup}
\end{figure}

A cuboidal 4H-SiC simulation cell of dimensions $9$\,nm $\times$ $10$\,nm $\times$ $21$\,nm, as shown in Fig.~\ref{fig:setup}, was constructed by stacking hexagonal close-packed layers along the $\langle 0001 \rangle$ direction following the ABCB stacking sequence characteristic of 4H-SiC, where A and C represent hexagonal layers and B corresponds to cubic layers. Periodic boundary conditions were applied in the in-plane ($x$ and $y$) directions, while the surface normal ($z$ direction) remained non-periodic.

\subsubsection{Implantation}
The system was first equilibrated in the isothermal-isobaric (NPT) ensemble using a Nosé-Hoover thermostat and barostat to relax residual stresses. 
For implantation, the simulation domain was divided into three regions (Fig.~\ref{fig:setup}): a Newtonian region, a thermostat region, and a fixed boundary layer. Atoms in the Newtonian region evolved in the micro-canonical ensemble (NVE) to allow for localized temperature increases due to ion bombardment for realistic cascade dynamics. The thermostat layer acted as a heat sink to dissipate excess energy, while atoms in the bottom boundary layer were fixed to prevent rigid-body motion of the cell.
Al ions were implanted sequentially with a kinetic energy of $2$\,keV at an incidence angle of $7^\circ$ relative to the surface normal of the $(0001)$ plane to avoid channeling. The implantation area shown in red in Fig.~\ref{fig:setup}(a) covered $8$\,nm $\times$ $8$\,nm, with ion impact points uniformly distributed within the surface plane. After each implantation event, the system was allowed to relax for 10\,ps to enable recombination of closely spaced Frenkel pairs and to allow the implantation-induced thermal spike to dissipate and the local temperature to return to the equilibrium wafer temperature.
Electronic stopping with an adaptive timestep was included via a velocity-proportional damping scheme, transferring a fraction of the ion kinetic energy to a fictional electron reservoir to avoid artificial overheating.
In total, seven implantation doses were applied: 
\mbox{$1\times10^{13}$}\,cm$^{-2}$, 
\mbox{$5\times10^{13}$}\,cm$^{-2}$, 
\mbox{$1\times10^{14}$}\,cm$^{-2}$, 
\mbox{$1.5\times10^{14}$}\,cm$^{-2}$,
\mbox{$2.5\times10^{14}$}\,cm$^{-2}$, 
\mbox{$5\times10^{14}$}\,cm$^{-2}$ and 
\mbox{$7.5\times10^{14}$}\,cm$^{-2}$. Two implantation temperatures, $500$\,K and $900$\,K, were selected following Ref.\,\cite{Michaud2013,ZANG2025}.

\subsubsection{Annealing}
After implantation, structures were equilibrated at $300$\,K in the NPT ensemble for $100$\,ps and subsequently heated to the target annealing temperature at a rate of $80$\,K/ps.
Annealing temperatures between $1500$\,K and $2500$\,K were investigated.
The annealing stage was carried out in the NPT ensemble for $100$\,ns to capture defect recombination, clustering, and dopant incorporation kinetics. 
Finally, samples were cooled to $300$\,K at $80$\,K/ps and equilibrated for an additional $100$\,ps prior to defect analysis.

\subsection{Defect Characterization}

Defect structures were analyzed using the Identify Diamond Structure (IDS) algorithm implemented in OVITO, in combination with a custom post-processing routine.
The IDS algorithm identifies atoms deviating from hexagonal
or cubic diamond coordination, which typically correspond to interstitials or atoms within locally amorphous regions.
In order to eliminate the highly defective surface in the upper $6$\,\AA\, of the crystal, the defect evaluation is limited to the bulk region. 

Antisites were identified by examining the four nearest neighbors of each lattice atom. A lattice site was classified as an antisite if it was occupied by an atom whose four nearest neighbors were all of the same chemical species as the atom itself.

Vacancy identification is more challenging, particularly within clusters or at crystalline-amorphous interfaces. Removal of an atom produces dangling bonds on neighboring lattice atoms of the opposite species, forming a cage-like configuration. Applying the coordination analysis in OVITO \cite{stukowski2009visualization}, an isolated vacancy was detected by identifying four threefold-coordinated atoms of identical species forming such a cage. For clustered or distorted environments, the presence of at least three dangling-bond atoms of identical species was used as a practical identification criterion. This heuristic preserves approximate equality between the total number of vacancies and interstitials expected from implantation-induced Frenkel-pair generation.

To distinguish isolated point defects from extended defect clusters, a connectivity-based clustering algorithm was implemented. Defects separated by less than the nearest-neighbor distance deduced from partial pair correlation functions were grouped into the same cluster. Cluster size was defined as the total number of constituent defects belonging to a connected defect network.

\section{Results and Discussion}
\label{sec:Results}

\subsection{As-implanted structures}
\begin{figure*}[ht!]
\centering
     \subfloat[]{\includegraphics[width=0.35\linewidth]{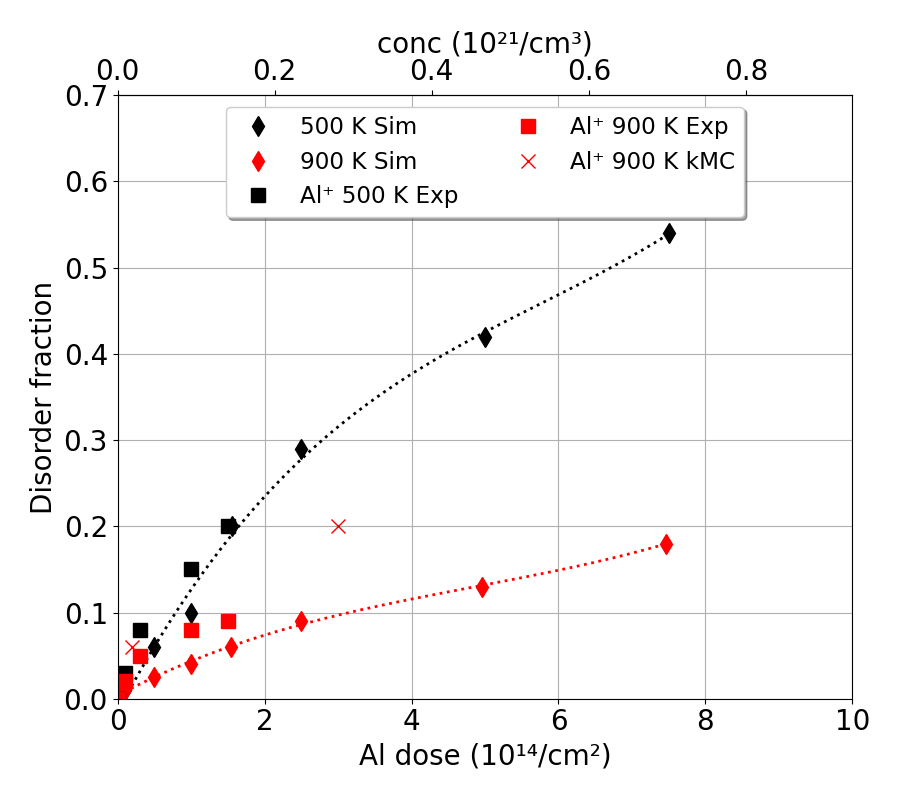}}
     \subfloat[]{\includegraphics[width=0.32\linewidth]{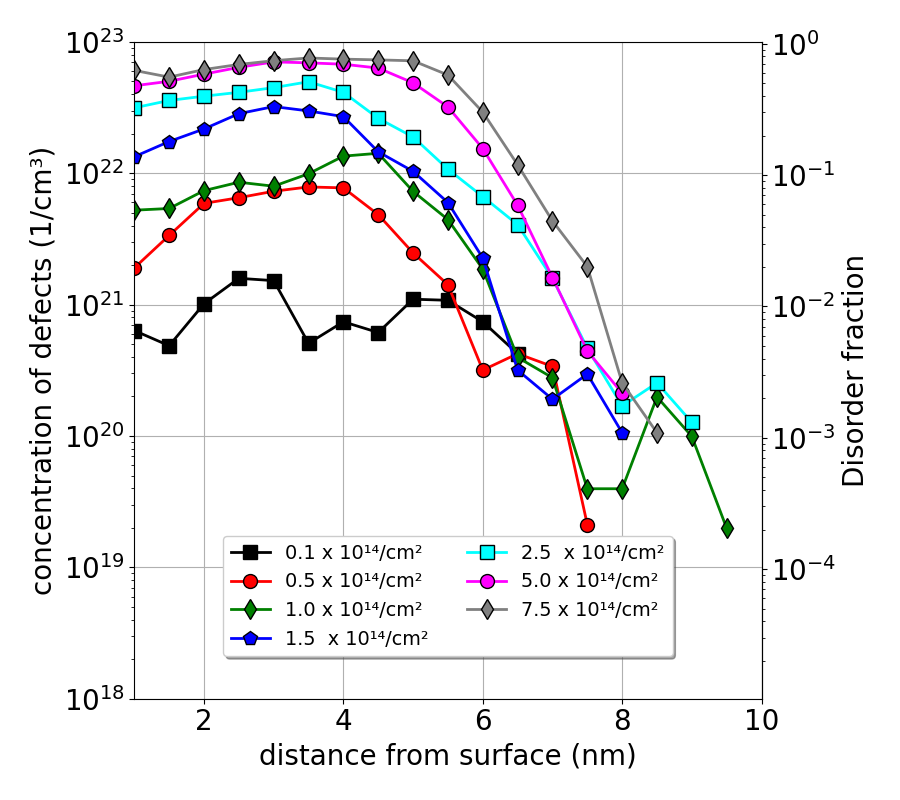}}
     \subfloat[]{\includegraphics[width=0.32\linewidth]{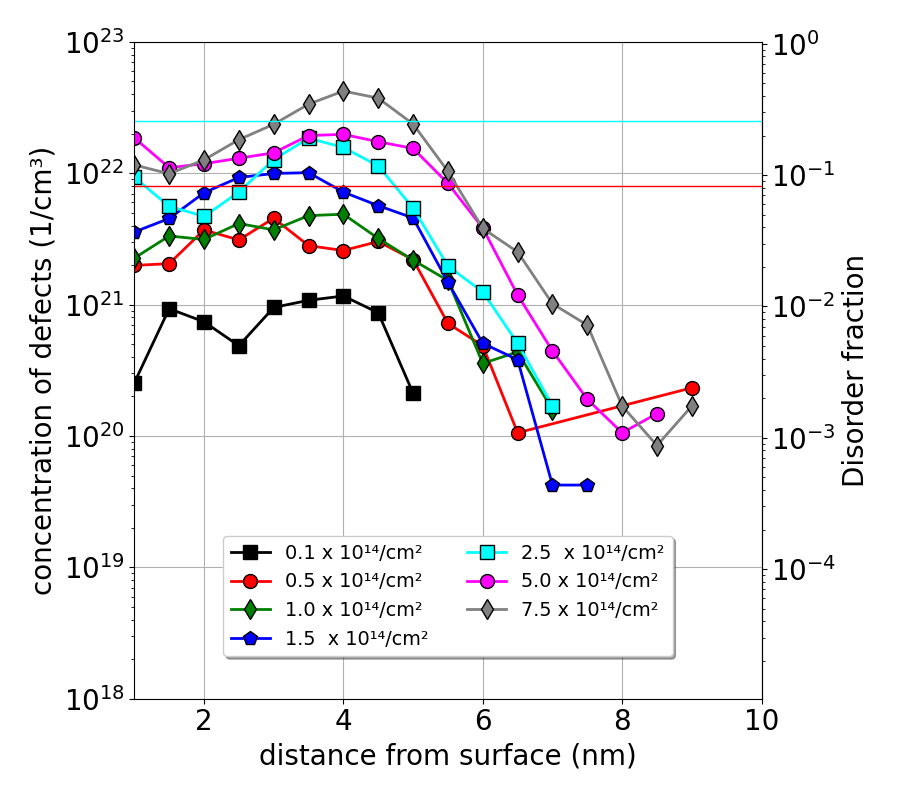}}
\caption{(a) Disorder fraction relative to the perfect crystal in the implanted volume, evaluated at half-width of the defect concentration profiles as a function of implantation parameters. Results are compared with experimental data from Hallen \emph{et al.}\,\cite{hallen2016} at an ion energy of $100$\,keV (squares); dotted lines serve as guides. (b-c) IDS defect concentration profiles immediately after implantation at $500$\,K (b) and $900$\,K (c) as a function of dose (from simulation). Solid horizontal lines indicate plateau concentration values from kinetic Monte Carlo simulations\,\cite{muting2020} for Al concentrations of $0.3$ and $3\times$ $10^{20}$/\,cm$^{-3}$ and Al implantation energies of $180$\,keV.} 
\label{fig:disorder_conc}
\end{figure*} 

As we have previously demonstrated\,\cite{leroch2023}, implantation of 4H-SiC at elevated temperatures leads to a pronounced reduction of disordered and amorphous regions compared to implantation at lower temperatures (near room temperature). 

This behavior, summarized in Fig.~\ref{fig:disorder_conc}(a) for the present implantation conditions, is consistent with experimental observations\,\cite{hallen2016} and further confirms the applicability of the present interaction potential. In this figure, black symbols denote disorder fractions obtained from simulations at an implantation temperature of $500$\,K, while red symbols correspond to implantation at $900$\,K. The disorder fraction is defined as the ratio of off-lattice atoms within a volume centered at the damage peak to the number of atoms occupying the same volume in the perfect crystal. Moreover, the MD results are compared with experimental data\,\cite{hallen2016} and kinetic Monte Carlo (kMC)  simulations\,\cite{muting2020}. 
To enable a meaningful comparison between simulations and experiments performed at different implantation energies, data are aligned using the implanted Al concentration near the damage peak rather than the nominal implantation dose. A comparison based solely on Al dose would yield different damage profiles because implantation depth and straggle depend on ion energy. In contrast, systems exhibiting similar Al plateau concentrations at the damage peak are expected to show comparable levels of local disorder.
Accordingly, the upper abscissa in Fig.~\ref{fig:disorder_conc}(a) indicates the corresponding experimental peak Al concentrations and the average Al concentration from MD deduced by dividing the number of defects by the implantation volume, assuming an average implantation depth of $9$\,nm. 

At low implantation doses (below $1\times 10^{14}$\,cm$^{-2}$), the substrate temperature has only a minor influence on the accumulated disorder. Even within the short MD time scale of $10$\,ps per cascade, closely spaced Frenkel pairs (FPs) can recombine efficiently during implantation, limiting defect accumulation. With increasing dose, the disorder fraction increases monotonically and follows the experimentally observed trend. Quantitatively, implantation at $900$\,K consistently yields lower disorder fractions than at $500$\,K, reflecting enhanced dynamic defect recovery at elevated temperature.
Fig.~\ref{fig:disorder_conc}(b,c) shows depth-resolved IDS defect concentration profiles immediately after implantation for $500$\,K (b) and $900$\,K (c). For comparison, plateau values obtained from kinetic Monte Carlo simulations\,\cite{muting2020} are included as solid horizontal lines in part (c) of the figure. Profiles and plateau values of identical color correspond to the same peak Al concentration.
The binary collision model (BCM)-based defect concentrations are systematically higher than those obtained from MD. This discrepancy is expected, since BCMs used in kMC do not account for the near-instantaneous recombination of closely spaced FPs during displacement cascades. In contrast, MD explicitly resolves cascade dynamics and therefore captures intracascade recombination, leading to lower surviving defect densities. The overestimation of defects in BCM relative to MD amounts to approximately a factor of $2$-$3$.
A critical disorder level of roughly $70\%$ marks the onset of amorphous inclusion formation in 4H-SiC\,\cite{zhang2002}. At $500$\,K, implantation doses corresponding to average Al concentrations above $2\times10^{20}$\,cm$^{-3}$ lead to the formation of a nanoscale amorphous pocket  around the damage peak at depths between $3$\,nm and $5$\,nm.

In contrast, implantation at $900$\,K maintains a largely crystalline structure across the implanted region, even at high doses (Fig.~\ref{fig:disorder_conc}(c)).

These observations confirm that elevated implantation temperature suppresses amorphization through dynamic defect recombination. However, as discussed in the following section, the reduction of immediate disorder does not necessarily translate into improved dopant activation after annealing.

\subsection{Annealed structures}

\subsubsection{Dose and temperature dependence}
\begin{figure*}[hbtp]
\centering
  \hfill
  \makebox[0.32\textwidth]{$\mathbf{5\times10^{13}}$\,\textbf{cm}$\mathbf{^{-2}}$} \hfill
  \makebox[0.32\textwidth]{$\mathbf{2\times10^{14}}$\,\textbf{cm}$\mathbf{^{-2}}$} \hfill
  \makebox[0.32\textwidth]{$\mathbf{5\times10^{14}}$\,\textbf{cm}$\mathbf{^{-2}}$} \hfill
  \vspace{-10pt}

  \rotatebox{90}{\makebox[0.28\textwidth]{\textbf{500\,K}}} \hspace{-5pt}
  \subfloat[]{\includegraphics[width=0.31\textwidth]{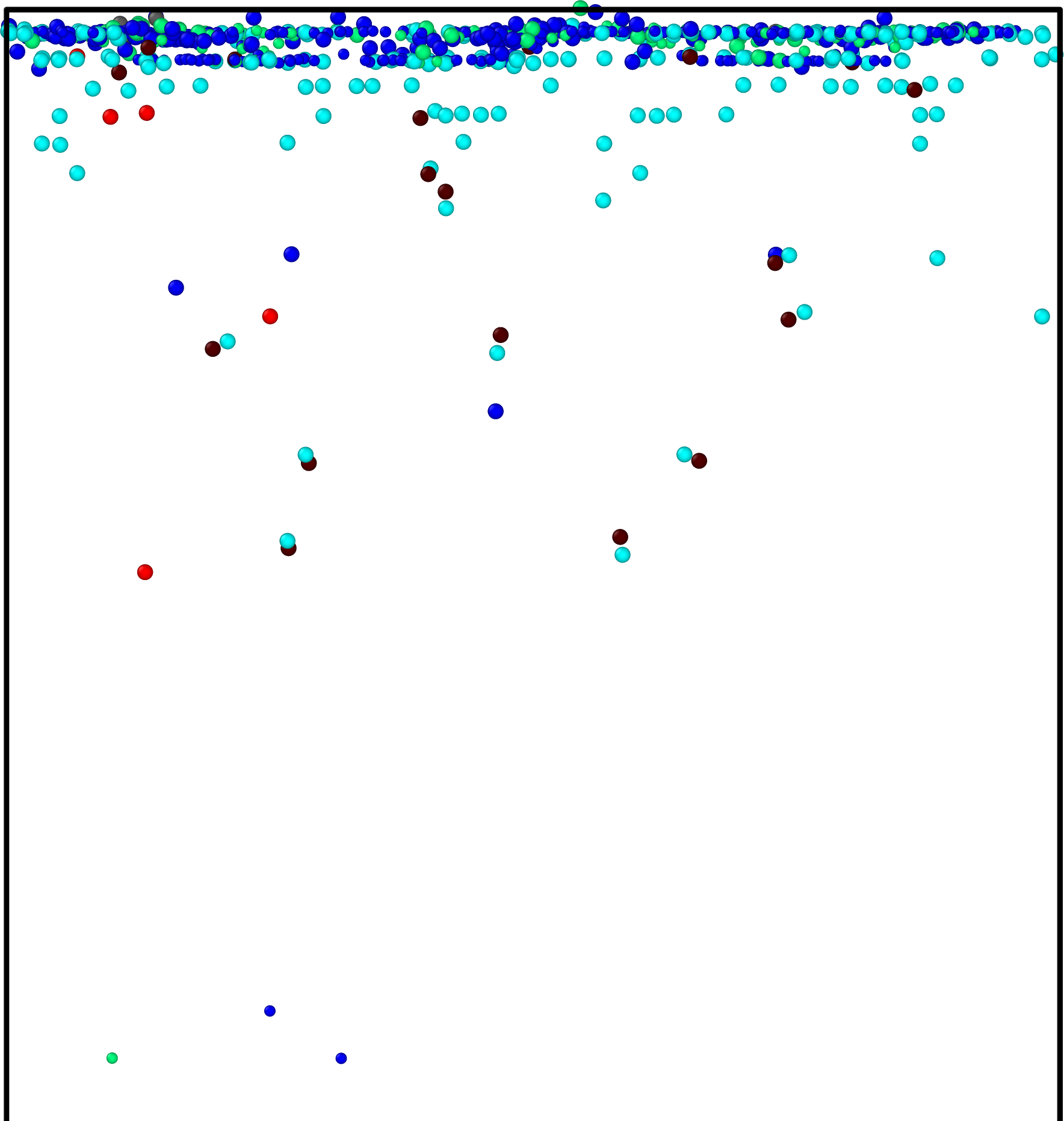}} \hfill
  \subfloat[]{\includegraphics[width=0.31\textwidth]{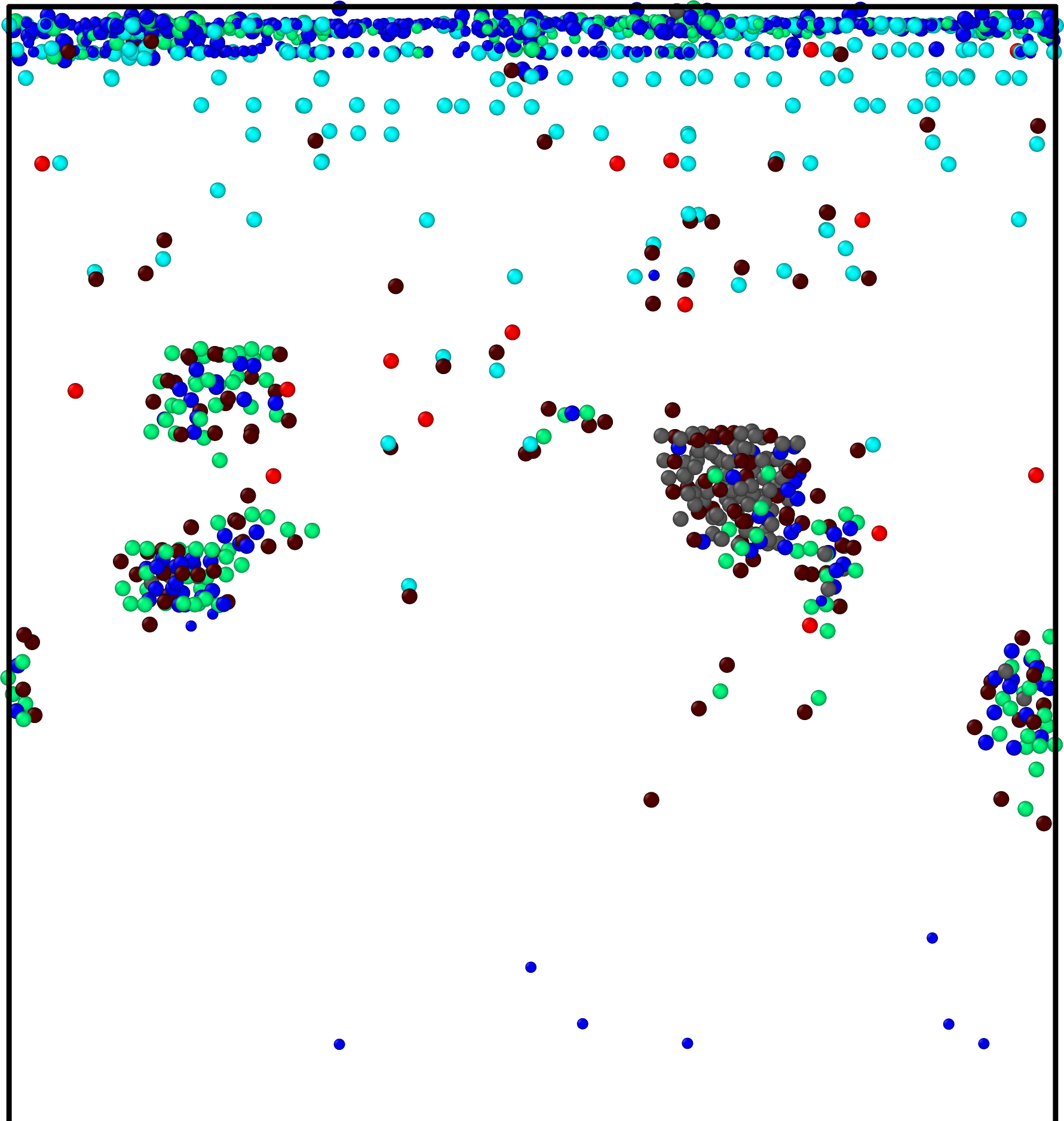}} \hfill
  \subfloat[]{\includegraphics[width=0.31\textwidth]{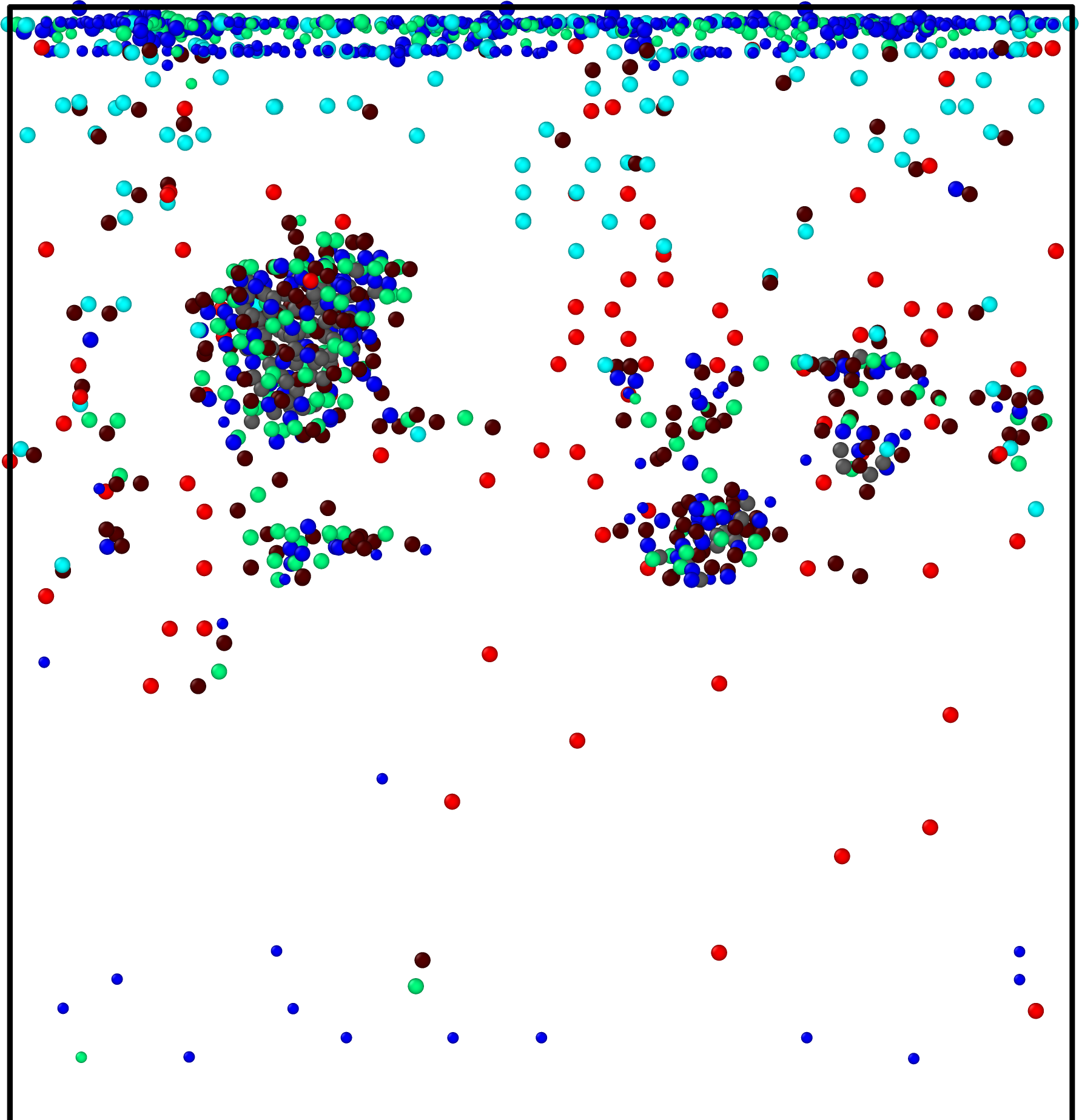}} \\

  \rotatebox{90}{\makebox[0.28\textwidth]{\textbf{900\,K}}} \hspace{-5pt}
  \subfloat[]{\includegraphics[width=0.31\textwidth]{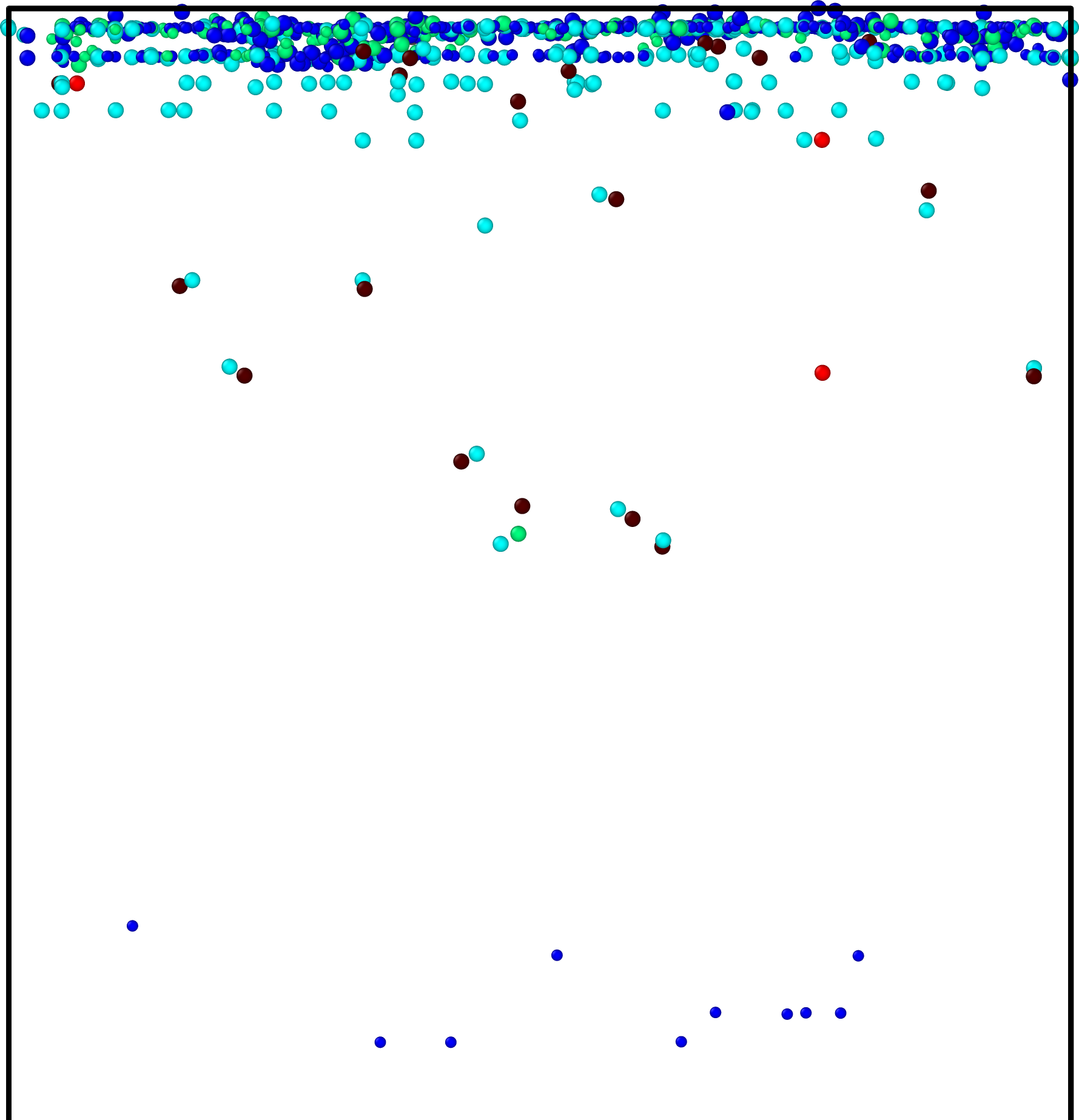}} \hfill
  \subfloat[]{\includegraphics[width=0.31\textwidth]{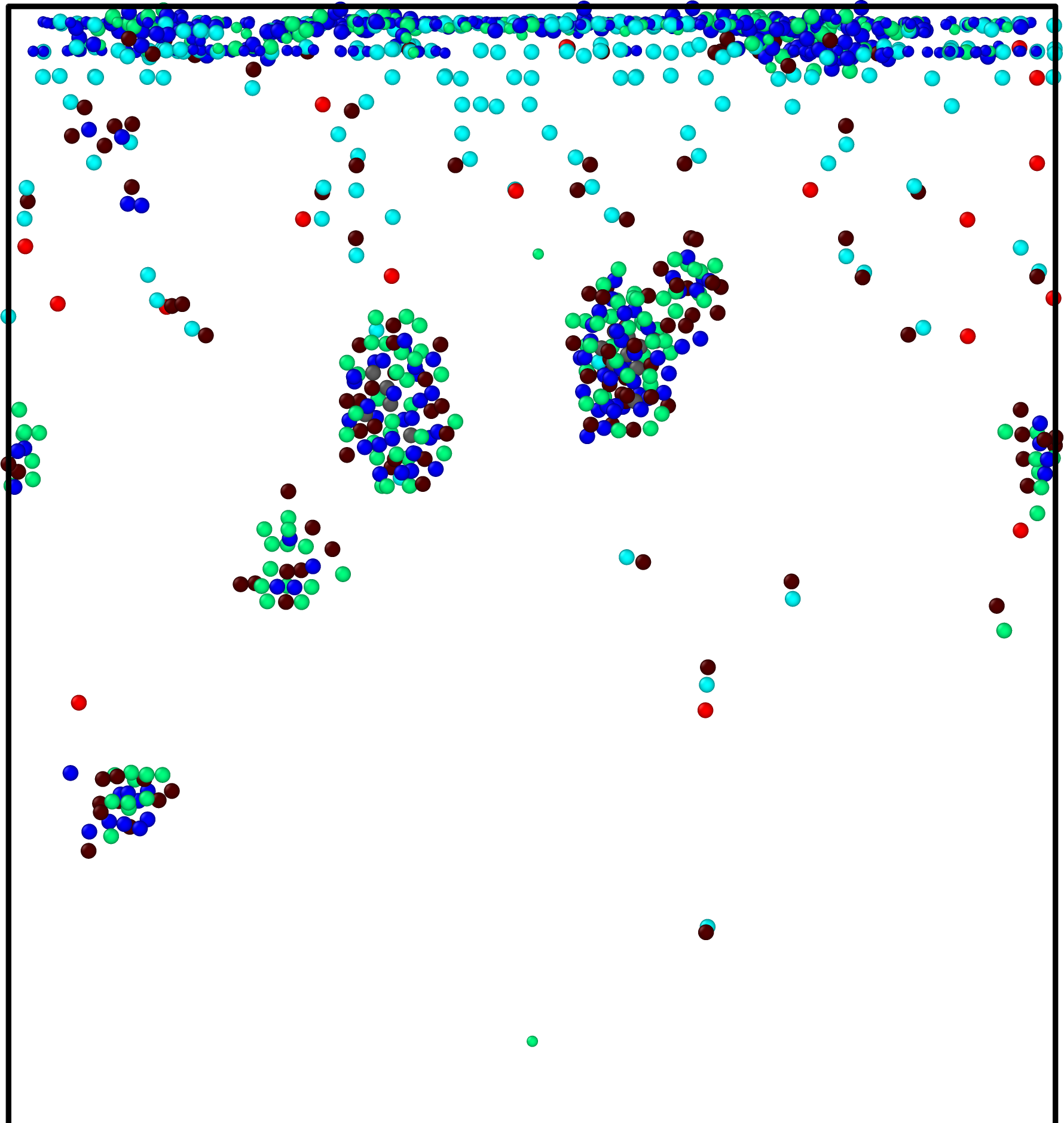}} \hfill
  \subfloat[]{\includegraphics[width=0.31\textwidth]{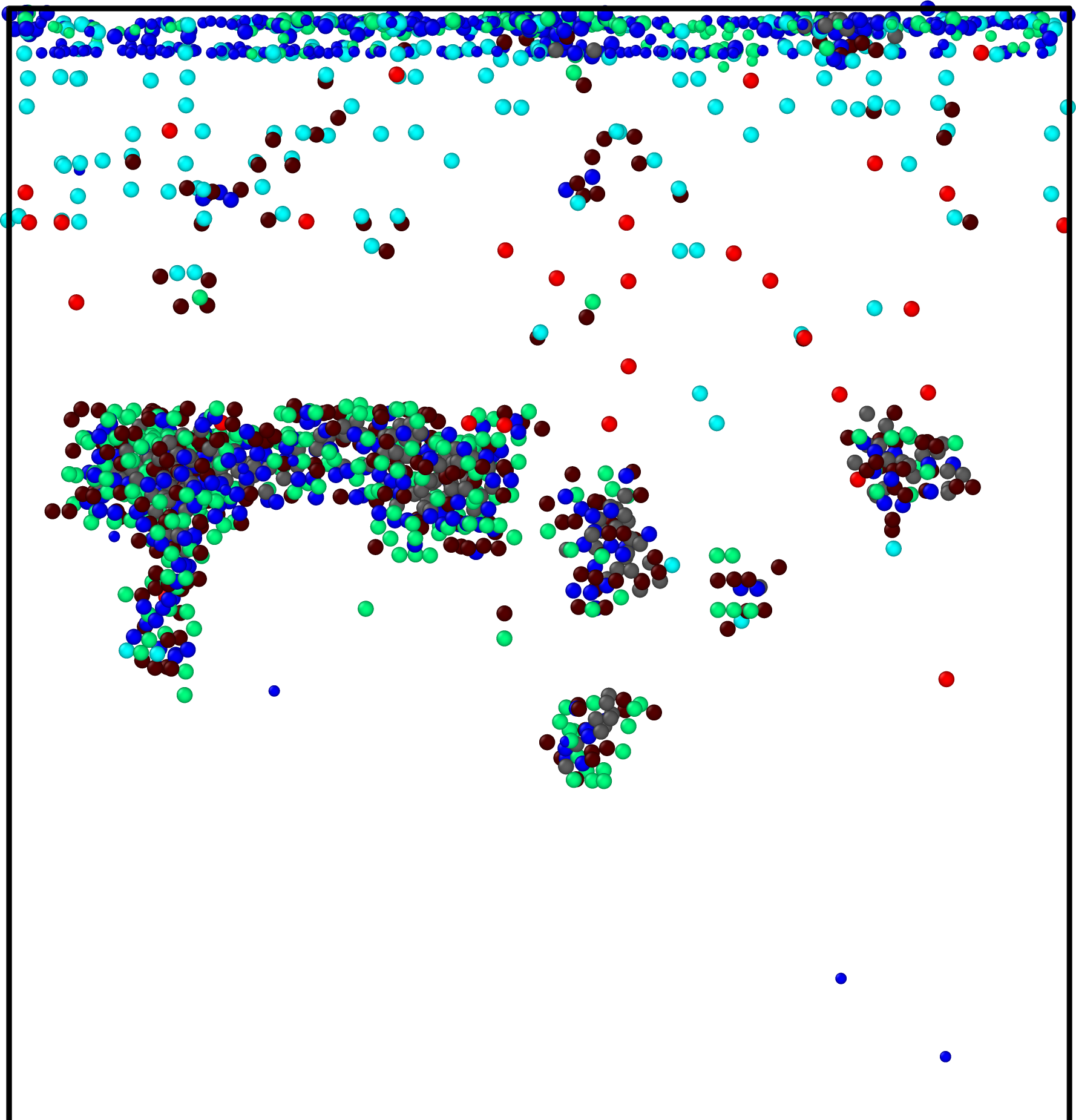}}

\caption{Representative defect morphologies after annealing at $2350$\,K (100\,ns) for three implantation doses and two implantation temperatures.
(a-c) Implantation at $500$\,K and (d-f) at $900$\,K for doses $5\times10^{13}$, $2\times10^{14}$, and $5\times10^{14}$\,cm$^{-2}$ (left to right).
Color code: gray = locally amorphous/disordered atoms (IDS), green = Si-related intrinsic defects, cyan/blue = C-related intrinsic defects, brown = Al interstitial, red = lattice-site Al counted as chemical activated (see text for activation definitions). Interstitials and vacancies are shown as enlarged and reduced spheres, respectively.}
\label{fig:defect_structures_dose_2350K}
\end{figure*}
In Fig.~\ref{fig:defect_structures_dose_2350K}, final defect structures after annealing at
$2350$\,K for $100$\,ns are shown for three selected doses:
(a,d) \mbox{$5\times10^{13}$}\,cm$^{-2}$,
(b,e) \mbox{$2\times10^{14}$}\,cm$^{-2}$, and
(c,f) \mbox{$5\times10^{14}$}\,cm$^{-2}$.
The lowest dose in (a,d) is well below the Al solubility limit in SiC of $\sim\,2\times 10^{20}$\,cm$^{-3}$, the intermediate dose (b,e) is close to this
 limit, and the highest dose (c,f) corresponds to strong oversaturation.
 
At the lowest dose, the residual damage after annealing consists primarily of isolated point defects and small defect complexes, with no pronounced dependence on implantation temperature. Near the solubility limit, numerous compact (globular) defect clusters remain after $100$\,ns annealing for both implantation temperatures, indicating incomplete recovery within MD-accessible time scales. At the highest dose, the implantation temperature strongly affects the resulting defect morphology. 
Comparing Fig.~\ref{fig:defect_structures_dose_2350K}(c) and Fig.~\ref{fig:defect_structures_dose_2350K}(f) shows that implantation at $900$\,K 
promotes the formation of an extended defect cluster in the basal plane and simultaneously reduces the fraction of chemically activated Al (red). Such extended clusters arise from agglomeration of (Al-Si-C) interstitials and/or attachment to larger planar defects such as stacking faults, forming efficient sinks for mobile species during annealing.

In summary, the effect of implantation temperature on the post-anneal microstructure reverses with increasing dose. Below the Al solubility limit, temperature has little influence on the residual defect state, whereas above it, elevated implantation temperature drives extended cluster nucleation and impedes recovery of the crystalline lattice, with direct consequences for dopant activation discussed below.

\subsubsection{Depth-resolved point defects and clustering}
\begin{figure*}[hbtp]
\centering
  \hfill
  \makebox[0.45\textwidth]{\textbf{Carbon-related Defects}} \hfill
  \makebox[0.45\textwidth]{\textbf{Al-related Defects}} \hfill \\
  \vspace{-10pt}

  \rotatebox{90}{\makebox[0.45\textwidth]{\textbf{500\,K}}} \hspace{-2pt}%
  \subfloat[]{\includegraphics[width=0.46\textwidth]{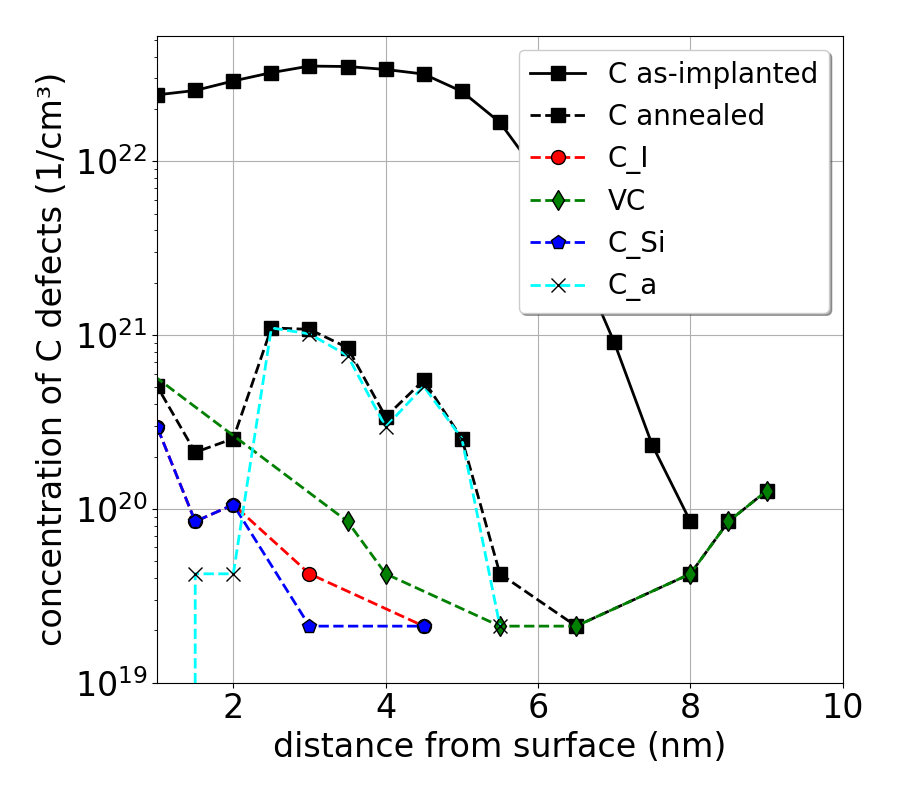}} \hfill
  \subfloat[]{\includegraphics[width=0.46\textwidth]{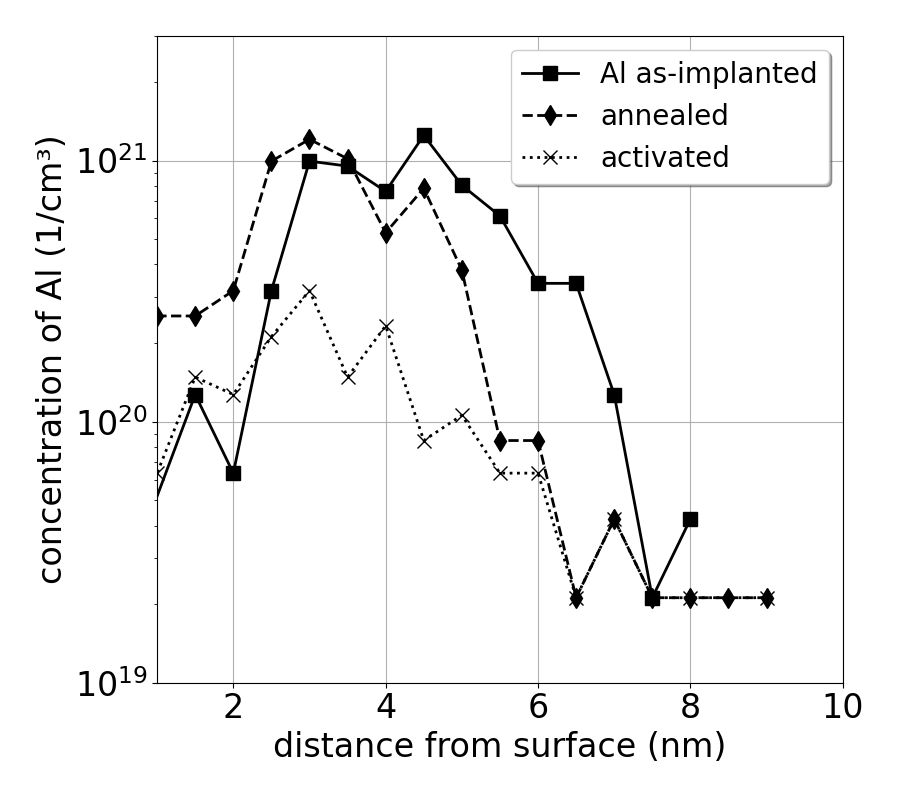}} \\

  \rotatebox{90}{\makebox[0.45\textwidth]{\textbf{900\,K}}} \hspace{-2pt}%
  \subfloat[]{\includegraphics[width=0.46\textwidth]{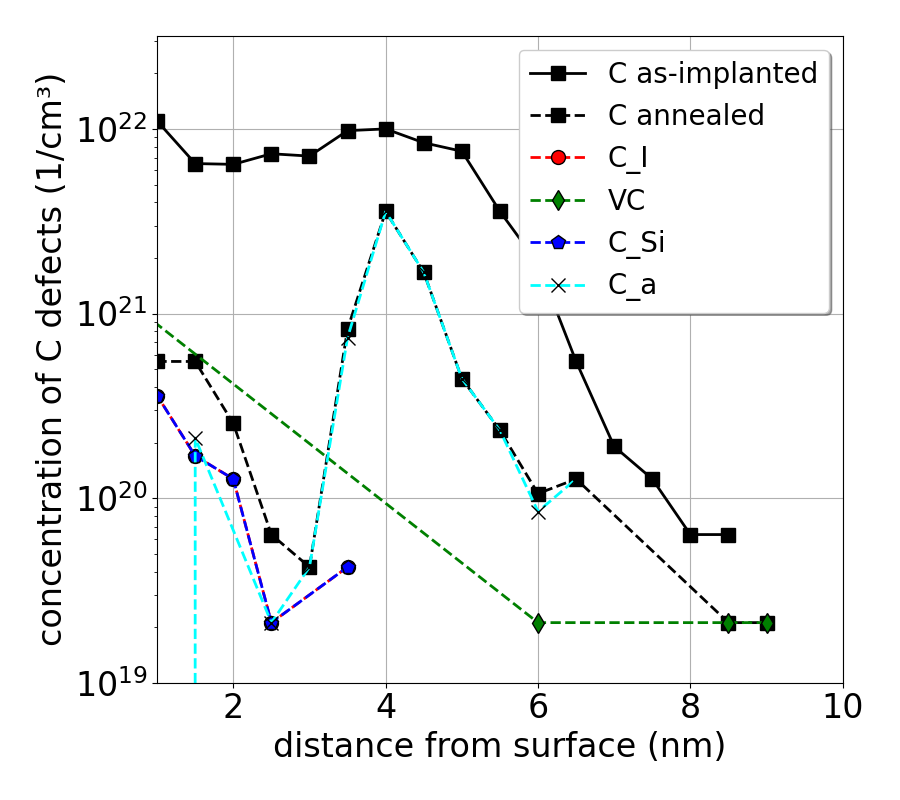}} \hfill
  \subfloat[]{\includegraphics[width=0.46\textwidth]{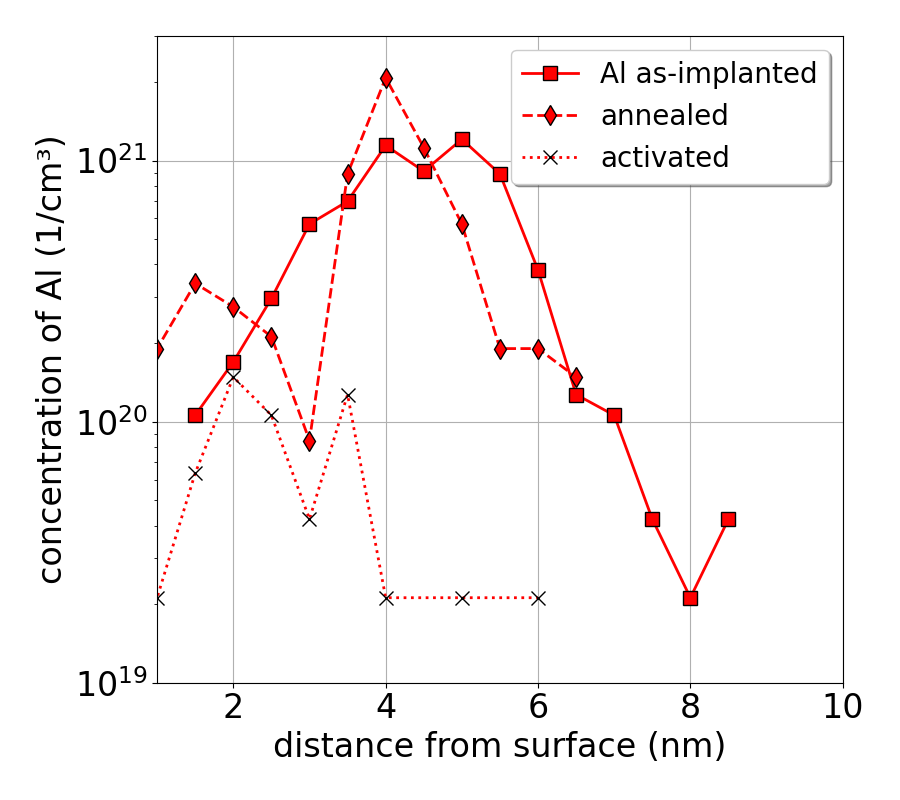}}

\caption{Depth-resolved defect concentrations for a dose of $5\times10^{14}$\,cm$^{-2}$.
(a,b) Implantation at $500$\,K; (c,d) implantation at $900$\,K.
Left column: C-related defects; right column: Al-related defects.
Solid curves: as-implanted; dashed curves: after annealing at $2350$\,K for 100\,ns.}
\label{fig:defect_histograms}
\end{figure*}
Depth-resolved defect concentrations for a representative high dose (\mbox{$5\times10^{14}$}\,cm$^{-2}$) are shown in Fig.~\ref{fig:defect_histograms} for implantation at $500$\,K and $900$\,K. Most intrinsic point defects (e.g., interstitials, C$_\mathrm{Si}$ and V$_\mathrm{C}$) are concentrated near the surface or diffuse beyond the primary implantation depth of around $8$\,nm. 
Despite their comparatively high diffusion barrier, Fig.~\ref{fig:defect_structures_dose_2350K}  indicates carbon vacancies (V$_\mathrm{C}$) in deeper regions outside the implanted zone, although they are absent immediately after implantation. Similar observations have been reported experimentally, where V$_\mathrm{C}$ diffusion at high temperature has been discussed as a possible explanation\,\cite{Wellmann2022}. DFT predicts a V$_\mathrm{C}$ migration barrier of $\sim\,3.3$\,eV at $0$\,K\,\cite{yan2020}; including entropic contributions may reduce the effective barrier to approximately $2.5$\,eV at elevated temperature\,\cite{Ayedh2017,Grossner2021}, which could enable V$_\mathrm{C}$ motion on the $100$\,ns time scale during high temperature annealing.

At these doses (\mbox{$5\times10^{14}$}\,cm$^{-2}$), interstitials predominantly agglomerate into clusters centered near the damage peak. Importantly, while the as-implanted defect density is higher at $500$\,K, the post-anneal state exhibits the opposite trend, where implantation at $900$\,K yields larger interstitial agglomerates that are kinetically more persistent, as they do not dissolve within the $100$\,ns annealing window.

\begin{figure*}[ht!]
\centering
    \subfloat[]{\includegraphics[width=0.32\textwidth]{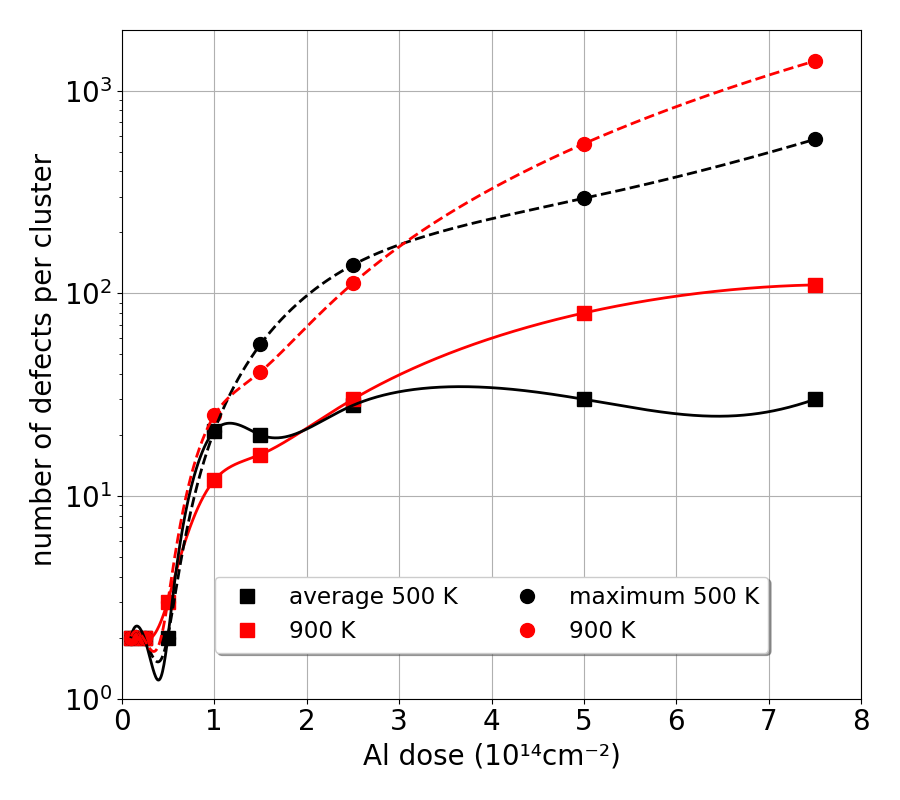}} \hfill
    \subfloat[]{\includegraphics[width=0.33\textwidth]{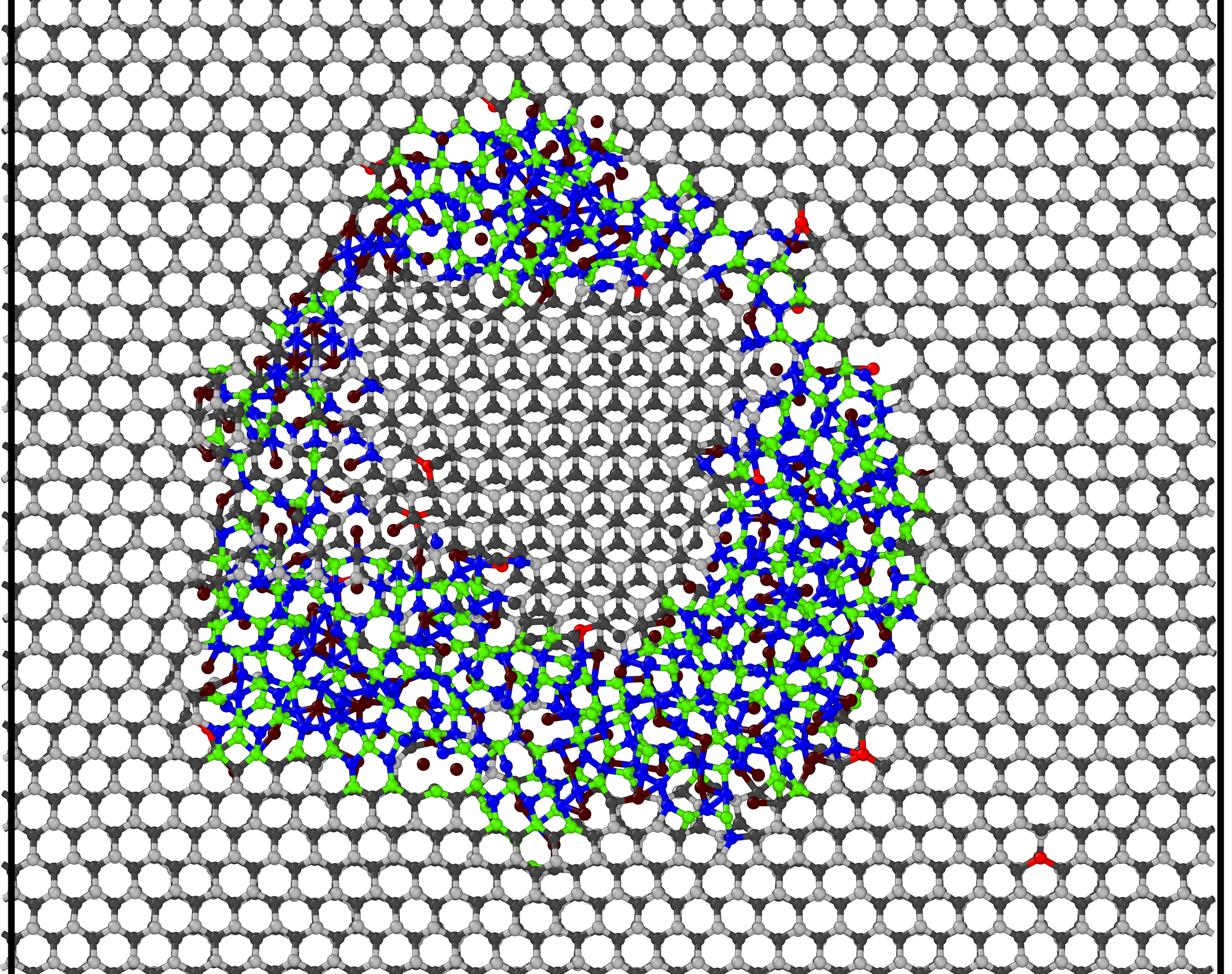}}\hfill
    \subfloat[]{\includegraphics[width=0.33\textwidth]{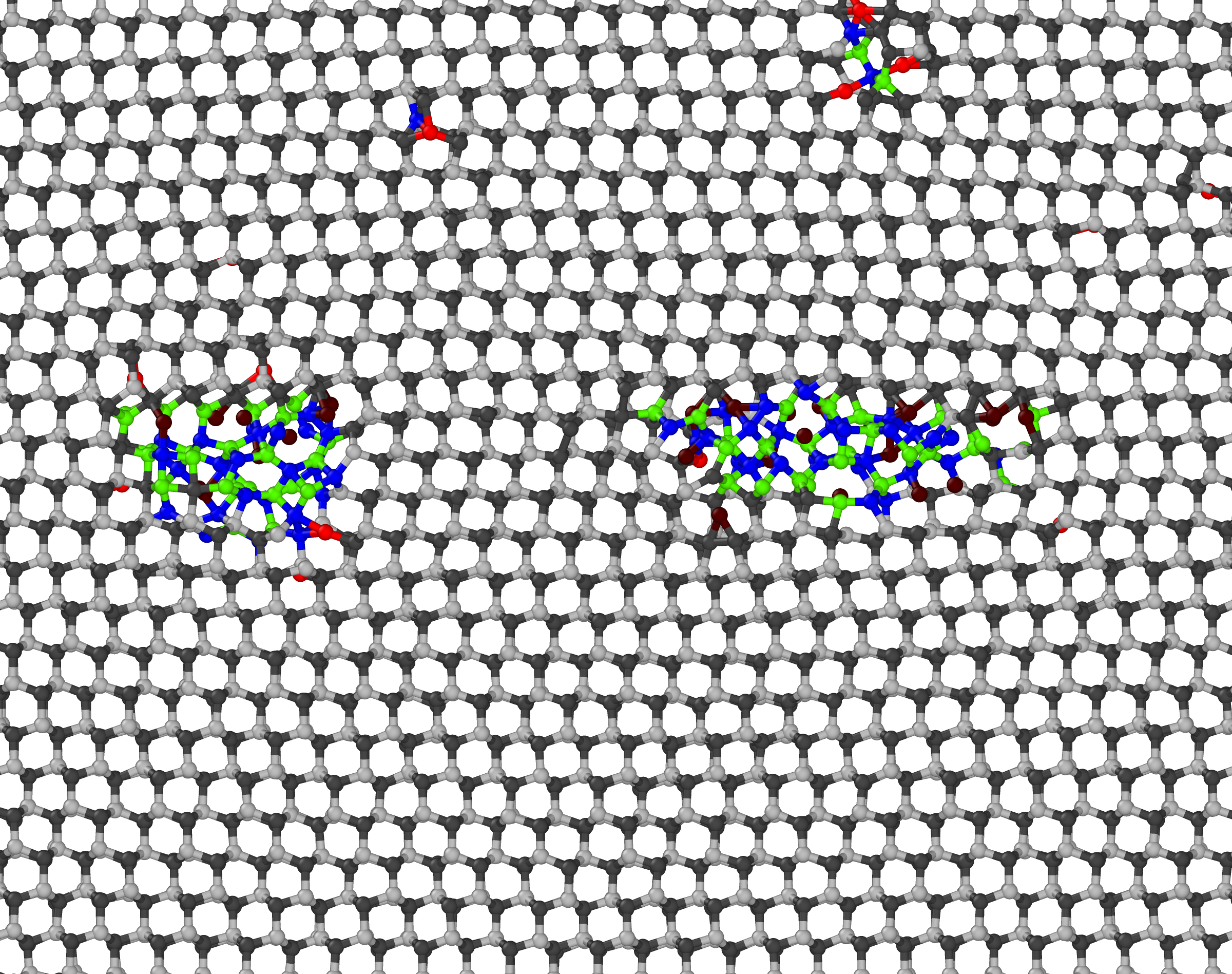}}
\caption{Cluster statistics and extended defect formation after annealing.
(a) Average (solid) and maximum (dashed) cluster size (number of defects per cluster) after annealing at $2350$\,K (100\,ns) as a function of implantation dose for implantation at $500$\,K and $900$\,K.
(b,c) Example Frank loop (diameter $\sim5$\,nm) formed after annealing for a dose of \mbox{$7.5\times10^{14}$}\,cm$^{-2}$ implanted at $900$\,K, shown (b) in the basal plane and (c) along the c-axis.}
\label{fig:cluster_size}
\end{figure*}

Consistent with Fig.~\ref{fig:defect_histograms}, implantation at $900$\,K produces substantially larger interstitial clusters after annealing (Fig.~\ref{fig:cluster_size}(a)). Both the average and maximum cluster sizes increase strongly with dose and begin to saturate at the highest concentrations, indicating that significant cluster ripening occurs within the $100$\,ns annealing window. At the maximum dose (\mbox{$7.5\times10^{14}$}\,cm$^{-2}$) implanted at $900$\,K, clustering culminates in the formation of a basal-plane faulted interstitial loop at all investigated annealing temperatures. The loop encloses an extrinsic stacking fault and is decorated by trapped interstitials as shown in Fig.~\ref{fig:cluster_size}(b,c). In contrast, no faulted loops emerged for the same implantation dose at $500$\,K; only small basal stacking faults enclosed by Shockley partials appear at annealing temperatures below $2000$\,K, and these are unstable and decay with increasing annealing temperature and time.
Overall, implantation at $900$\,K promotes the accumulation of clustered interstitials at the damage peak, driving the transition from compact defect complexes to extended planar defect precursors. This temperature dependence is consistent with experiments by Wang \emph{et al.}\,\cite{WANG2023} and Zang \emph{et al.}\,\cite{ZANG2025}, who reported enhanced planar clustering and increased dislocation loop densities with rising  implantation temperatures.

\subsubsection{Al--C complexes}
A key observation in Fig.~\ref{fig:defect_structures_dose_2350K} is the high abundance of carbon-related defects, in particular carbon antisites C$_\mathrm{Si}$ (cyan) and Al-C complexes (brown-cyan), with only weak dependence on implantation temperature.
During implantation, Frenkel pairs are generated along the ion trajectory. During annealing, mobile interstitials migrate and recombine with vacancies. Because carbon interstitials are typically more abundant than silicon interstitials, carbon antisites form as a common recombination product. While the isolated carbon antisite does not introduce levels in the band gap\,\cite{kobayashi2019}, the dicarbon antisite (C$_2$)$_\mathrm{Si}$, also known as P-T PL center \cite{gali2007}, is a thermodynamically stable defect with deep levels and a binding energy of $\sim\,3.5$\,eV\,\cite{mattausch2004,Li2023}. Both can trap diffusing Al and thus promote the formation of Al-C complexes. The IDS-based analysis cannot unambiguously distinguish 
whether an observed Al-C signature corresponds to a split interstitial on a Si site (Al-C)$_\mathrm{Si}$, to substitutional aluminum bound to a carbon interstitial (Al$_\mathrm{Si}$C$_I$) or an Al interstitial bound to a carbon antisite (Al$_\mathrm{I}$C$_{Si}$). In the following, we therefore group these configurations under the collective label "Al-C" complexes.

\begin{figure*}[hbtp]
\centering
     \subfloat[]{\includegraphics[width=0.23\textwidth]{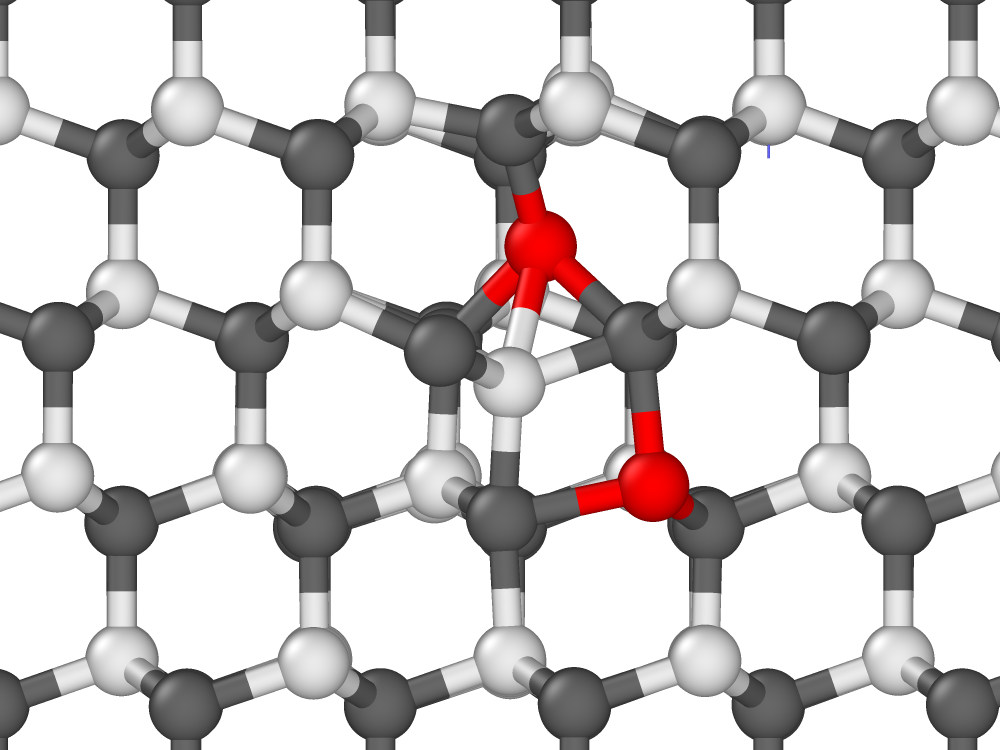}}
     \hspace{2pt}
     \subfloat[]{\includegraphics[width=0.23\textwidth]{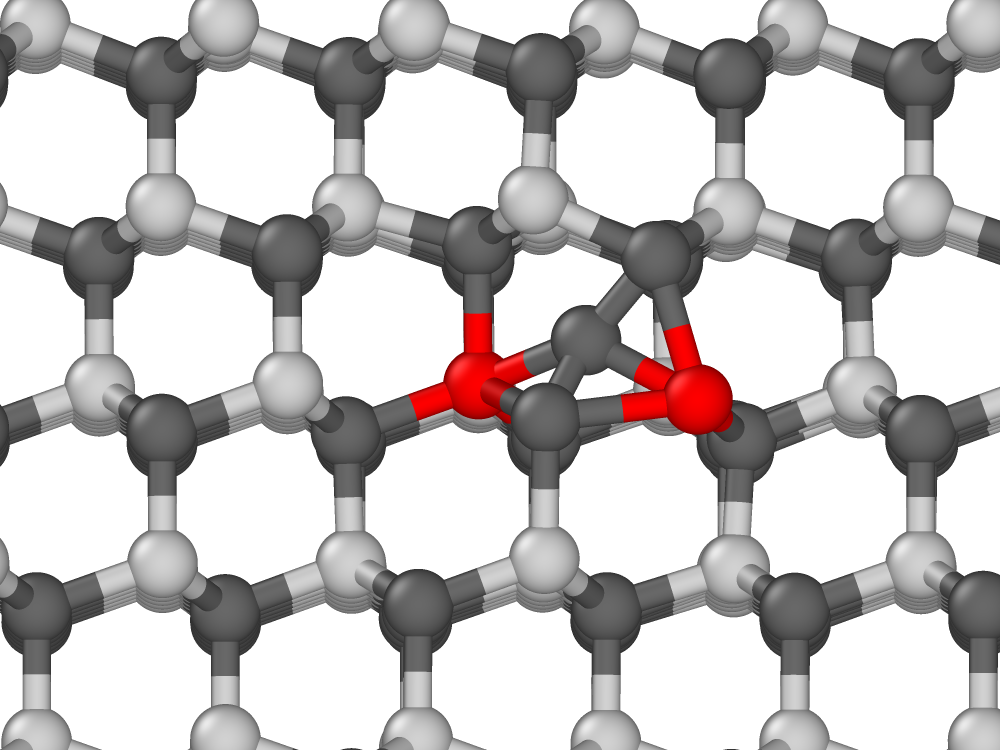}}
     \hspace{2pt}
     \subfloat[]{\includegraphics[width=0.23\textwidth]{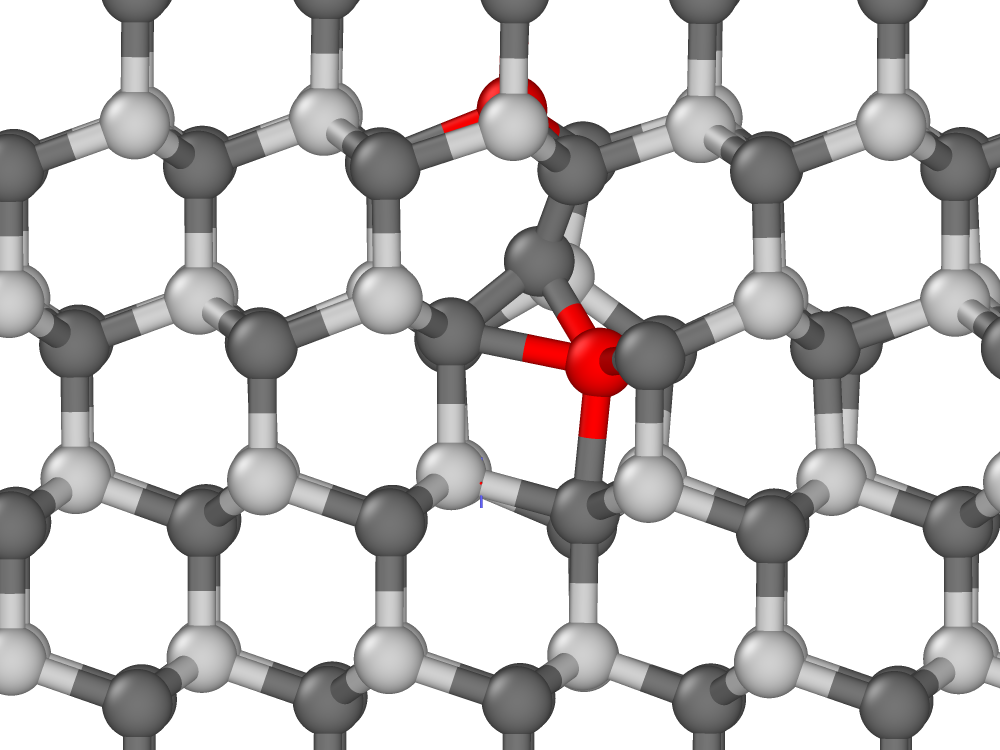}}
     \hspace{2pt}
     \subfloat[]{\includegraphics[width=0.23\textwidth]{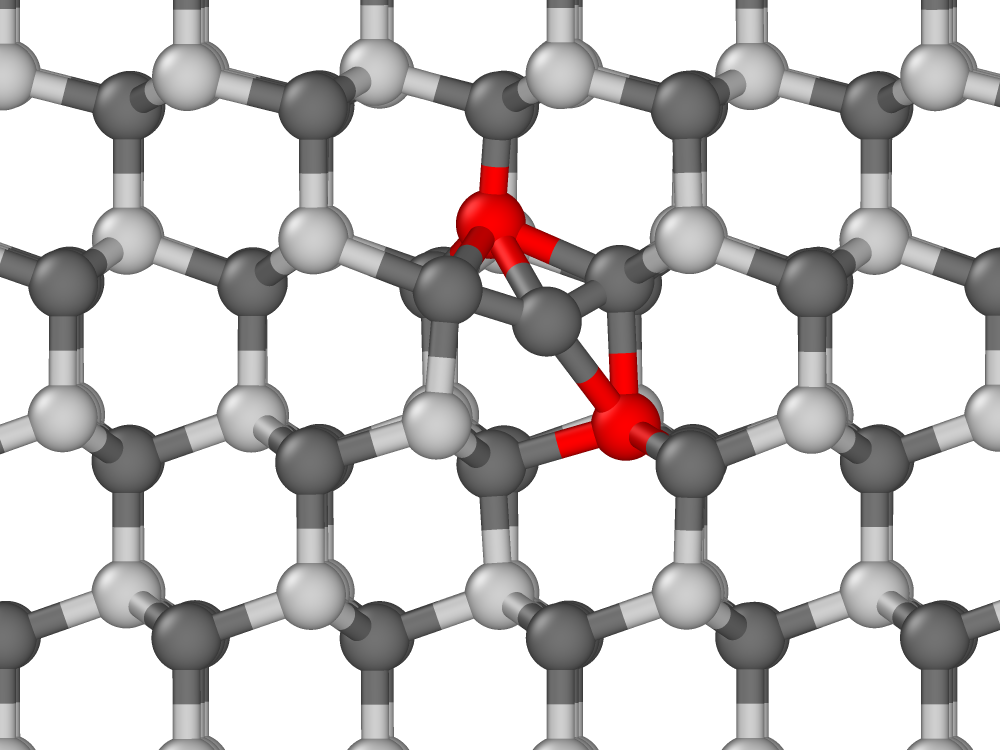}} \\
   
     \subfloat[]{\includegraphics[width=0.23\textwidth]{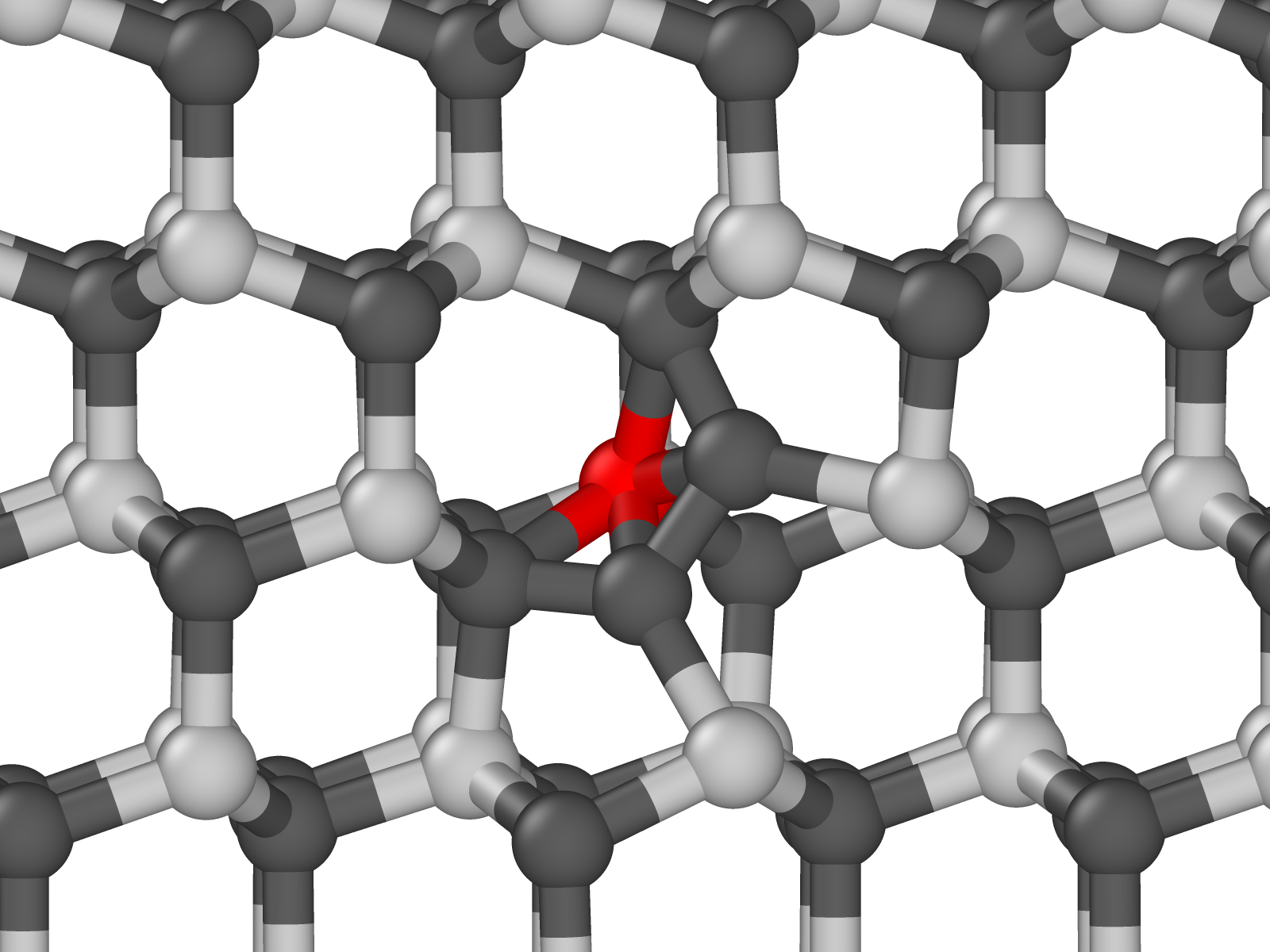}}
     \hspace{2pt}
     \subfloat[]{\includegraphics[width=0.23\textwidth]{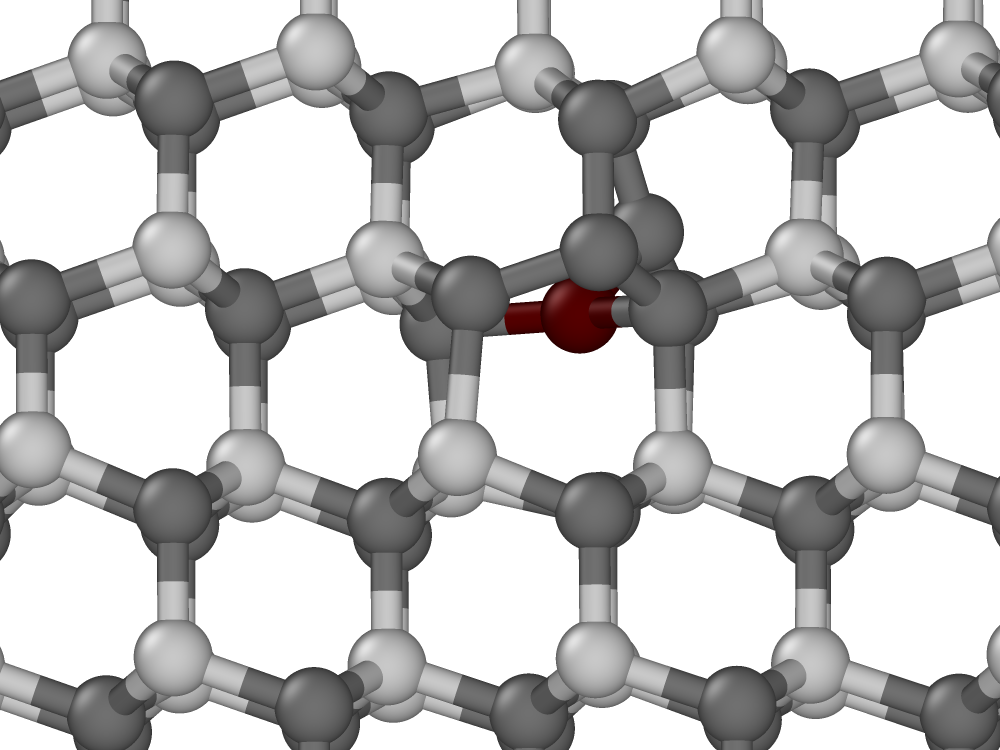}} 
     \hspace{2pt}
     \subfloat[]{\includegraphics[width=0.23\textwidth]{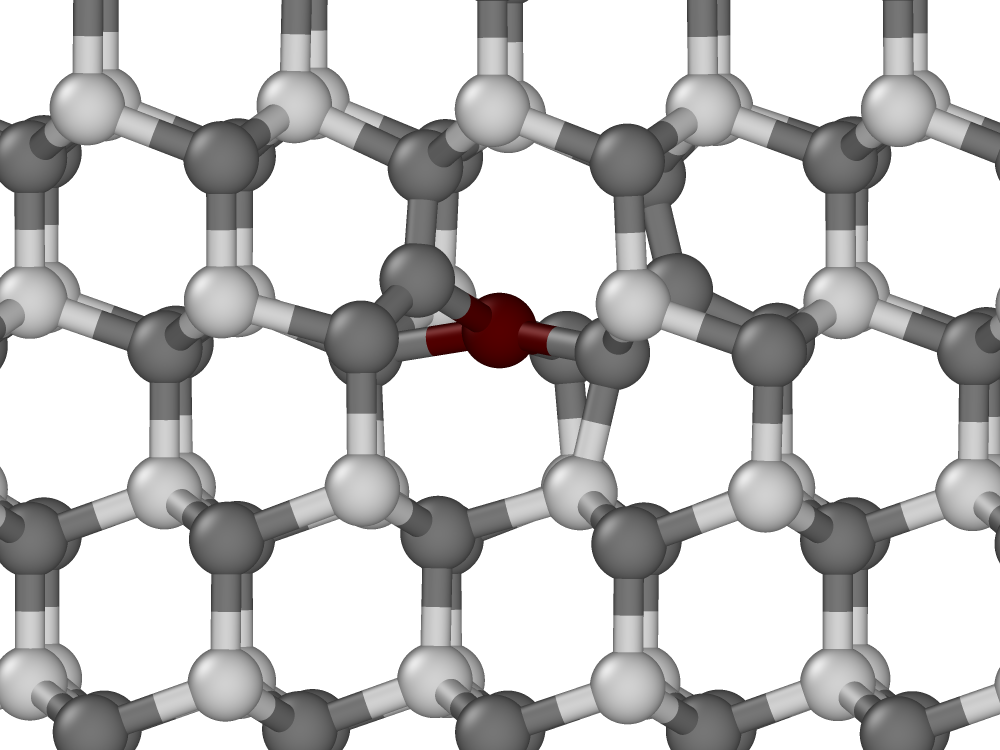}}  
     \hspace{2pt}
     \subfloat[]{\includegraphics[width=0.23\textwidth]{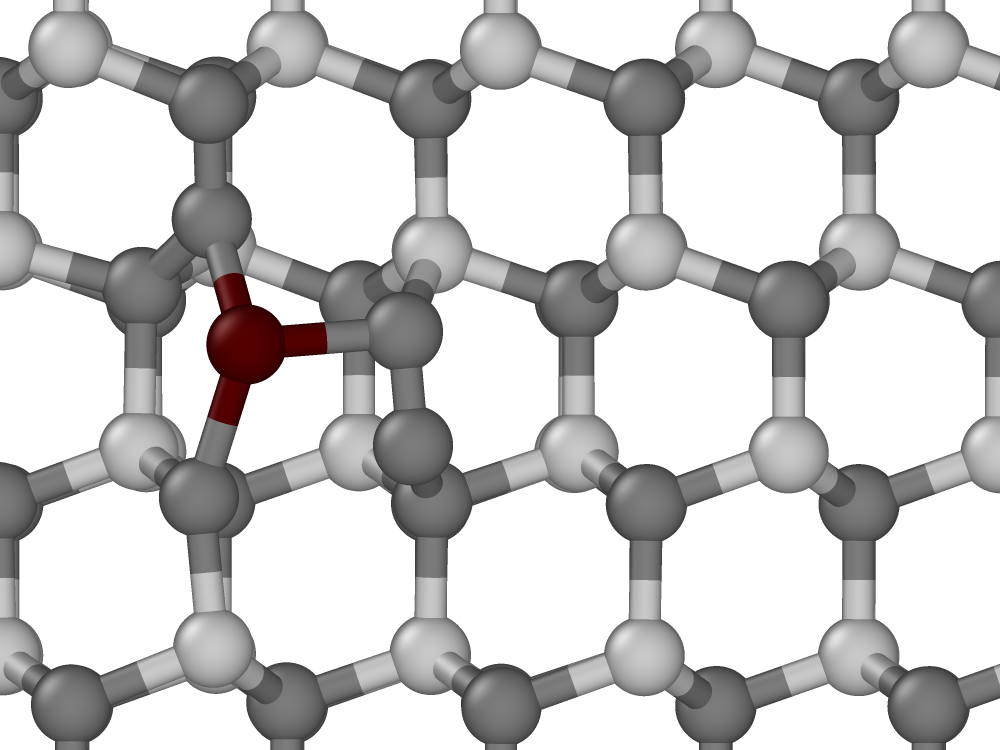}} 
\caption{Thermodynamically stable neutral Al-related 3-mer defect complexes observed in MD, after structural equilibration in DFT.
(a) Silicon interstitial in the hexagonal plane sandwiched by two substitutional Al along the c-axis Al$_\mathrm{Si,h}$Al$_\mathrm{Si,c}$Si$_\mathrm{I,h}$, (b) two substitutional Al bound to a carbon interstitial in the basal hexagonal plane in  $\langle0\overline110\rangle$ direction 2Al$_\mathrm{Si,h}$C$_\mathrm{I,h}$, (c) Separation of out-of-plane defect Al$_\mathrm{Si,h}$Al$_\mathrm{Si,c}$C$_\mathrm{I,c}$ into a substitutional Al and an (Al-C)$_{Si}$ split interstitial, (d) two substitutional Al bound to a carbon interstitial along the c-axis Al$_\mathrm{Si,h}$Al$_\mathrm{Si,c}$C$_\mathrm{I,h}$, (e) Ladder-like Al$_{Si}$(C$_I$)$_\mathrm{2}$ complex, (f) two carbon antisites bound to an Al interstitial in the basal cubic plane in $\langle\overline12\overline10\rangle$ direction Al$_\mathrm{I}$2(C$_\mathrm{Si,c})$, (g) Separation of the defect Al$_\mathrm{I,c}$2C$_\mathrm{Si,c}$ in the cubic basal plane in $\langle0\overline110\rangle$ direction into an isolated antisite and an (Al-C)$_{Si}$ split interstitial (h) Al interstitial sandwiched by two carbon antisites along the c-axis Al$_\mathrm{I,h}$C$_\mathrm{Si,c}$C$_\mathrm{Si,h}$. Host atoms: C (dark gray) and Si (light gray), Al: substitutional (red) interstitial (brown).}
\label{fig:3mer_complexes}
\end{figure*}

This picture is consistent with recent experimental spectroscopy studies. Kumar \emph{et al.}\,\cite{Kumar2024} combined low-energy muon spin rotation (LE-$\mu$SR) and deep-level transient spectroscopy (DLTS) to investigate defects in Al-implanted 4H-SiC. They reported persistent Al- and C-related defect complexes after annealing, with signal intensity increasing with implantation dose. However, the exact atomistic structure of these complexes remained unclear.  Gali and Hornos \emph{et al.}\,\cite{gali2007,hornos2008} investigated Al-C complexes using DFT to identify the origin of electrically deep DLTS centers in 4H-SiC known as A$1$, A$2$, and A$3$.  They focused particularly on complexes consisting of a single acceptor Al or Al interstitial and one or two C interstitials, which form stable compounds in various charge states. The most stable structure among them is formed by substitutional Al bound to two carbon interstitials (Al$_\mathrm{Si}$(C$_\mathrm{I}$)$_2$) which is stable in the negative state around mid-gap. The defect atoms adopt a ladder-like structure in which the two carbon interstitials are connected to two silicon and two carbon atoms, as shown in Fig.~\ref{fig:3mer_complexes}(a). The complex was proposed to form either through kick-out of one carbon atom from a dicarbon antisite by a diffusing Al interstitial or through sequential capture of two diffusing carbon interstitials by an acceptor Al, involving a kick-out reaction. The defect shows negative U-behavior with a transition level ($1/-1$) around mid-gap and has an additional level in the valence band. The neutral state exists but is not in the ground state, and thus much less stable. That could explain why the neutral ladder configuration was much less abundant in the current MD simulations.  
A second complex, Al$_{Si}$C$_I$, investigated by Gali \textit{et al.} \cite{gali2007} and found in larger numbers in the current MD simulations, consists of an Al acceptor that has captured a single diffusing C interstitial
to form a tilted split configuration of C$_I$ near Al$_{Si}$. This configuration has occupied levels in the band gap and was found to be unstable in the negative charge state \cite{gali2007} but stable in the neutral and positive state, in agreement with Table~S5 and fig. S14 in the supplementary information. 
In general, complexes involving Al acceptors were shown to be considerably more stable than purely interstitial Al$_\mathrm{I}$C$_\mathrm{I}$ complexes that exist in positive and neutral states. The 
metastable neutral Al$_\mathrm{I}$C$_\mathrm{I}$ behaves similarly to isolated Al$_\mathrm{I}$, which has unpaired electrons in the conduction band \cite{hornos2008}. In agreement with these DFT trends, purely neutral interstitial complexes (Al$_\mathrm{I}$C$_\mathrm{I}$) have been identified as metastable in the present MD simulations whereas complexes involving Al$_{Si}$ are among the most stable Al-C complexes. 

Moreover, in implanted SiC, acceptor Al can also bind to carbon vacancies. The neutral complex could be assigned to the A$_3$ center \cite{gali2007}. The (Al$_{Si}$V$_C$) complex has deep acceptor levels in the band gap but no level in the conduction band, as in 3C-SiC, consistent with hybrid-functional DFT calculations for Al-vacancy complexes in 4H-SiC\,\cite{Igumbor2019}. 

MD simulations revealed that the stable neutral Al$_{Si}$C$_I$ forms not only through the capture of a carbon interstitial by Al$_{Si}$ as proposed by Gali et al \cite{gali2007} but also through the kick-in reaction of an aluminum interstitial with a carbon antisite. We will therefore discuss the kick-in process together with the Al$_{Si}$C$_I$ complex in more detail in Section 3.2.4.
Besides dimers, also trimer complexes were also identified in the MD simulations, such as the thermodynamically meta-stable neutral ladder-like complex Al$_\mathrm{Si}$(C$_\mathrm{I}$)$_2$ shown in Fig.~\ref{fig:3mer_complexes}(a), which has already been discussed by Hornos \cite{hornos2008}. Moreover, other complexes in the basal plane and along the c-axis exist. They are formed by two acceptor Al atoms bound to one carbon or one silicon interstitial and occurring in all investigated MD systems at high Al dose regardless of temperature. A second class of trimer complexes consists of two carbon antisites and one Al interstitial. Both types of complexes are displayed in Fig.~\ref{fig:3mer_complexes}. 
All trimer carbon-related complexes with two substitutional Al atoms, and further analyzed by DFT in the supplementary information, are thermodynamically stable in the positive charge state. The negative complexes are unstable, and in the neutral state, only the $2\mathrm{Al}_\mathrm{Si,h}\mathrm{C}_\mathrm{I,h}$ complex exists. In contrast, the carbon complexes with interstitial Al are very stable in both the neutral and charged states. The Si-related complex with two substitutional Al atoms is also thermodynamically stable across the entire band gap. While the stability of complexes with substitutional Al decreases with rising Fermi level, the Al interstitial complexes become most stable in the n-type regime. 
Where changes in the binding energy in dependence of Fermi level are caused by the absorption or emission of electrons during complex dissolution, since reactants and products in a chemical reaction typically do not have the same charge state as shown in Figures S5-S11 and S14b and S16b in the supplementary information. Along with the Fermi-level-dependent binding energies, the neutral binding energies from DFT and the GW potential are also shown in the figures. These can deviate significantly from the actual binding energies, particularly for the Al–C complexes. While the neutral DFT and GW energies agree well for the dimer complexes, they provide conflicting statements regarding the stability of the trimer complexes as discussed in the supplementary information. This is probably because the Morse potential does not include three-body interactions to better represent the structure of trimer complexes.
Moreover, stability is determined not only by the stoichiometry of the compound but also by the relative positions of the atoms inside the defect and by whether the central atom is located in the cubic or hexagonal plane. For example, the Al$_\mathrm{Si,h}$Al$_\mathrm{Si,c}$C$_\mathrm{I,c}$ complex is less stable than the Al$_\mathrm{Si,h}$Al$_\mathrm{Si,c}$C$_\mathrm{I,h}$ complex, in which just the carbon interstitials are located in different layers. While the initial MD structure of Al$_\mathrm{Si,h}$Al$_\mathrm{Si,c}$C$_\mathrm{I,h}$ remains intact during DFT equilibration, Al$_\mathrm{Si,h}$Al$_\mathrm{Si,c}$C$_\mathrm{I,c}$ decays into two separate defects, as shown in Figure~\ref{fig:3mer_complexes}(c). 
The binding energies—or the energies required to remove either the intrinsic (Al$_{Si}$-complex) or the Al (Al$_I$-complex) interstitial from the complex are summarized in Table S5 in the supplementary information. Depending on the kinetic barriers, the neutral and negative acceptor Al complexes will dissolve during annealing, similar to the neutral and negative Al$_{Si}$C$_I$ complex, and contribute to chemical activation. The interstitial Al complexes, on the other hand, are very stable, especially in the intrinsic to n-type regime. The defect has many shallow and deep compensating levels. 
In heavily ion-implanted Al doped SiC a thermally very stable (up to $2100$\,K) shallow level at $E_V + 0.37$\,eV, which was found in DLTS experiments and has already been discussed by Gali et al \cite{gali2007} but could not be assigned, was detected. The signal also appeared in implanted samples exposed to very low implantation energies to ensure that only the C atoms in SiC are displaced from their sites. This suggests that the shallow trap levels are caused by Al-C-related defects. A possible candidate for a shallow trap level combined with high binding energies and barrier for Al diffusion is the defect level ($+2/+1$)$: 0.37$\,eV of the Al$_\mathrm{I,h}$C$_\mathrm{Si,c}$C$_\mathrm{Si,h}$ complex. Whether the defect actually caused the discussed trap level requires more detailed investigations, which is not the subject of this paper.

In summary, the most abundant or thermally persistent Al-related defect species identified across all simulated conditions are the Al$_\mathrm{Si}$C$_\mathrm{I}$ complex, the ladder-like Al$_\mathrm{Si}$(C$_\mathrm{I}$)$_2$ and Al interstitial related trimers. They introduce compensating or deep levels in the band gap and are continuously regenerated during annealing, making them the dominant Al trapping channel independent of implantation temperature.
\subsubsection{Chemical activation and kinetic interpretation}
\paragraph{Activation via kick-in to the carbon antisite}
\begin{figure*}[hbtp]
\begin{center}
      \subfloat[]{\includegraphics[width=0.35\textwidth]{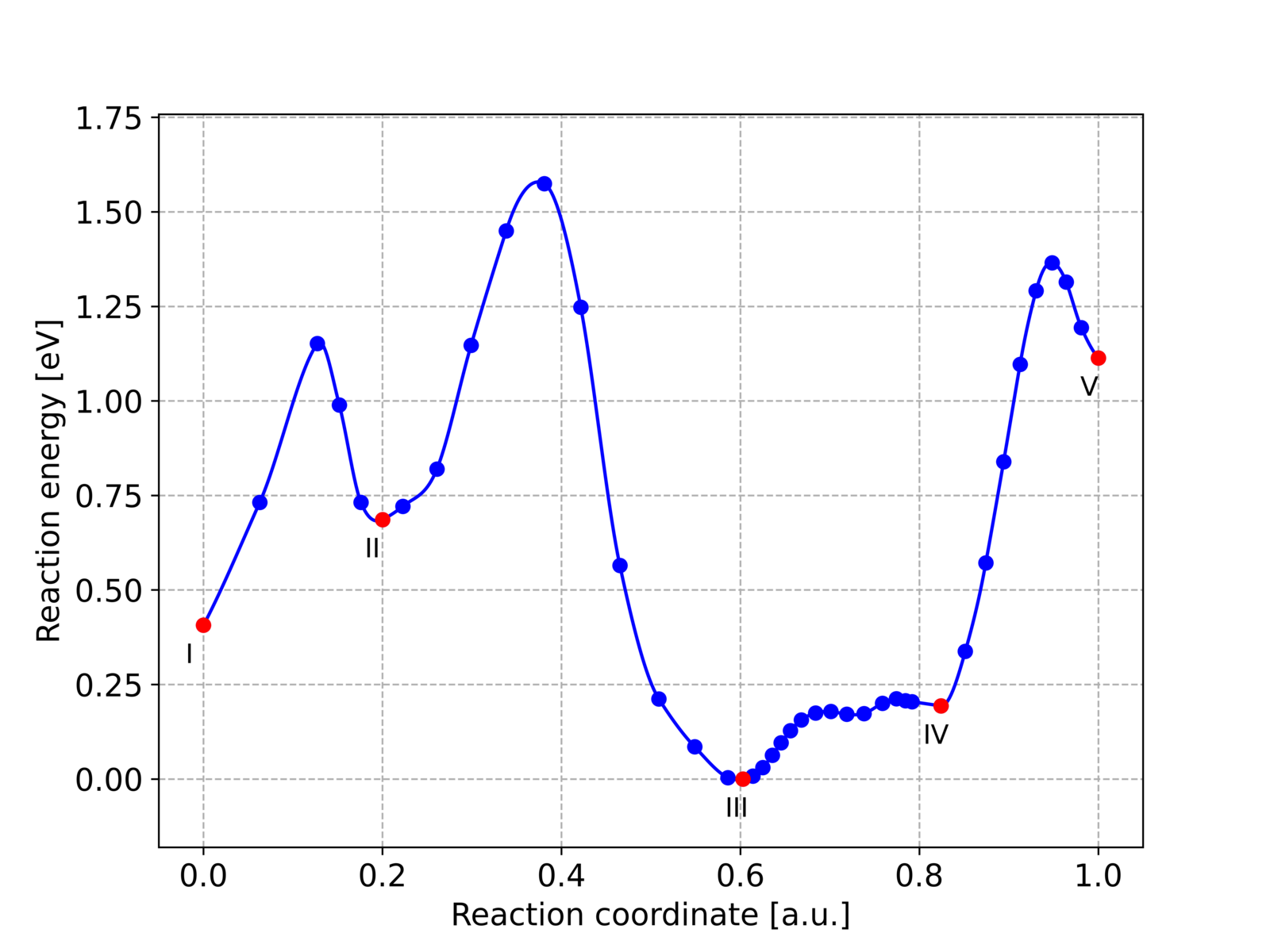}}
      \subfloat[V]{\includegraphics[width=0.3\textwidth]{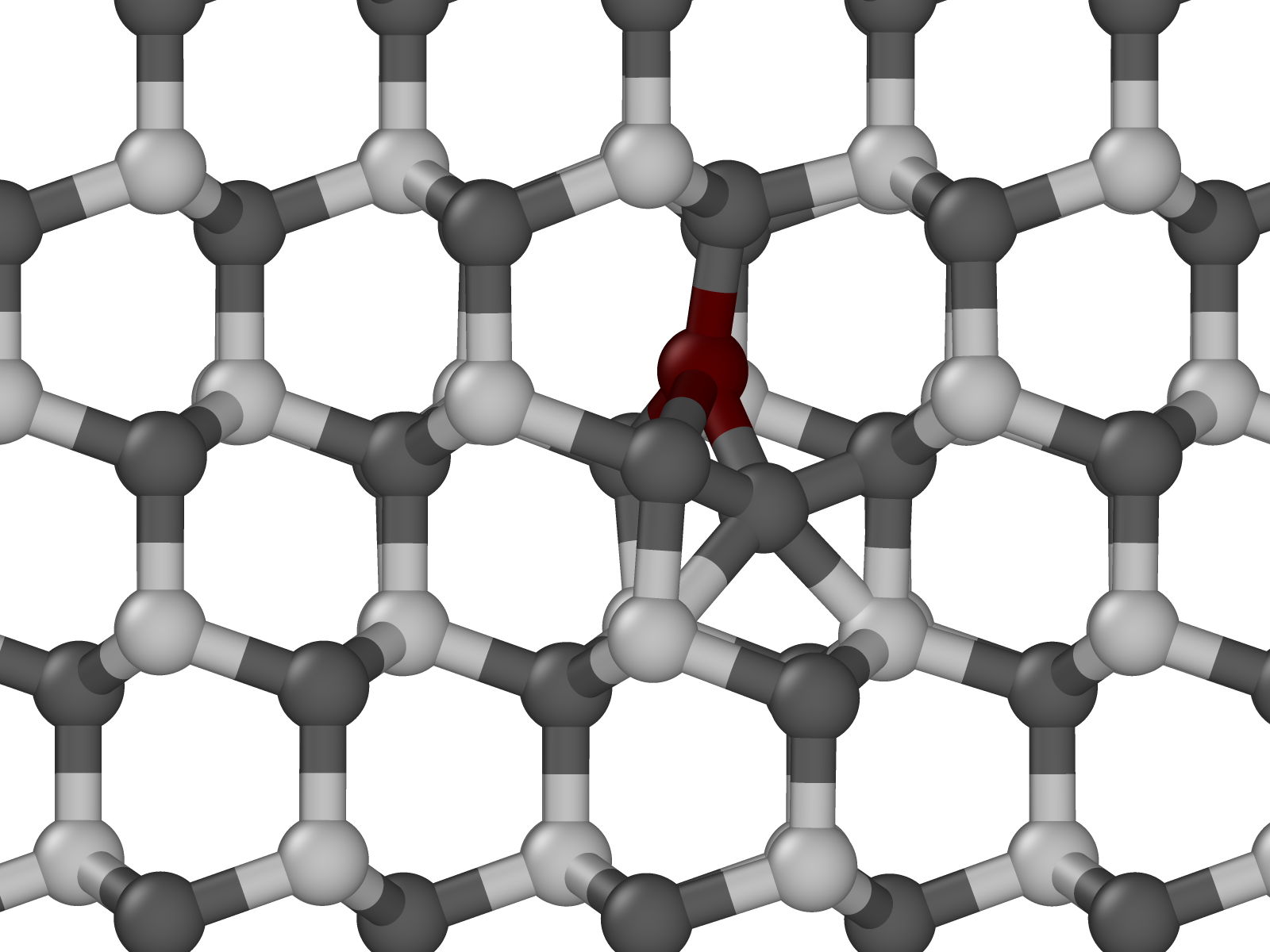}}
      \hspace{2pt}
      \subfloat[IV]{\includegraphics[width=0.3\textwidth]{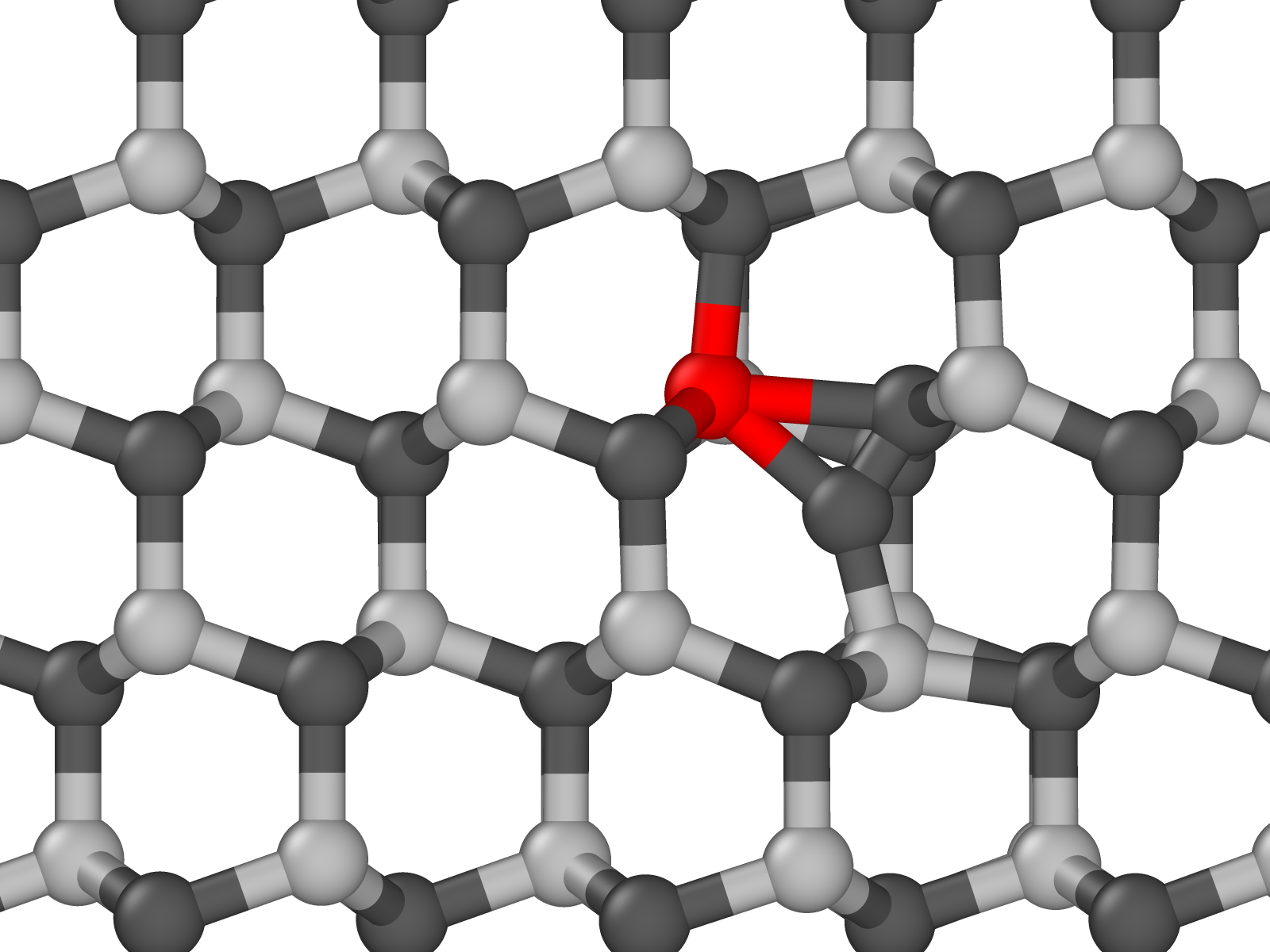}}\\
      \subfloat[III]{\includegraphics[width=0.3\textwidth]{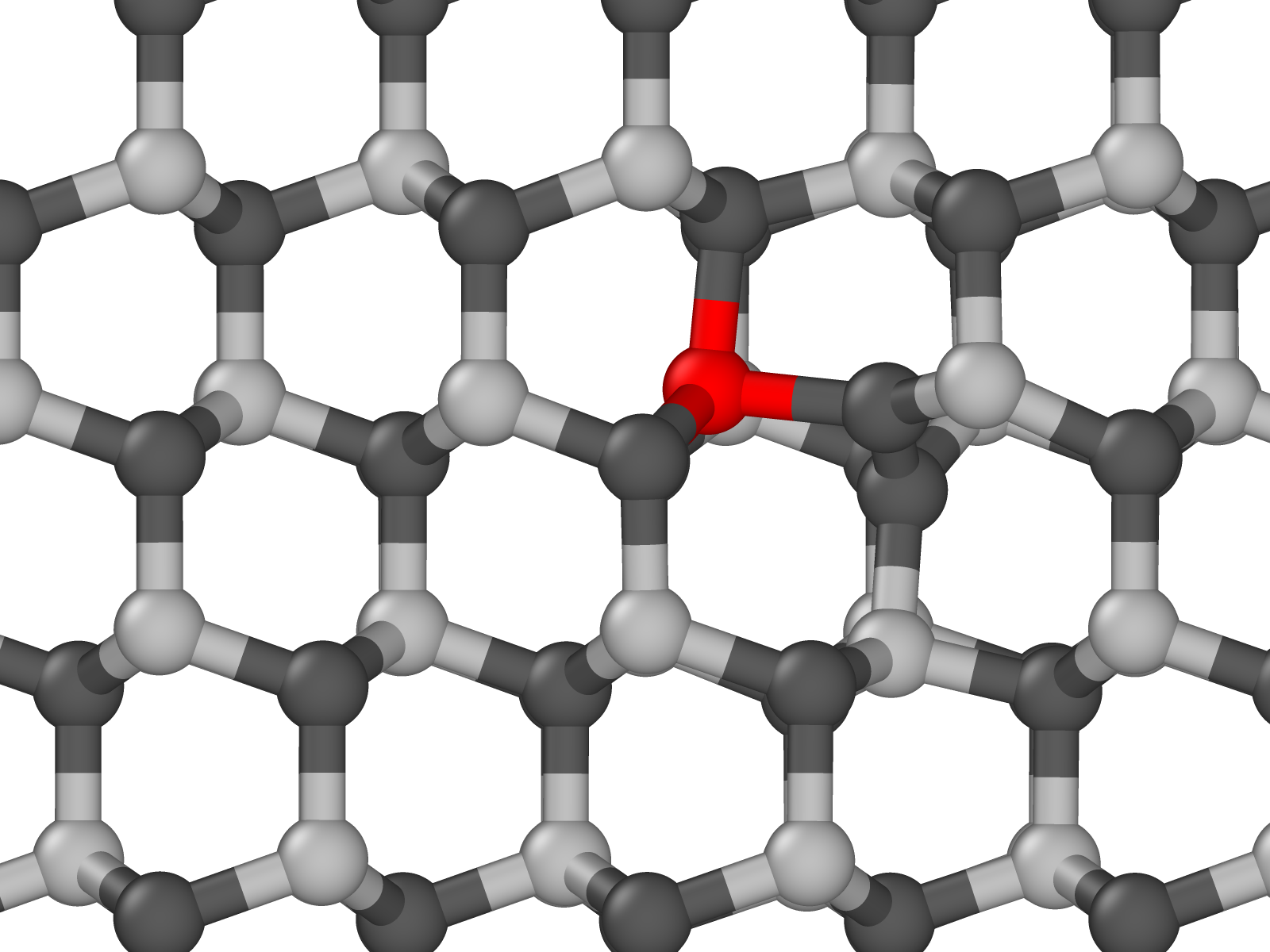}}
      \hspace{4pt}
      \subfloat[II]{\includegraphics[width=0.3\textwidth]{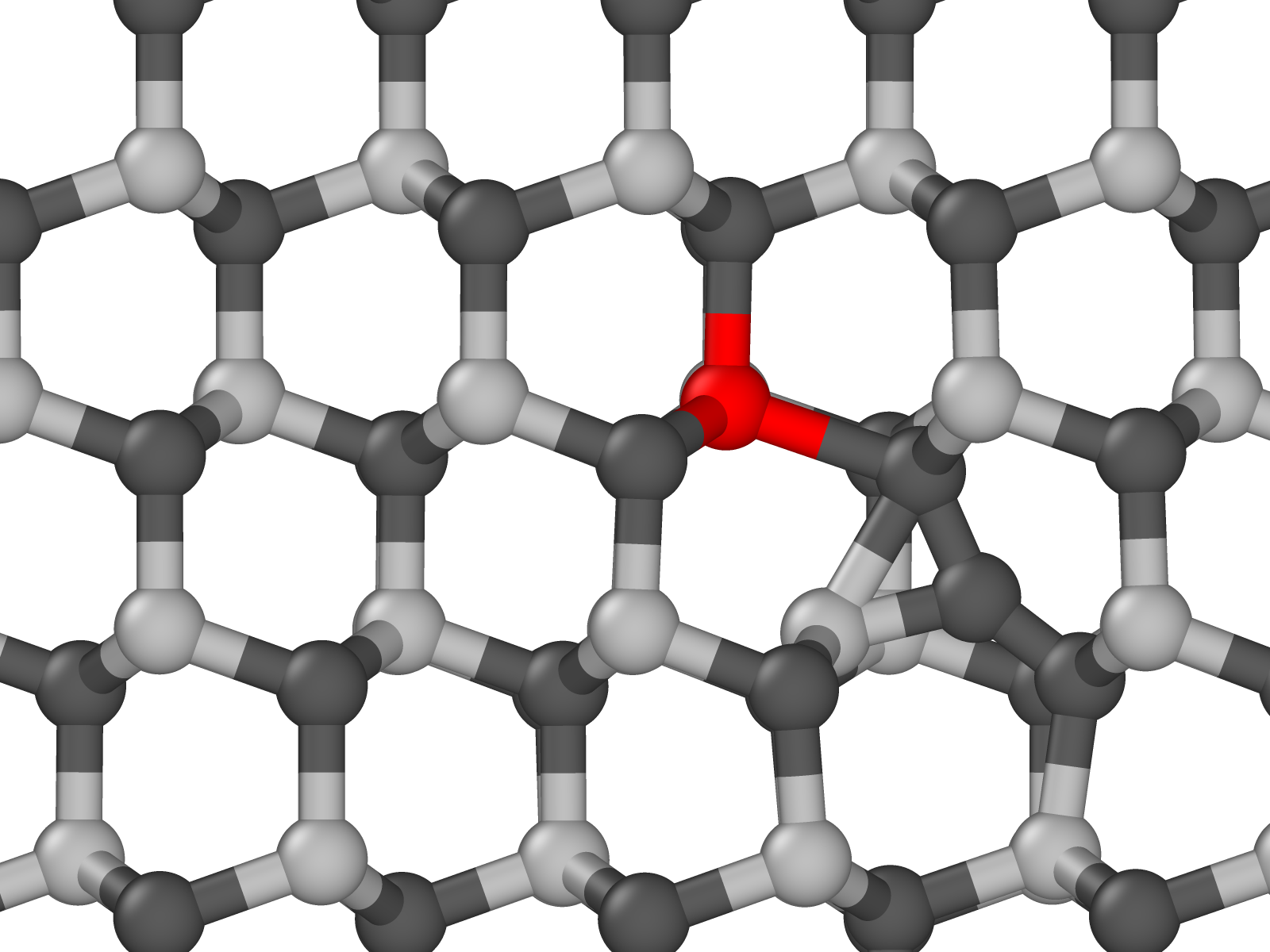}}
      \hspace{4pt}
      \subfloat[I]{\includegraphics[width=0.3\textwidth]{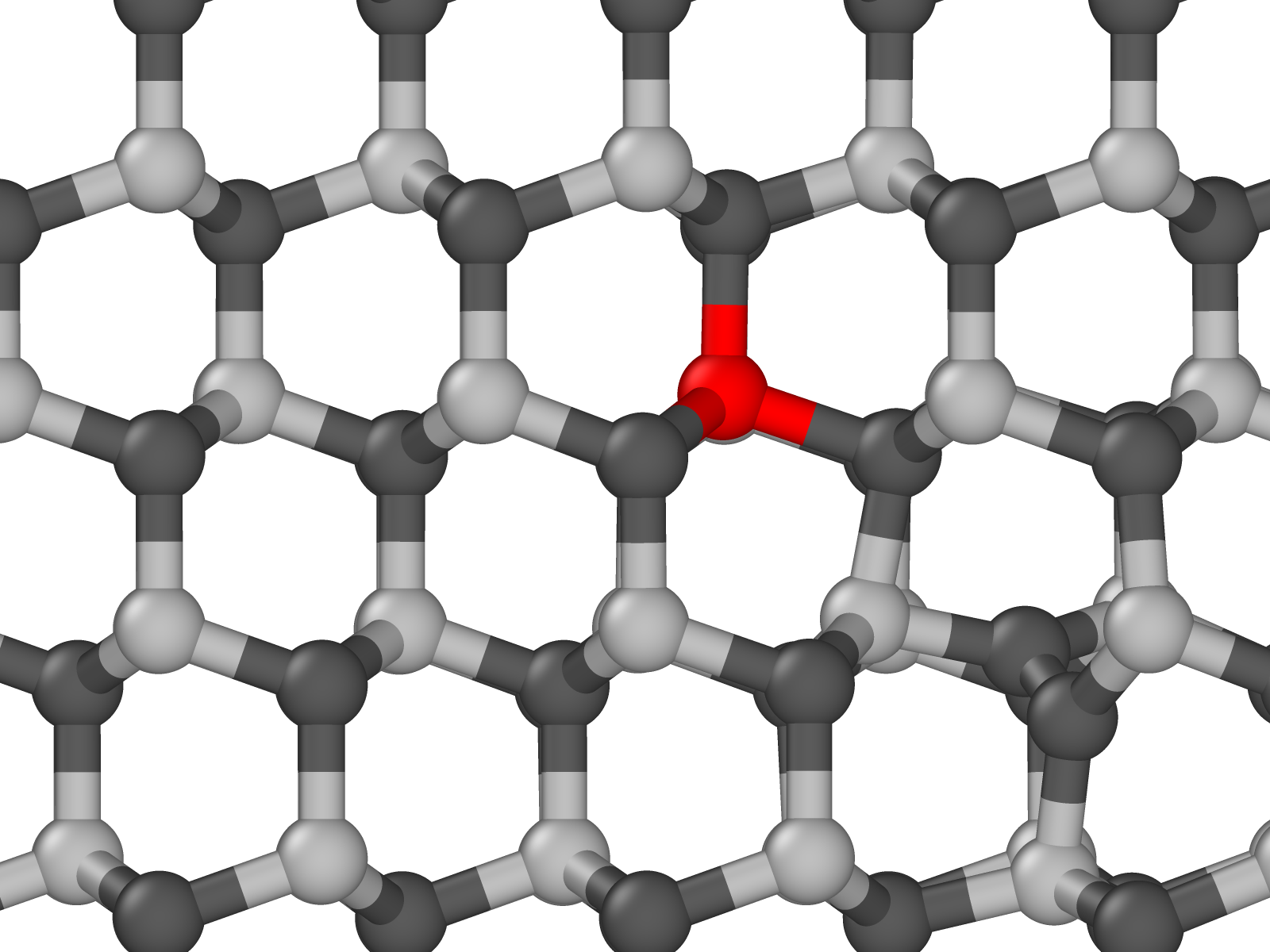}}
\end{center}
\caption{Kick-out reaction of the carbon antisite by the diffusing Al interstitial applying DFT NEB for the neutral state.
(a) Activation barriers passed in the course of the reaction. Displayed configurations are marked by red dots. (b) Spontaneous formation of an (Al-C)$_{Si}$ split interstitial when Al approaches the nearest neighbor distance of the carbon antisite at $V$. (c) Bond breakage followed by upwards rotation of the carbon interstitial ultimately forming a carbon split interstitial next to Al$_{Si}$ in $IV$. (d) Relaxation towards the most stable Al$_{Si}$C$_I$ complex configuration in $III$. (e) Detachment of the carbon split interstitial from Al$_{Si}$ in $II$. (f) Carbon interstitial diffuses away, Al is chemically activated in $I$.}
\label{fig:AlCSi_kickin}
\end{figure*}
Before we study the effect of extended defect clusters on Al activation, we will first discuss the new activation mechanism (besides the well known recombination with V$_{Si}$) via the carbon antisite, as revealed by the MD simulations.
The observed kick-in mechanism was therefore recalculated for the neutral state using DFT-NEB in Fig.~\ref{fig:AlCSi_kickin}. Several barriers occur along the energetically most favorable reaction path. The kick-in/recombination barrier ($0.3-0.5$\,eV) as shown in Fig.~S13 is considerably lower than the migration barrier of the neutral Al interstitial which is destabilized by the capture of electrons during complex formation. Thus, as soon as the neutral Al interstitial approaches the neighborhood of the carbon antisite during high temperature annealing, an (Al-C) split interstitial  spontaneously forms at the silicon site at point $V$ shown in part (a) of the figure. With a barrier as low as $0.25$\,eV, the split interstitial transforms finally into the stable Al$_{Si}$C$_I$ complex, as shown at points IV and III. The complex is only $0.3$\,eV thermodynamically more stable than the isolated defects in I. The defect's high stability is primarily due to the kinetic barriers. The neutral complex constitutes a trapping state that returns to the metastable split configuration with a barrier of $1.2$\,eV or decays into isolated defects with a desorption barrier of $1.6$\,eV. Furthermore, when Al is activated in this way by releasing a carbon interstitial, this carbon can diffuse to another substitutional Al site to form a new thermodynamically stable Al$_{Si}$C$_I$ complex with a recombination barrier of $0.9$\,eV (I$\rightarrow$III). This is also observed in the current MD simulations. The persistence of the Al$_{Si}$C$_I$ complex is not primarily due to its thermodynamic stability, but rather due to the continuous formation of new complexes, which keeps their concentration relatively constant over long times. The activation process via the carbon antisite is moreover investigated in the supplementary information as a function of charge state and summarized in Table S7. The binding energies in Table S5 show that the stability of the complex decreases with Fermi level. In p-type the formation of the positively charged complex during annealing is limited due to the high Al migration barrier, whereas in n-type, the negatively charged complex is unstable and decays, thereby increasing Al activation as soon as the diffusing Al interstitial approaches the carbon antisite.

Moreover, our DFT-NEB calculations have shown that the neutral (Al-C)$_{Si}$ split configuration 
 is metastable in the $\langle001\rangle$ direction, which is also confirmed by DFT calculations of Matsushima \textit{et al.} \cite{matsushima2019} 
showing that the neutral out-of-plane $\langle001\rangle$(Al-C)$_{Si}$ split interstitial is metastable in 3C-SiC.
\paragraph{Chemical activation versus Al compensation}
\begin{figure*}[ht!]
\centering
     \subfloat[]{\includegraphics[width=0.48\textwidth]{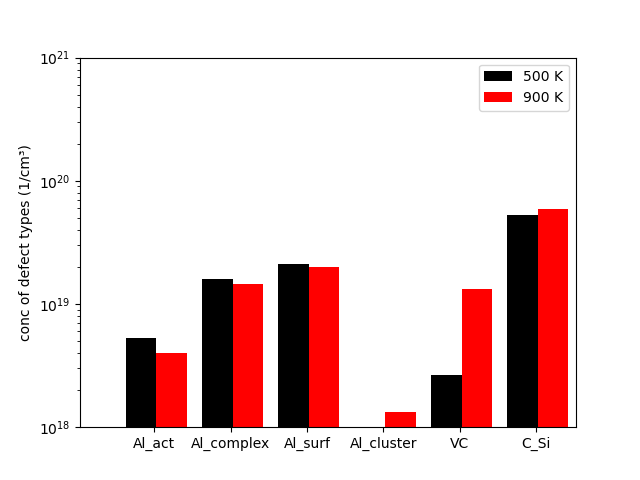}} \hfill
     \subfloat[]{\includegraphics[width=0.48\textwidth]{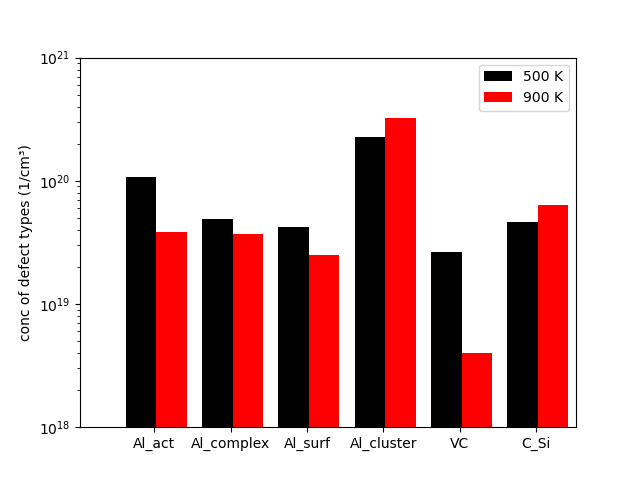}}
\caption{Bulk concentrations of dominant defect classes after annealing at $2350$\,K for implantation at $500$\,K and $900$\,K.
(a) Dose $5\times10^{13}$\,cm$^{-2}$.
(b) Dose $5\times10^{14}$\,cm$^{-2}$.
Al populations are partitioned into perfectly activated Al$_\mathrm{Si}$, Al in small complexes (size $\le3$), and clustered Al (size $\ge4$). Intrinsic defect concentrations (C$_\mathrm{Si}$, V$_\mathrm{C}$, etc.) are shown for comparison.}
\label{fig:defect_types}
\end{figure*}
\begin{figure*}[ht!]
\centering
     \subfloat[]{\includegraphics[width=0.32\textwidth]{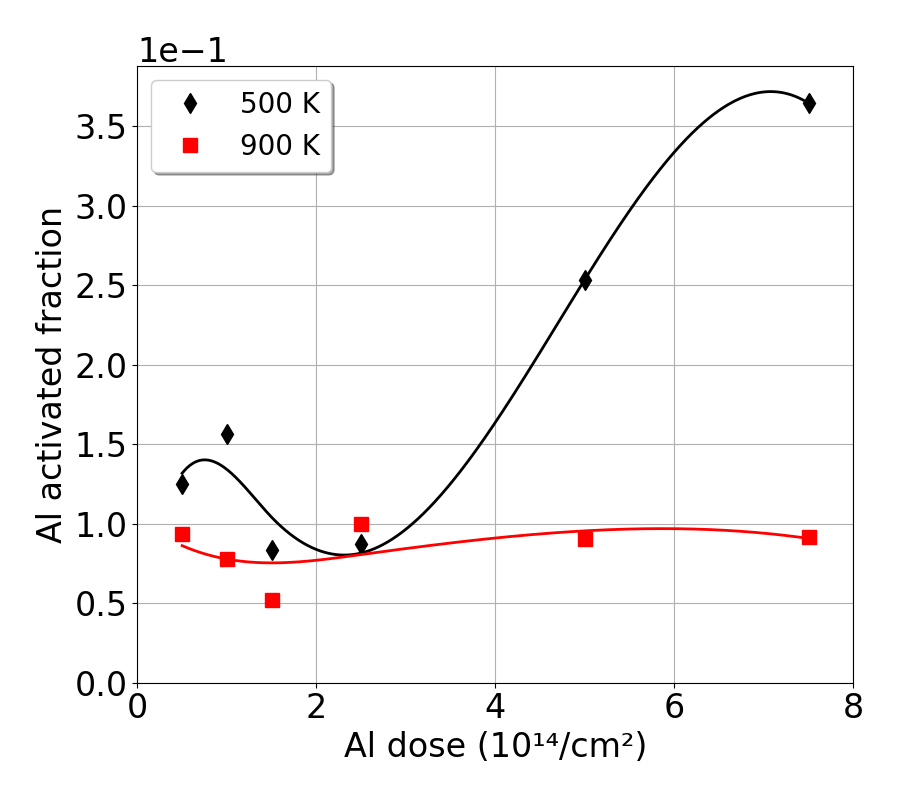}} \hfill
     \subfloat[]{\includegraphics[width=0.32\textwidth]{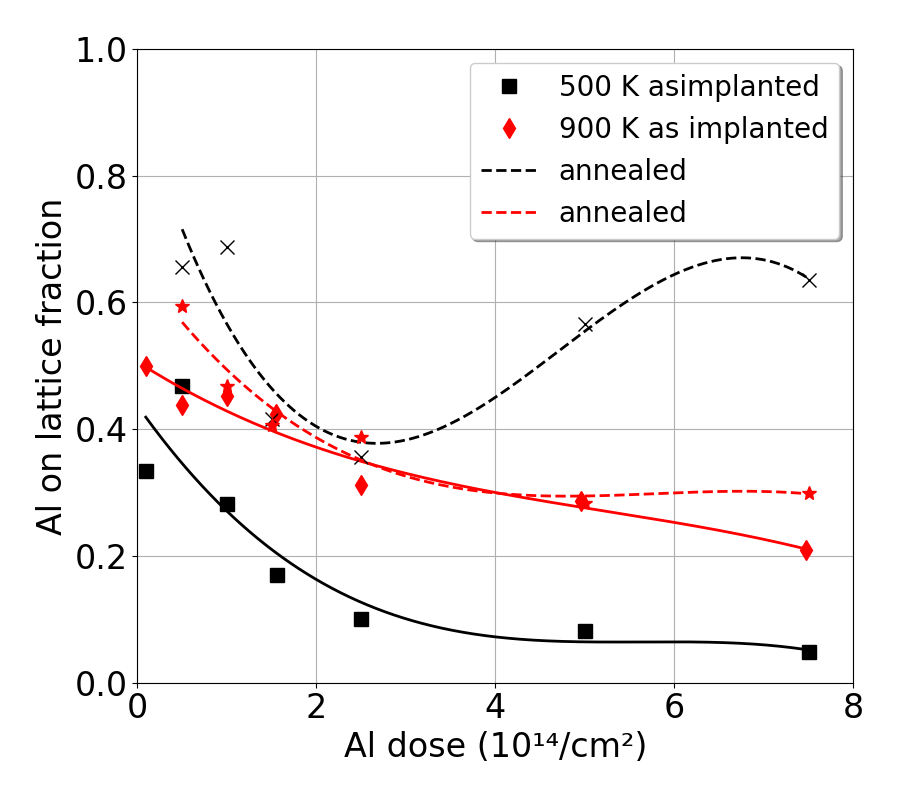}}\hfill
     \subfloat[]{\includegraphics[width=0.35\textwidth]{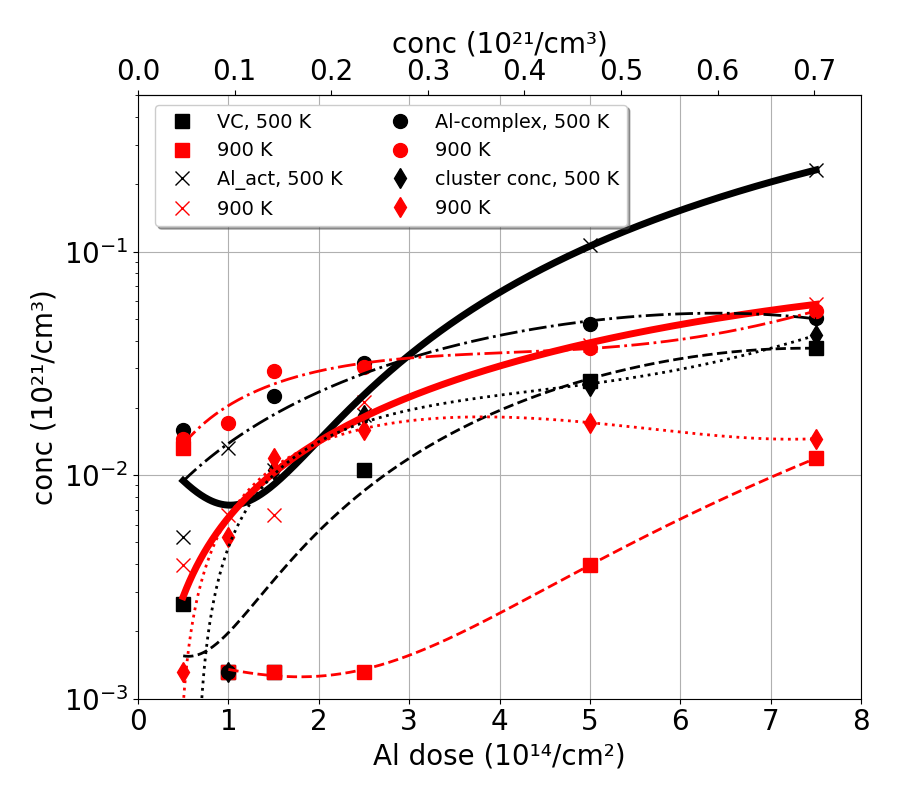}}
\caption{Activation trends after annealing at $2350$\,K (100\,ns).
(a) Perfect activation as a function of implantation dose and temperature.
(b) General activation for as-implanted (solid) and annealed (dashed) structures.
(c) Concentration of perfectly activated Al (solid) together with representative compensating or trapping defects: carbon vacancies V$_\mathrm{C}$ (dashed), Al-C complexes (dash-dotted), and clustered defects (dotted). Lines serve as visual guides.} 
\label{fig:Al_activation}
\end{figure*}
To quantify how Al is distributed among different defect environments after annealing, the bulk concentrations of dominant defect classes are summarized in Fig.~\ref{fig:defect_types} for representative low and high doses. 
At low dose (\mbox{$5\times10^{13}$}\,cm$^{-2}$), most Al resides on lattice sites  or is bound in Al complexes after annealing.  In contrast, at high dose (\mbox{$5\times10^{14}$}\,cm$^{-2}$), a substantial fraction of Al becomes incorporated into clusters (cluster size $>3$), particularly for implantation at $900$\,K. The fraction of clustered Al is significantly larger for the $900$\,K implantation than for the $500$\,K implantation, consistent with the enhanced interstitial agglomeration discussed above. Carbon antisite concentrations remain comparatively insensitive to implantation temperature, whereas V$_\mathrm{C}$ concentrations vary more strongly due to diffusion toward the surface and incorporation into defect clusters. In shallow implantation, a significant fraction of Al diffuses to the surface and is therefore lost to activation, which is particularly pronounced at low doses.
Fig.~\ref{fig:Al_activation} summarizes Al activation as a function of implantation dose and temperature using two structural metrics. \emph{Perfect activation} counts substitutional Al$_\mathrm{Si}$ without other defects within the next-nearest-neighbor distance, thereby excluding Al bound in complexes or clusters. \emph{General activation} includes all Al located on lattice sites, independent of local defect environment (including a small fraction of Al$_\mathrm{C}$). 
About $10\%$ of lattice-site Al counted in the general activation metric occupies C lattice sites (Al$_\mathrm{C}$), producing severe local lattice distortions. Consistent with DFT, Al$_\mathrm{C}$ is metastable, exhibits high formation energy, and introduces deep levels in the band gap\,\cite{matsushima2019,Igumbor2019}; it therefore does not contribute to acceptor activation.

At high dose, Fig.~\ref{fig:Al_activation}(c) shows an apparent saturation of perfectly activated Al, in qualitative agreement with experimental trends~\cite{Nipoti2018}. Absolute saturation values are not expected to match steady-state experimental data because the MD simulations access early-stage annealing. However, the relative trends are robust and correlate directly with the defect inventory: V$_\mathrm{C}$, Al complexes, and extended clusters represent compensating and trapping centers that limit the population of electrically active acceptors.
Because the metrics in Fig.\ref{fig:Al_activation}(c) do not explicitly include charge states or compensation effects, they represent chemical incorporation rather than direct electrical activation. Although explicit charge states are not treated within classical MD, DFT studies presented in the supplementary information, especially in Table S5, indicate that several of the Al–C complexes and intrinsic defect configurations identified here introduce deep or compensating levels in 4H-SiC, suggesting that the trends observed in chemical activation are qualitatively correlated with electrical activation behavior.

Although the as-implanted activation at $500$\,K is initially lower than at $900$\,K due to increased partial amorphization, annealing reverses this trend at high dose. For \mbox{$5\times10^{14}$}\,cm$^{-2}$ and \mbox{$7.5\times10^{14}$}\,cm$^{-2}$, perfect activation as shown in Fig.~\ref{fig:Al_activation}(a) after annealing for samples implanted at $500$\,K exceeds that for samples implanted at $900$\,K by a factor of $2-3$. At intermediate doses 
(\mbox{$1.5\times10^{14}$}\,cm$^{-2}$ and \mbox{$2.5\times10^{14}$}\,cm$^{-2}$), activation differences between implantation temperatures are comparatively small. Under identical annealing conditions, the general activation curve in part (b) of the figure exhibits a minimum near \mbox{$2.5\times10^{14}$}\,cm$^{-2}$, consistent with the experimentally observed transition from near-complete activation at low dose to saturation behavior at high dose\,\cite{Nipoti2011}.
\paragraph{Kinetic Interpretation}

The defect inventories provide the quantitative basis for understanding the temperature- and dose-dependent activation trends. The same qualitative behavior is observed across the entire annealing temperature range investigated here ($1500 - 2500$\,K).
The simulations indicate that reduced as-implanted disorder at elevated temperature does not necessarily translate into improved activation at high dose. Implantation at $900$\,K enhances Frenkel-pair separation and interstitial mobility, which, besides desired recombinations, promotes interstitial-interstitial encounters and cluster nucleation during implantation and early annealing. As a result, larger interstitial clusters that persist throughout the annealing window form near the damage peak, acting as trapping centers for Al. This clustering reduces the fraction of isolated substitutional Al$_\mathrm{Si}$ and thereby lowers perfect chemical activation. In contrast, implantation at $500$\,K kinetically preserves a more disordered structure, including nanoscale amorphous pockets/layers. During annealing, solid-phase epitaxial regrowth provides an efficient pathway for regrowth-assisted incorporation of Al$_\mathrm{Si}$ while annihilating vacancies. 

In shallow implantation geometries, mobile interstitials also have relatively short diffusion paths to the surface, which acts as an efficient sink. The amorphous pockets are sufficiently small (nanometer scale) to recrystallize completely within the simulated annealing time, facilitating dopant incorporation rather than cluster ripening. Noticeably, the highest activation levels coincide with implantation conditions that produce intermediate disorder fractions of approximately $40$–$60$\,$\%$ at the damage peak (Fig.~\ref{fig:disorder_conc}(a)). 
Under the present implantation settings, the level of disorder (disorder fraction) 
on the one hand, remains below the threshold of about $70\%$ \cite{zhang2002} for irreversible amorphization, and on the other hand, it is sufficient to cause a high density of vacancies and short diffusion pathways for regrowth-assisted substitution while simultaneously preventing the formation of extended defects.

Taken together, these results support a kinetic interpretation of the experimentally observed activation window \cite{WANG2023}. Moderate implantation temperatures suppress excessive clustering while allowing subsequent regrowth-assisted incorporation, whereas elevated temperatures favor defect agglomeration and planar defect precursor formation at high dose, ultimately reducing dopant activation.
\section{Conclusion}
Despite the limited annealing times accessible to molecular dynamics, the present simulations using the combined Gao-Weber-Morse potential  reproduce key experimental trends and provide atomistic insight into shallow Al implantation and early-stage annealing in 4H-SiC.
 
Two distinct regimes emerge as a function of Al concentration. In the low-concentration regime, below the Al solubility limit ($\sim2\times10^{20}$\,cm$^{-3}$ at $2000$\,K\,\cite{Linnarsson2001}), annealing leaves predominantly isolated point defects, small complexes, and compact (Al-Si-C) clusters, with only minor dependence on implantation temperature. This aligns well with experimental findings \cite{kimoto2014} according to which, at concentrations below Al saturation, implantation at temperatures close to room temperature is sufficient to achieve good dopant activation. In the high-concentration regime above saturation, defect clustering dominates the microstructural evolution. Large interstitial agglomerates form and can evolve into extended planar defects, including stacking faults and dislocation loops, depending on implantation temperature.
Cluster size increases strongly with implantation dose and is significantly larger for elevated implantation temperatures in agreement with \cite{WANG2023,ZANG2025}. At high doses, implantation at $900$\,K enhances Frenkel-pair separation and interstitial mobility, promoting cluster nucleation and ripening that ultimately stabilizes extended planar defects. In contrast, implantation at $500$\,K kinetically preserves a partially disordered state. During subsequent annealing, solid-phase epitaxial regrowth couples vacancy annihilation to substitutional Al incorporation, thereby enhancing chemical activation and suppressing segregation into large clusters. 
These findings support common practice in shallow implantation experiments \cite{kyungwon2015}: amorphizing the wafer prior to implantation to  prevent the formation of extended defects and achieve better activation through recrystallization.

The stable defect populations identified after annealing are consistent with experimental observations \cite{Kumar2024} and DFT predictions \cite{gali2007,hornos2008}. Residual defects include V$_\mathrm{C}$, C$_\mathrm{Si}$, and small stable Al-C complexes, as well as, at high Al supersaturation, large (Al-Si-C) clusters and basal-plane faulted interstitial loops observed in experiments \cite{Persson2003, Nipoti2018-2,Wong2005}. 
Where the interstitial loops form spontaneously and in higher concentration with increasing implantation temperature, in agreement with experiments\,\cite{Persson2003,WANG2023,ZANG2025}. These Frank loops are sessile and generate local strain fields, acting as permanent defect sinks that degrade carrier transport, mobility, and lifetime. Their high thermal stability makes them difficult to remove completely by subsequent annealing, underscoring the importance of controlling implantation conditions to avoid extended defect formation.

Among the smaller defects present in all the samples irrespective of the implantation conditions, Al-C complexes are particularly abundant and thermally stable\,\cite{hornos2008}, and several configurations introduce compensating levels in p-type semiconductors. 
Carbon antisites and acceptor Al serve as efficient trapping centers that promote Al-C complex formation during annealing. The neutral Al$_{Si}$C$_I$ complex, 
complexes formed by two substitutional Al atoms with a single carbon or silicon interstitial, or complexes consisting of two carbon antisites and an Al interstitial have been identified as thermodynamically stable (Al-C) complexes by MD simulations. DFT evaluations of the MD generated defect structures confirmed the thermodynamic stability of the complexes in several charge states. However, regarding the binding energies of the trimer complexes, there were significant discrepancies between the neutral DFT and MD energies, which assigned too high a stability to the Al$_{Si}$-related complexes and too low a stability to the Al$_I$-related complexes. This indicates that the simple two-body term of the Morse potential is insufficient to correctly describe trimer structures.
Moreover, DFT calculations have also shown that the stability of the complexes depends on the Fermi level when the decay is accompanied by the absorption or emission of electrons. Approximations of binding energies based on neutral formation energies—as are often used in kinetic Monte Carlo simulations—can therefore result in over- or underestimates of the cluster stability. \\
In the current MD simulations, a new diffusion mechanism for the neutral Al interstitial was discovered, which proceeds via jumps between Al interstitial and basal (Al-Si) split interstitial configurations with barriers of $0.6$ in the hexagonal and $0.8$\,eV in the cubic plane. The migration barrier in the GW potential, at $1.1$\,eV, lies slightly above this range and thus covers the Fermi level region from conduction band minimum down to mid-gap, where the ($2+$) charged Al interstitial in the hexagonal plane diffuses with a barrier of $1$\,eV.  
Moreover, the new Al activation mechanism involving the carbon antisite identified through MD simulations, could be confirmed by DFT calculations. The kick-in reaction produces the Al$_{Si}$C$_I$ complex, which is stable around mid gap in the neutral charge state where it represents a trapping state; it  returns to the metastable (Al-C)$_{Si}$ split configuration with roughly the same probability as it decays, thereby contributing to Al activation. The negative defect becomes unstable at high Fermi levels and spontaneously decays into Al$_{Si}$ and C$_I$. 
The MD results were consistent with experimental observations regarding temperature- and dose-dependent extended defec formation. 

Overall, the simulations are consistent with experimental reports of higher Al incorporation on lattice sites and reduced clustering for moderate or near-room-temperature implantation compared to elevated temperatures\,\cite{Negoro2004,Michaud2013,Wendler1998}. The results suggest that optimal dopant activation occurs within an intermediate damage regime characterized by local disorder fractions on the order of $40-60$\,\% at the damage peak which has to be provided by the implantation conditions. 
These findings coincide with high-resolution TEM images in \cite{zhang2002}, which show that in as-implanted 4H-SiC with disorder fractions of up to $40-60$\,\%, the over all lattice structure is highly strained but still crystalline. It thus retains sufficient lattice memory for the original polytype structure to be recovered during annealing.
Accordingly, the simulations define a kinetic processing window that can guide the choice of implantation temperature and dose for p-type doping in 4H-SiC.

\section*{Acknowledgement}
We would like to express our special thanks to Dominic Waldhör for fruitful discussions and his guidance in performing the DFT calculations done in this work.
Moreover, financial support by the Federal Ministry of Labour and Economy, the National Foundation for Research, Technology and Development and the Christian Doppler Research Association is gratefully acknowledged.

\bibliography{SiC.bib}

\end{document}


\begin{frontmatter}
\title{Supplementary Material\\The influence of implantation conditions on dopant activation in Al-implanted 4H-SiC:\\
A MD study applying an Al potential fitted to DFT barriers.}
\author{}
\begin{abstract}
Supplementary figures, tables, and analysis supporting the main manuscript.
\end{abstract}
\end{frontmatter}

\section{Additional Results}
\subsection{Formation Energy and transition levels of point defects and Al-complexes}
While PBE underestimates the band gap and yields charge-transition levels that differ quantitatively from those obtained using hybrid functionals, the predicted stable defect charge states under p-type, intrinsic, and n-type conditions are consistent with previous HSE06 studies as can be seen for the intrinsic defects in \cite{yan2020} and for the Al-Si complexes in \cite{huang2022} which are recalculated in Table~\ref{tab:levels} within PBE to allow a qualitative comparison. Since the present analysis is concerned primarily with identifying the dominant charge states, for binding energy and barrier calculations, rather than accurately determining transition-level positions, the computationally less demanding PBE calculations provide a suitable framework for this purpose.
The experimental band gap is $3.2$\,eV instead of $2.4$\,eV calculated with DFT on PBE level.  In this work  we assume Fermi level positions of $1.2 - 2.0$\,eV,  $<0.5$\,eV and $>2.7$\,eV to describe the intrinsic, p-type and n-type regime, respectively. 
For PBE the following mapping was applied: 
\begin{itemize}
\item p-type ($0$ - $0.5$\,eV) 
\item intrinsic: ($1.2$ - $2.0$\,eV) -> ($0.8$ - $1.6$\,eV)
\item n-type ($2.7$ - $3.2$\,eV) -> ($1.9$ - $2.4$\,eV)
\end{itemize}
The formation energy $E_f$ of a defect D is given as
\[
E_f(D) = E_{tot}-E_{bulk}-\sum_in_i\mu_i+qE_F,
\]
with $E_{tot}$ the total energy of the system including the defect, $E_{bulk}$ the pristine bulk energy, $\mu_i$ the chemical potential for adding or removing species, charge $q$ and $E_F$ the current Fermi energy. 
If one wants to calculate the binding energy $\Delta E$ of two defects forming a complex one has to subtract the formation energies of the isolated defects and the complex as follows.
\[
\Delta E = -E_f(D_1)-E_f(D_2)+E_f(D_1D_2)
\]
where $E_f(D_i)$ is the formation energy of an isolated defect and $E_f(D_1D_2)$ is the formation energy of the complex. The binding energy is then defined as $E_B=-\Delta E$ with  $E_B > 0$ indicating a thermodynamically stable complex.

\begin{figure}[htbp]
\subfloat[]{\includegraphics[width=0.9\columnwidth]{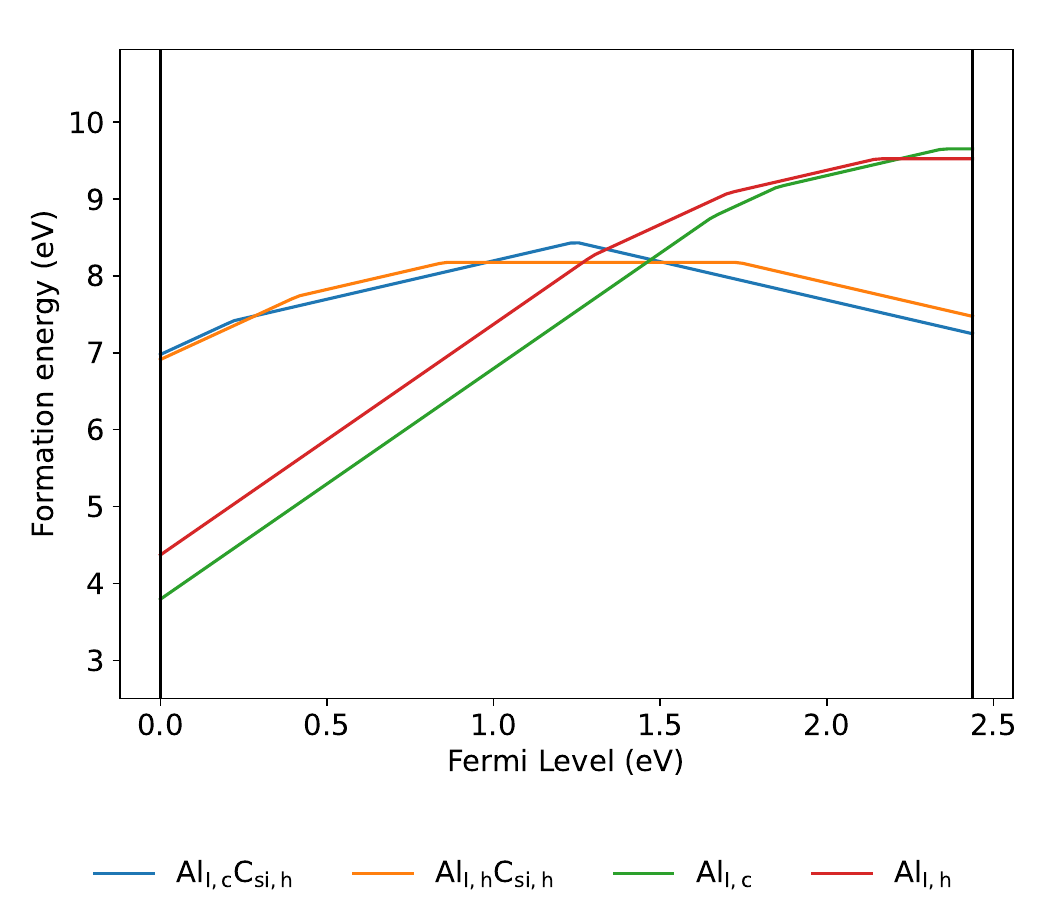}}
\caption{Formation energies of Al point defects.}
\label{fig:point_defects}
\end{figure}
The intrinsic regime is highlighted in bold in  Tables~\ref{tab:levels},~\ref{tab:levels_complexes} and ~\ref{tab:levels_cluster}.
\begin{table}[htbp]
\centering
\caption{Charge transition levels for intrinsic point defects and complexes, where complexes marked with *, \textasciitilde and ' are from \cite{Wang2013}, \cite{hornos2008}  and \cite{Li2023}. The stable charge state in the intrinsic regime is highlighted in bold, units in eV.}
\label{tab:levels}
\begin{tabular}{llllll}
\hline
  config &  &  &  &  &  \\
\hline
$\mathrm{Si}_\mathrm{I,c}$ & +4/{\bf 0: 1.16}   & -- & -- & -- & -- \\
$\mathrm{Si}_\mathrm{I,h}$ & +3/+2: 0.28   & {\bf +2}/0: {\bf 1.68  } & -- & -- & -- \\
$\mathrm{C}_\mathrm{I,c}$ & +2/+1: 0.53   & {\bf +1/0: 1.38  } & 0/-1: 1.73   & -1/-2: 2.29   & -- \\
$\mathrm{C}_\mathrm{I,h}$ & +2/+1: 0.37   & {\bf +1/0: 1.15  } & 0/-1: 2.32   & -- & -- \\
$\mathrm{V}_\mathrm{Si}$ & 0/-1: 0.6   & {\bf -1/-2: 1.5} & -2/-3: 2.4   & -- & -- \\
$\mathrm{V}_\mathrm{C}$ & {\bf +1/0: 1.5}   & 0/-1: 1.8 & -1/-2: 2.4   & -- & -- \\
\hline
 $\mathrm{V}_\mathrm{Si}\mathrm{V}_\mathrm{C}^*$ & +2/+1: 0.3 & +1/0: 0.4  & {\bf 0/-1: 1.2} & -1/-2: 1.5  &-2/-3: 2.4\\
$\mathrm{V}_\mathrm{C}\mathrm{C}_\mathrm{Si}^*$ & {\bf +2/+1: 1.5} & +1/-1: 1.8  & -- & --  & --\\
$\mathrm{Si}_{\mathrm{I}}\mathrm{C}_{\mathrm{I}}^{\sim}$ & {\bf +2/0: 1.9} & 0/-2: 2.43  & -- & --  & --\\
$(\mathrm{C}_\mathrm{I})_\mathrm{2hh}^{'}$ & {\bf +1/0: 0.3} & --  & -- & --  & --\\
$(\mathrm{C}_\mathrm{I})_\mathrm{3hh}^{'}$ & {\bf +1/0: 0.4} & --  & -- & --  & --\\
$(\mathrm{C}_\mathrm{2})_\mathrm{Si,h}^{'}$ & { +2/+1: 0.5} &{\bf  +1/0: 1.5}  & (0/-1) 2.5 & --  & --\\
\hline
\end{tabular}
\end{table}
The silicon interstitials exhibit donor-like behavior. The hexagonal cage configuration $(\mathrm{Si}_{\mathrm{I,h}})$ introduces a shallow donor transition at $(+3/+2)$ located at (E$_V+0.28$) eV, making it a potential compensating defect under p-type conditions. Both cage configurations display pronounced negative-(U) behavior with deep $(+2/0)$ and $(+4/0)$ transitions above mid-gap, indicating direct multi-electron trapping in agreement with HSE06 \cite{yan2020}.

Carbon interstitials are stable in positive and neutral charge states over a wide Fermi-level range and become acceptor-like under n-type conditions.

The silicon vacancy acts as a deep acceptor with stable $-1$ and $-2$ charge states in the intrinsic regime, while the carbon vacancy remains predominantly donor-like with stable (+1) and neutral charge states both in qualitative agreement with \cite{yan2020}.

For completeness, moreover the most stable intrinsic complexes in SiC with transition levels from literature \cite{yan2020,hornos2008,Li2023} are listed. 
The most stable carbon complexes the di and tri carbon complex ($5.28 - 5.23$\,eV) with the carbon interstitials in the hexagonal plane $(\mathrm{C}_\mathrm{I})_\mathrm{2hh}$ and $(\mathrm{C}_\mathrm{I})_\mathrm{3hh}$ are neutral over most of the band gap, introducing shallow donor levels at $E_V+0.3/0.4$.
The stable di-carbon antisite $(\mathrm{C}_\mathrm{2})_\mathrm{Si,h}$ ($3.8$\,eV) is mostly neutral and positively charged, and exhibits a shallow and a deep donor level, in n-type it shows acceptor-like behavior.
\begin{figure}[htbp]
\subfloat[]{\includegraphics[width=0.9\columnwidth]{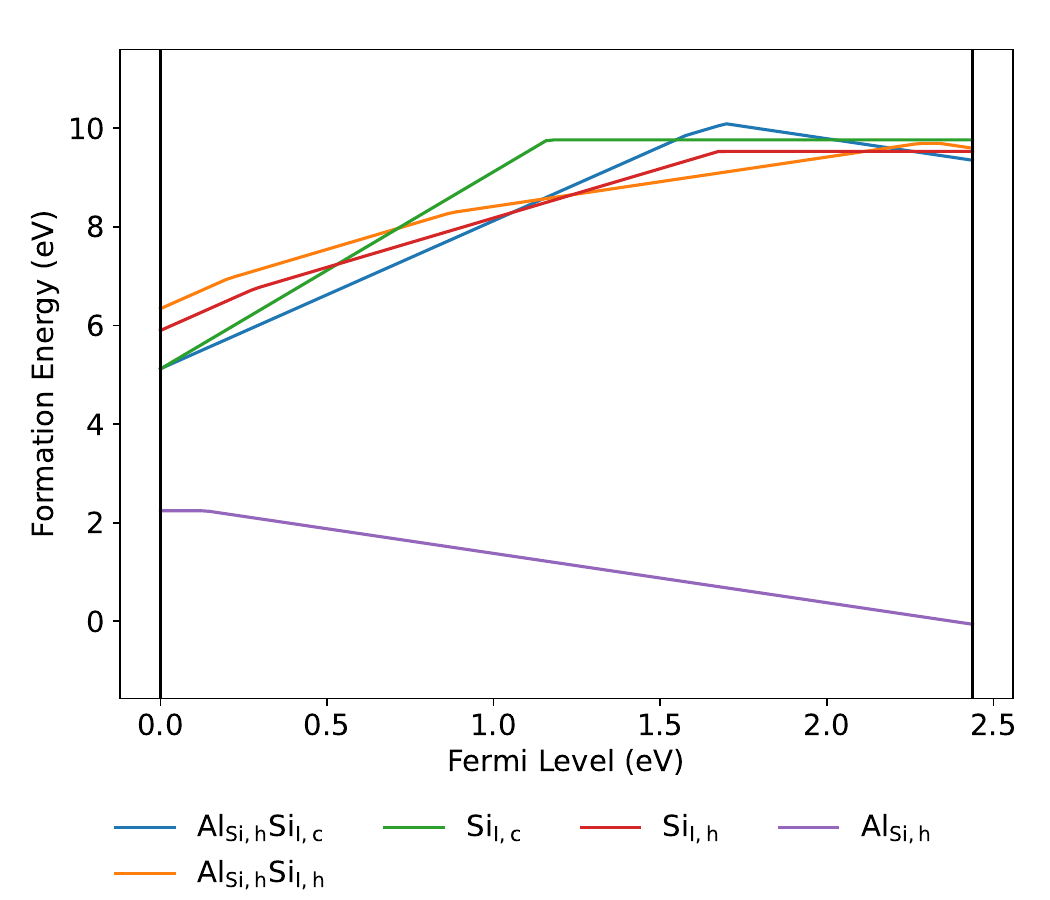}}
\caption{Formation energies of Al-Si complexes}
\label{fig:AlSi}
\end{figure}
\begin{figure}[htbp]
\subfloat[]{\includegraphics[width=0.9\columnwidth]{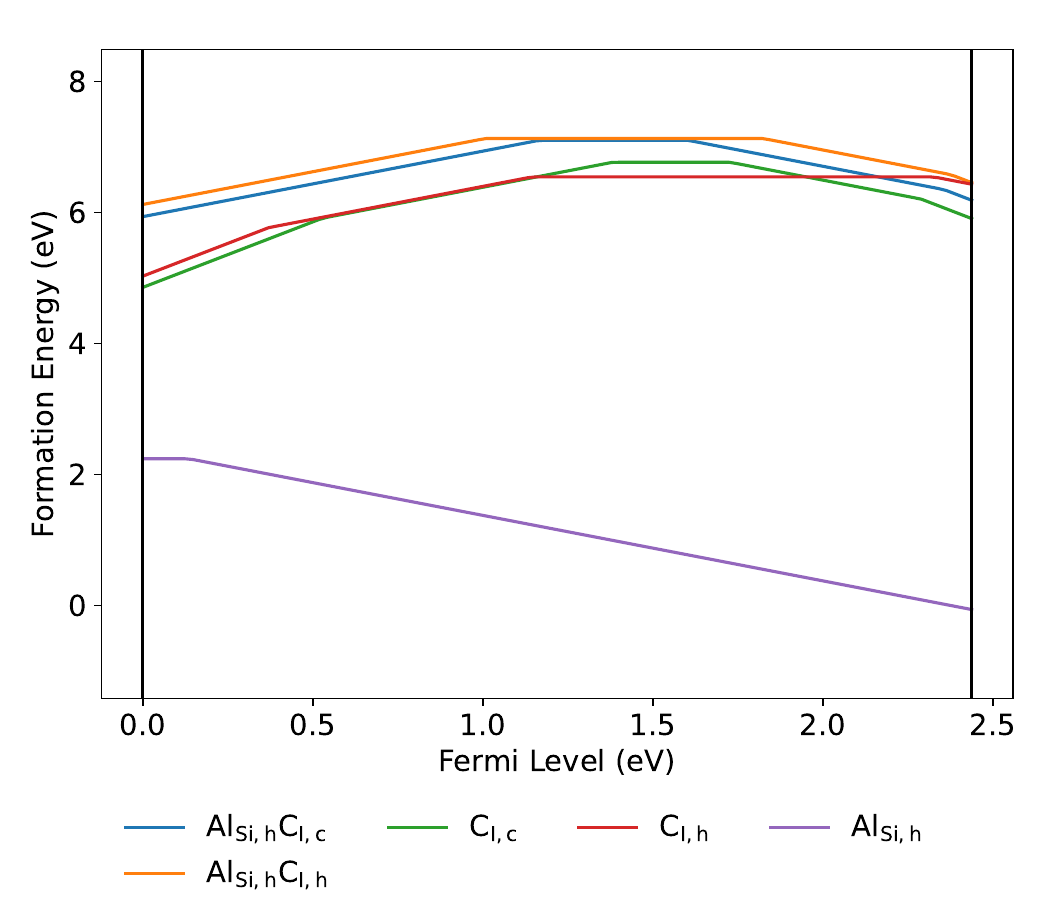}}
\caption{Formation energies of Al-C complexes}
\label{fig:AlC}
\end{figure}
\begin{table}[htbp]
\centering
\caption{Charge transition levels for Al-point defects and  Al-complexes where the (Al--V) complex is from \cite{hornos2008}. The stable charge state in the intrinsic regime highlighted in bold, units in eV.}
\label{tab:levels_complexes}
\begin{tabular}{llllll}
\hline
 config &  &  &  &  &  \\
\hline
 $\mathrm{Al}_\mathrm{Si,h}\mathrm{Si}_\mathrm{I,c}$ & {\bf +3/+2: 1.58  } & +2/-1: 1.70   & -- & -- & -- \\
 $\mathrm{Al}_\mathrm{Si,h}\mathrm{Si}_\mathrm{I,h}$ & +3/+2: 0.20   & {\bf +2/+1: 0.87  } & +1/0: 2.28   & 0/-1: 2.34   & -- \\
 $\mathrm{Al}_\mathrm{Si,h}\mathrm{C}_\mathrm{I,c}$& {\bf +1/0: 1.17  } & 0/-1: 1.60   & -1/-2: 2.36   & -- & -- \\
 $\mathrm{Al}_\mathrm{Si,h}\mathrm{C}_\mathrm{I,h}$& {\bf +1/0: 1.01  } & 0/-1: 1.83   & -1/-2: 2.38   & -- & -- \\
 $\mathrm{Al}_{\mathrm{Si}}\mathrm{V}_{\mathrm{C}}$ & {\bf +1/0: 1.5} & 0/-1: 1.61  & -1/-3: 2.33 & --  &-- \\
$\mathrm{Al}_\mathrm{I,c}$ & {\bf +3/+2: 1.66  } & +2/+1: 1.85   & +1/0: 2.35   & -- & -- \\
$\mathrm{Al}_\mathrm{I,h}$ & { +3/+2: 1.30  } & {\bf +2/+1: 1.70  } & +1/0: 2.15   & -- & -- \\
$\mathrm{Al}_\mathrm{I,c}\mathrm{C}_\mathrm{si,h}$ & +2/+1: 0.22   & {\bf +1/-1: 1.24  } & -- & -- & -- \\
$\mathrm{Al}_\mathrm{I,h}\mathrm{C}_\mathrm{Si,h}$ & +2/+1: 0.41   & {\bf +1/0: 0.85  } & 0/-1: 1.73   & -- & -- \\
\hline
\end{tabular}
\end{table}
Al interstitials in cage configurations are strong donors with stable charge states ranging from $3+$ to $1+$. Their ($+1/0$) transitions occur close to the conduction-band edge, whereas the split-interstitial configurations introduce donor transitions near the valence-band edge and therefore may contribute to compensation under p-type conditions. In particular, $(\mathrm{Al}_{\mathrm{I,c}}\mathrm{C}_{\mathrm{Si,h}})$ exhibits negative-(U) behavior through the ($+1/-1$) transition near mid-gap, suggesting that it may act as an efficient carrier trapping or recombination center.

The Al--Si complexes exhibit predominantly donor-like behavior. The cubic-layer configuration is stable in the (3+) charge state around mid-gap and introduces a deep donor transition $(+3/+2)$ in good qualitative agreement with HSE06 \cite{huang2022}. 
In addition, it displays pronounced negative-(U) behavior through the $(+2/-1)$ transition at (E$_V+1.70$)\,eV. The hexagonal-layer configuration is most stable in the (1+) charge state under intrinsic conditions. It contains a shallow donor level, 
while near the conduction-band edge, the defect changes from donor-like $(+1/0)$ to acceptor-like $(0/-1)$ in qualitative agreement with \cite{huang2022}.
\begin{figure}[htbp]
\subfloat[]{\includegraphics[width=0.9\columnwidth]{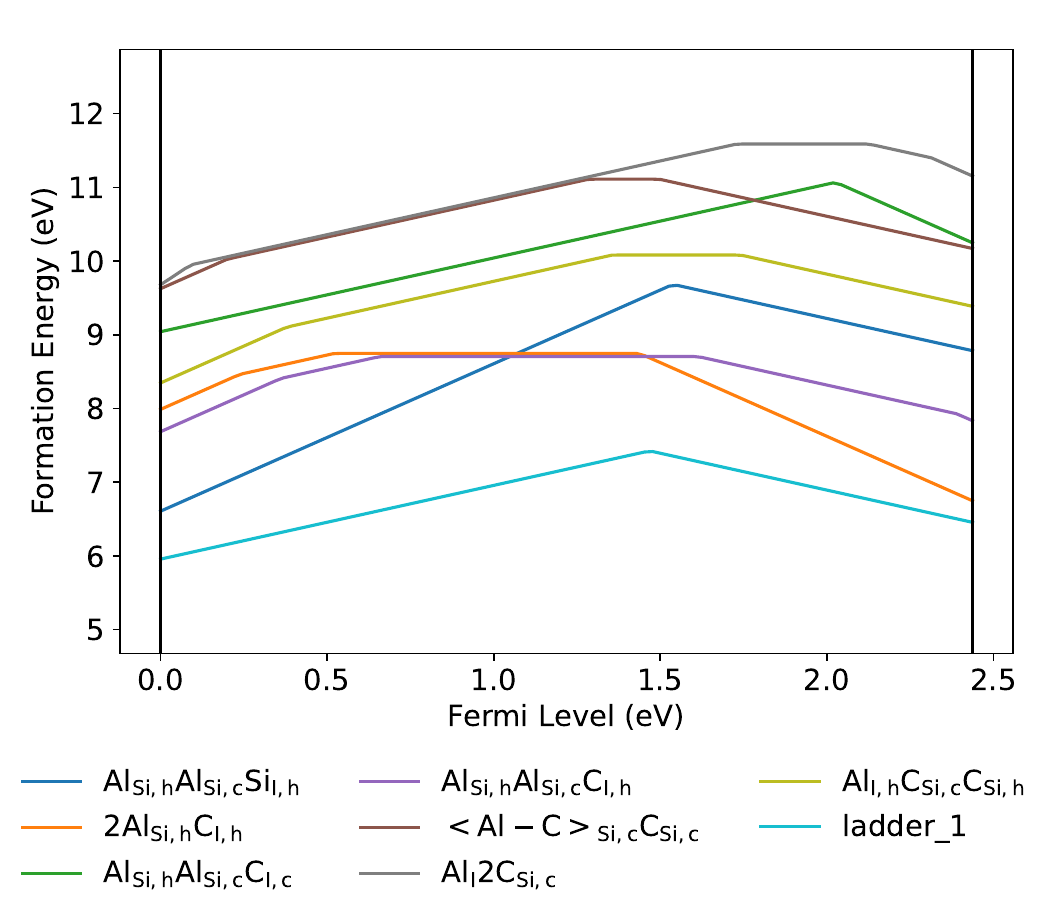}}
\caption{Formation energies of small Al-C clusters.}
\label{fig:clusters}
\end{figure}
\begin{table}[htbp]
\centering
\caption{Transition levels for Al-trimers at different charge states, the stable charge state in the intrinsic regime is highlighted in bold, units in eV.}
\label{tab:levels_cluster}
\begin{tabular}{llllll}
\hline
 config &  &  &  &  &  \\
\hline
$2\mathrm{C}_\mathrm{Si,c}$ & -- & -- & -- & -- & -- \\
$2\mathrm{C}_\mathrm{Si,h}$ & -- & -- & -- & -- & -- \\
$\mathrm{C}_\mathrm{Si,c}\mathrm{C}_\mathrm{Si,h}$ & -- & -- & -- & -- & -- \\
$2\mathrm{Al}_\mathrm{Si,h}$ & 0/-1: 0.17   & -1/{\bf -2: 0.60  } & -- & -- & -- \\
$\mathrm{Al}_\mathrm{Si,h}\mathrm{Al}_\mathrm{Si,c}$ & 0/-1: 0.16   & -1/{\bf -2: 0.56  } & -- & -- & -- \\
$2\mathrm{Al}_\mathrm{Si,h}\mathrm{C}_\mathrm{I,h}$ & +2/+1: 0.24   & +1/0: 0.52   & {\bf 0/-2: 1.44  } & -- & -- \\
$\mathrm{Al}_\mathrm{Si,h}\mathrm{Al}_\mathrm{Si,c}\mathrm{C}_\mathrm{I,c}$ & {\bf 1}/-2: 2.03   & -- & -- & -- & -- \\
$\mathrm{Al}_\mathrm{Si,h}\mathrm{Al}_\mathrm{Si,c}\mathrm{C}_\mathrm{I,h}$ & +2/+1: 0.36   & +1/0: 0.66   & {\bf 0/-1: 1.61  } & -1/-2: 2.39   & -- \\
$\mathrm{Al}_\mathrm{Si,h}\mathrm{Al}_\mathrm{Si,c}\mathrm{Si}_\mathrm{I,h}$ & {\bf +2/-1: 1.54  } & -- & -- & -- & -- \\
$\mathrm{Al}_\mathrm{I}2\mathrm{C}_\mathrm{Si,c}$ & +3/+1: 0.09   & {\bf +1}/0: {\bf 1.73  } & 0/-1: 2.13   & -1/-2: 2.31   & -- \\
$\mathrm{Al}_\mathrm{I,h}\mathrm{C}_\mathrm{Si,c}\mathrm{C}_\mathrm{Si,h}$ & +2/+1: 0.38   & {\bf +1/0: 1.36  } & 0/-1: 1.74   & -- & -- \\
$\mathrm{<Al-C>}_\mathrm{Si,c}\mathrm{C}_\mathrm{Si,c}$ & +2/+1: 0.20   & {\bf +1/0: 1.29  } & {\bf 0/-1: 1.50  } & -- & -- \\
$\mathrm{Al}_\mathrm{Si,h}\mathrm{C}_\mathrm{2}$& +1/{\bf -1: 1.47}   & -- & --    & -- & -- \\
\hline
\end{tabular}
\end{table}

The carbon antisites and their complexes shown in Figure~\ref{fig:clusters} do not introduce electronic states within the band gap. In contrast, complexes containing substitutional Al exhibit a rich charge-state behavior. The paired substitutional defects ($2\mathrm{Al}_{\mathrm{Si,h}}$) and ($\mathrm{Al}_{\mathrm{Si,h}}\mathrm{Al}_{\mathrm{Si,c}}$) behave as shallow acceptors and are stable in the (2-) charge state around mid-gap, similar to the isolated ($\mathrm{Al_{Si}}$) defect.
Most of the dimer and trimer complexes studied in Tables~\ref{tab:levels_complexes} and ~\ref{tab:levels_cluster} have deep and shallow compensating levels. Thus, they are summarized together with the most stable charges states and binding energies in dependence of Fermi level in Table~\ref{tab:clusters}.

\subsection{Binding energies of Al complexes}
\begin{figure}[htbp]
\includegraphics[width=0.9\columnwidth]{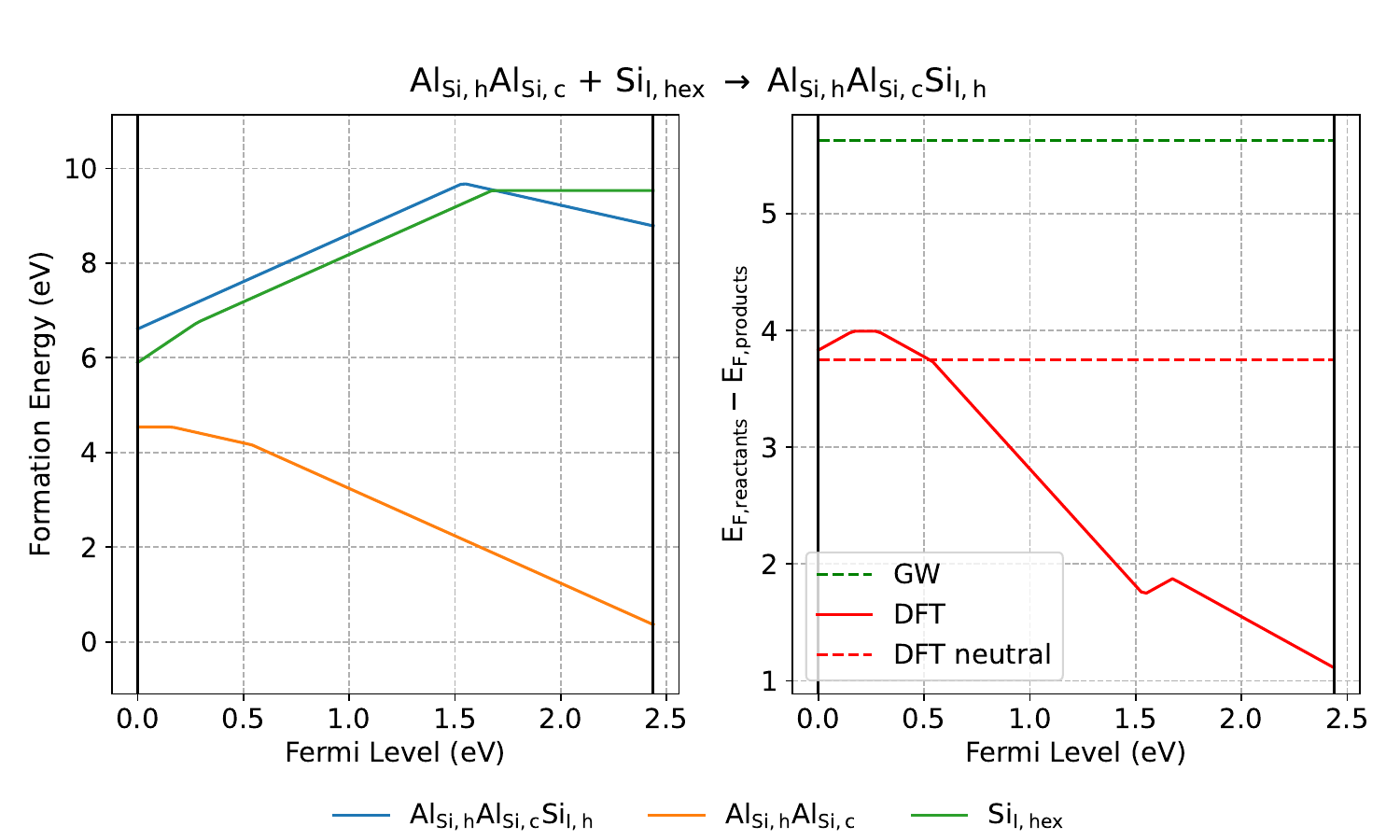}
\caption{Formation energy and binding energy of the $\mathrm{Al}_\mathrm{Si,h}\mathrm{Al}_\mathrm{Si,c}\mathrm{Si}_\mathrm{I,h}$ complex along the c-axis in the hexagonal plane. A positive energy means that the complex is more stable than its reactants.}
\label{fig:AlAlSi}
\end{figure}

\begin{figure}[htbp]
\includegraphics[width=0.9\columnwidth]{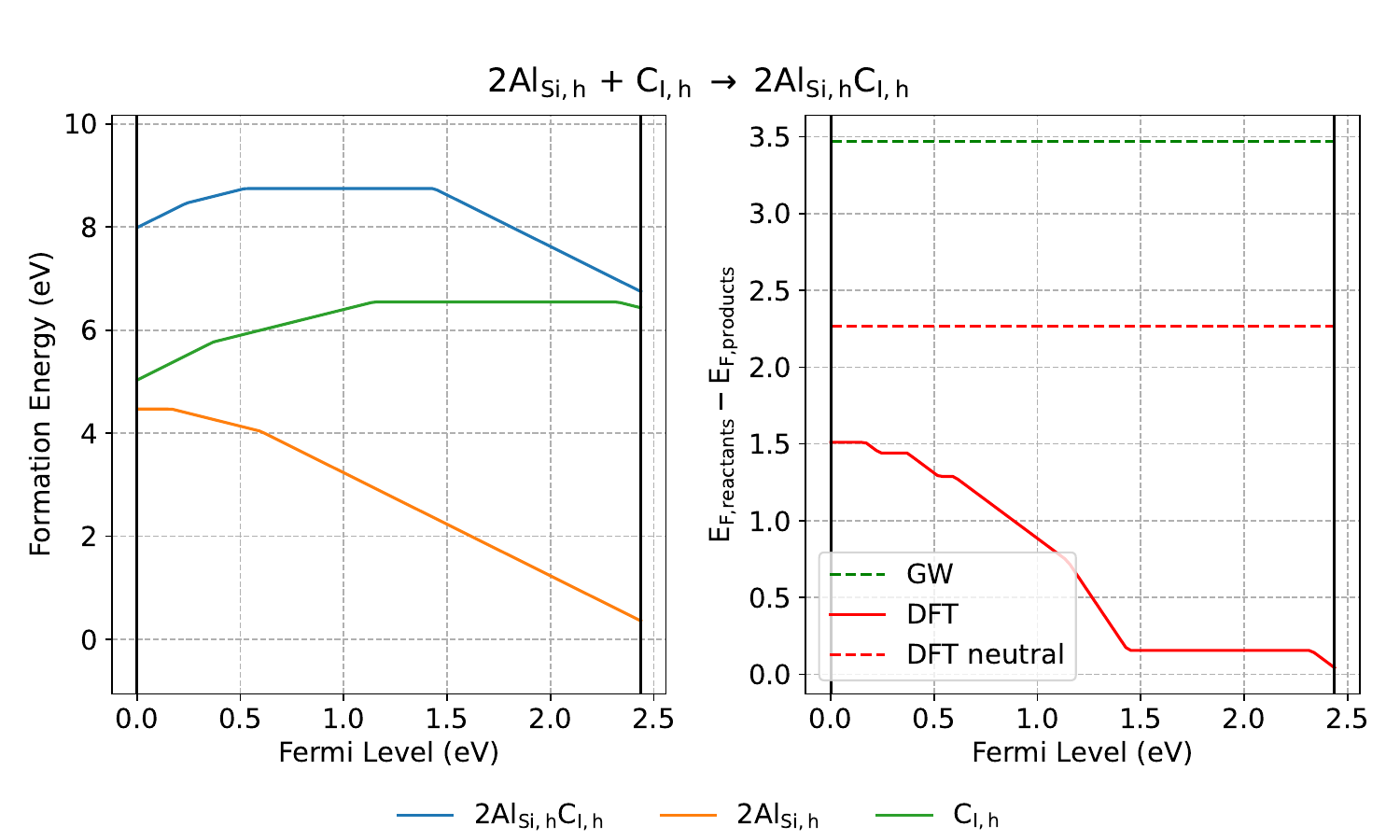}
\caption{Formation energy and binding energy of the $2\mathrm{Al}_\mathrm{Si,h}\mathrm{C}_\mathrm{I,h}$ complex in the hexagonal basal plane in $\langle0\overline110\rangle$ direction. A positive energy means that the complex is more stable than its reactants.}
\label{fig:2AlCh}
\end{figure}

\begin{figure}[htbp]
\includegraphics[width=0.9\columnwidth]{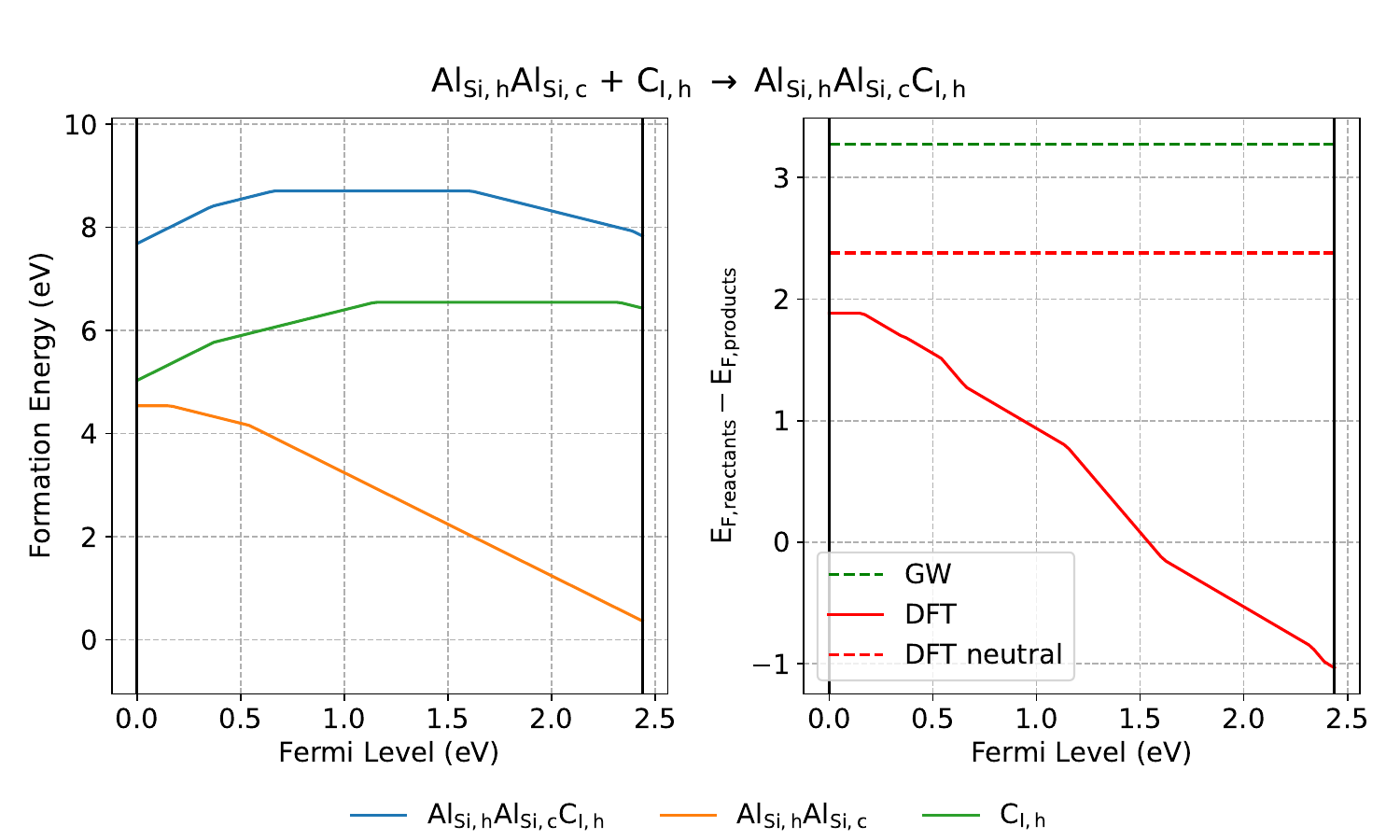}
\caption{Formation energy and binding energy of the $\mathrm{Al}_\mathrm{Si,h}\mathrm{Al}_\mathrm{Si,c}\mathrm{C}_\mathrm{I,h}$ complex along the c-axis in the hexagonal plane. A positive energy means that the complex is more stable than its reactants.}
\label{fig:AlAlCh}
\end{figure}

\begin{figure}[htbp]
\includegraphics[width=0.9\columnwidth]{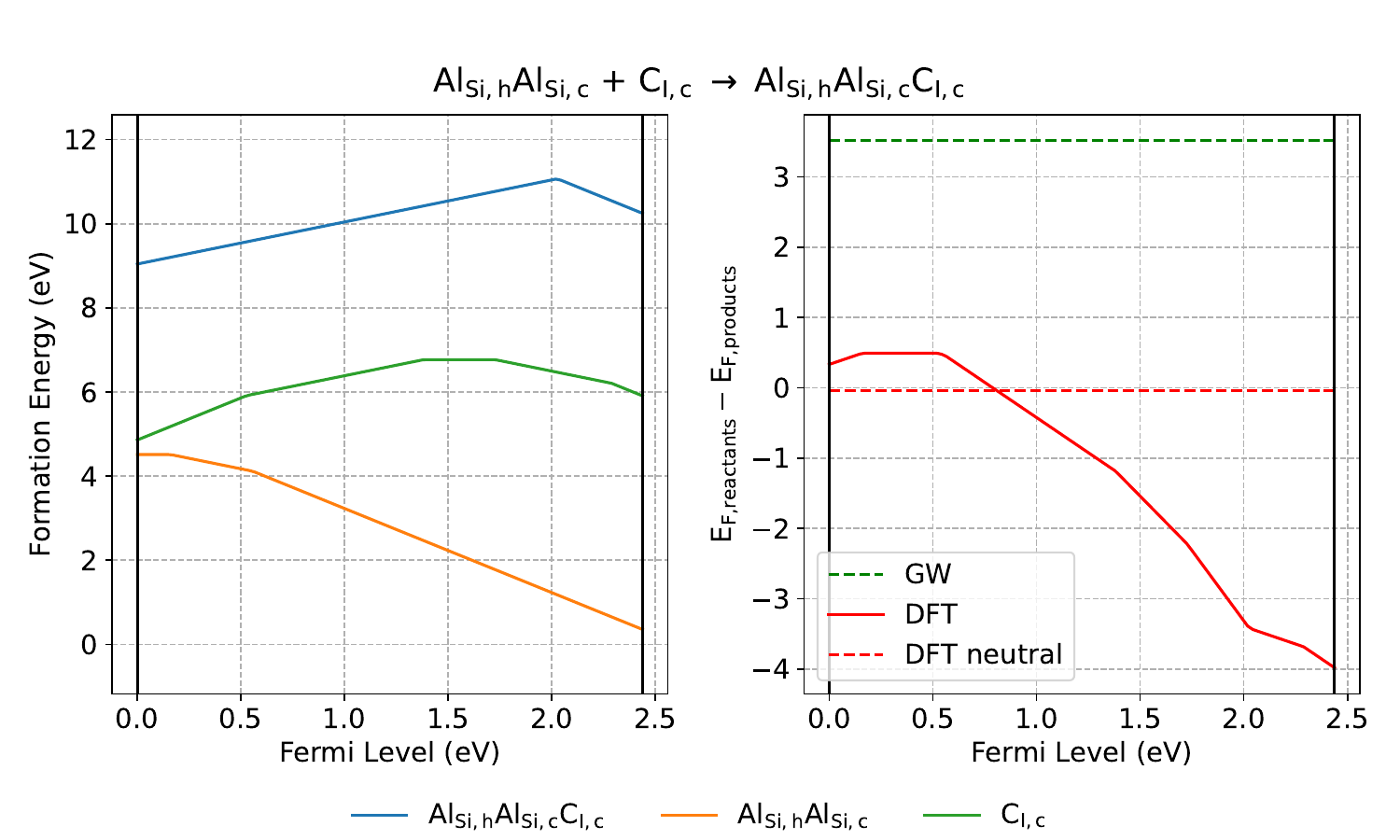}
\caption{Formation energy and binding energy of the $\mathrm{Al}_\mathrm{Si,h}\mathrm{Al}_\mathrm{Si,c}\mathrm{C}_\mathrm{I,c}$ complex along the c-axis in the cubic plane. A positive energy means that the complex is more stable than its reactants.}
\label{fig:AlAlCc}
\end{figure}

\begin{figure}[htbp]
\includegraphics[width=0.9\columnwidth]{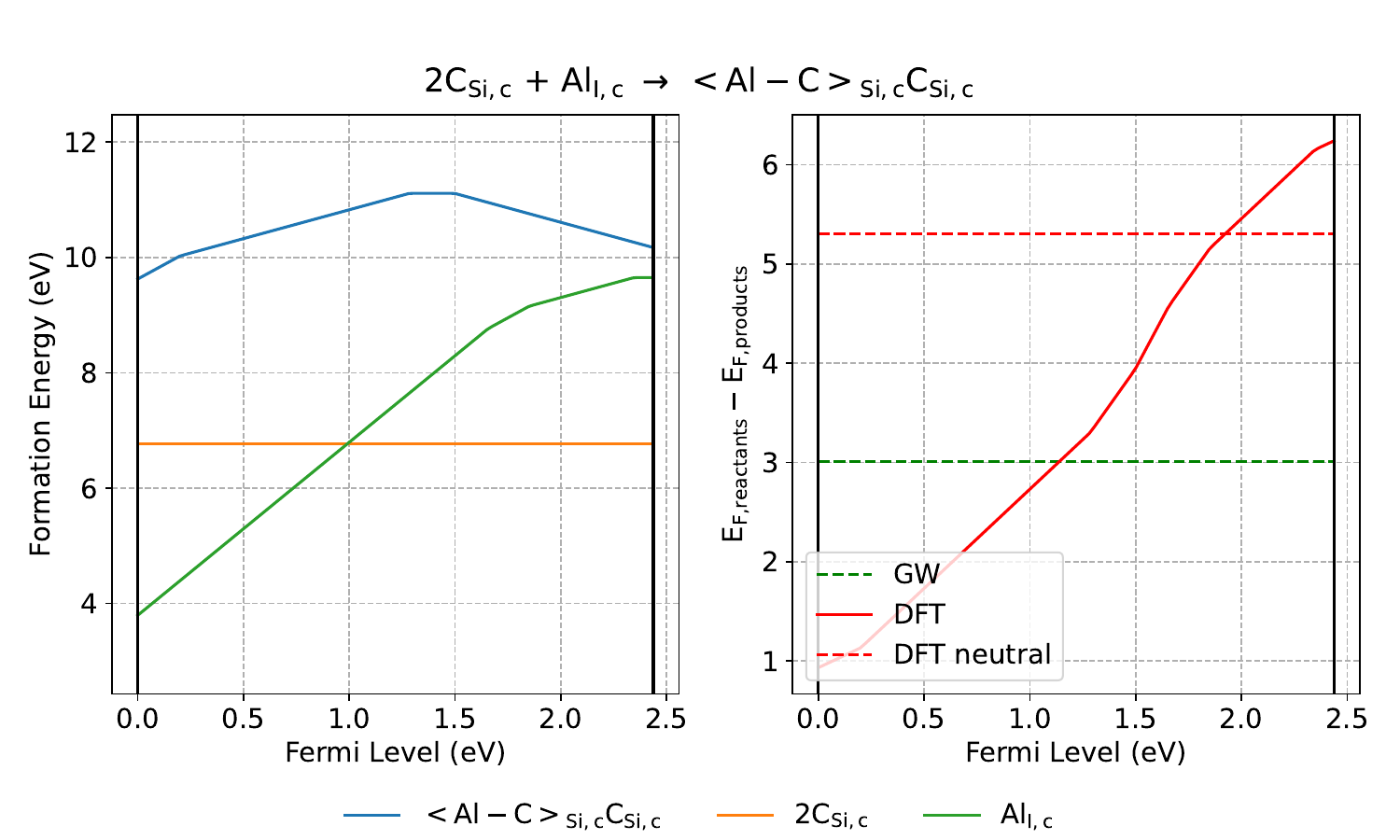}
\caption{Formation energy and binding energy of the $\mathrm{Al}_\mathrm{I,c}2\mathrm{C}_\mathrm{Si,c}$ complex in the cubic basal plane along the $\langle0\overline110\rangle$ direction. A positive energy means that the complex is more stable than its reactants.}
\label{fig:Al2CSi}
\end{figure}

\begin{figure}[htbp]
\includegraphics[width=0.9\columnwidth]{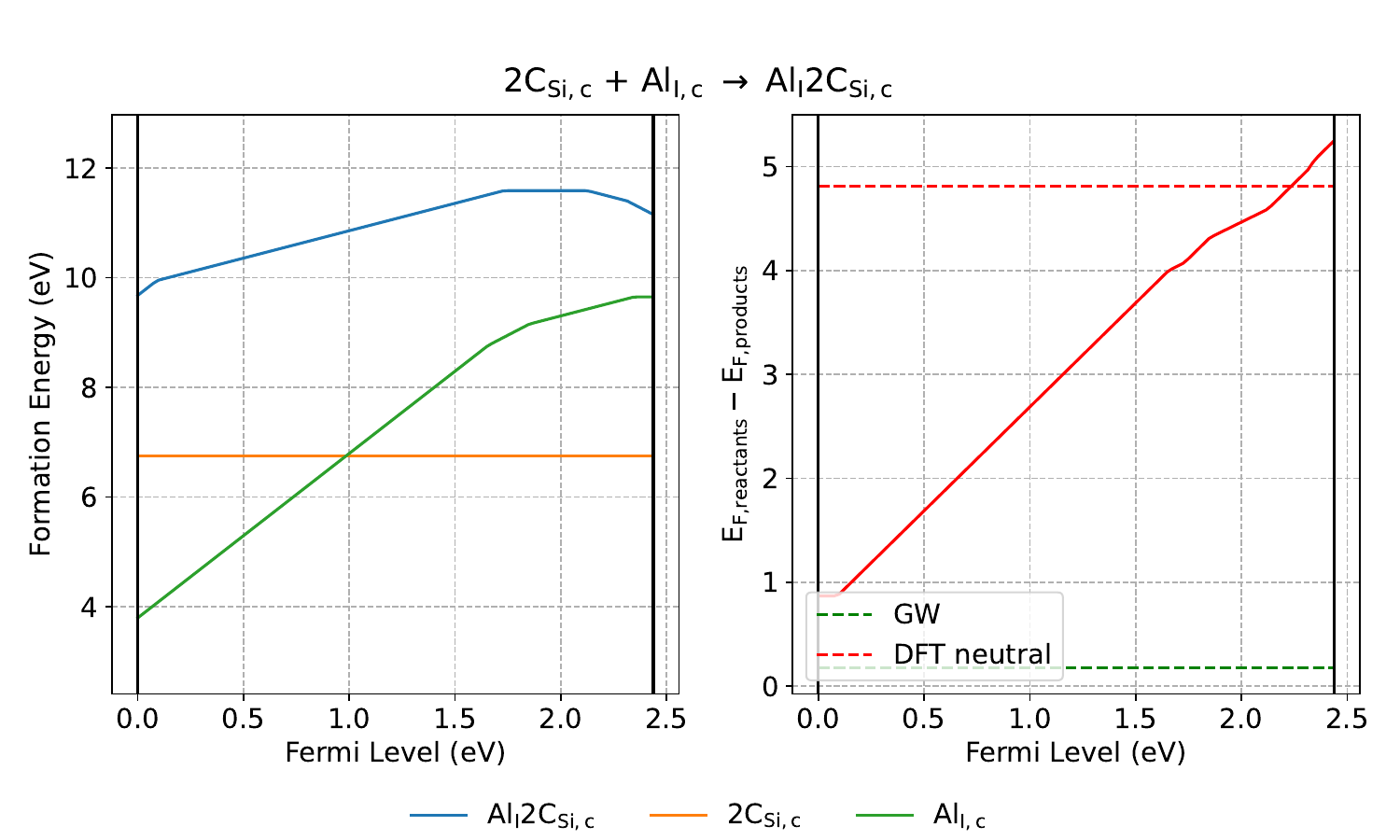}
\caption{Formation energy and binding energy of the $\mathrm{Al}_\mathrm{I}2\mathrm{C}_\mathrm{Si,c}$ complex in the cubic basal plane along the  $\langle\overline12\overline10\rangle$ direction. A positive energy means that the complex is more stable than its reactants.}
\label{fig:Al2CSi_1210}
\end{figure}

\begin{figure}[htbp]
\includegraphics[width=0.9\columnwidth]{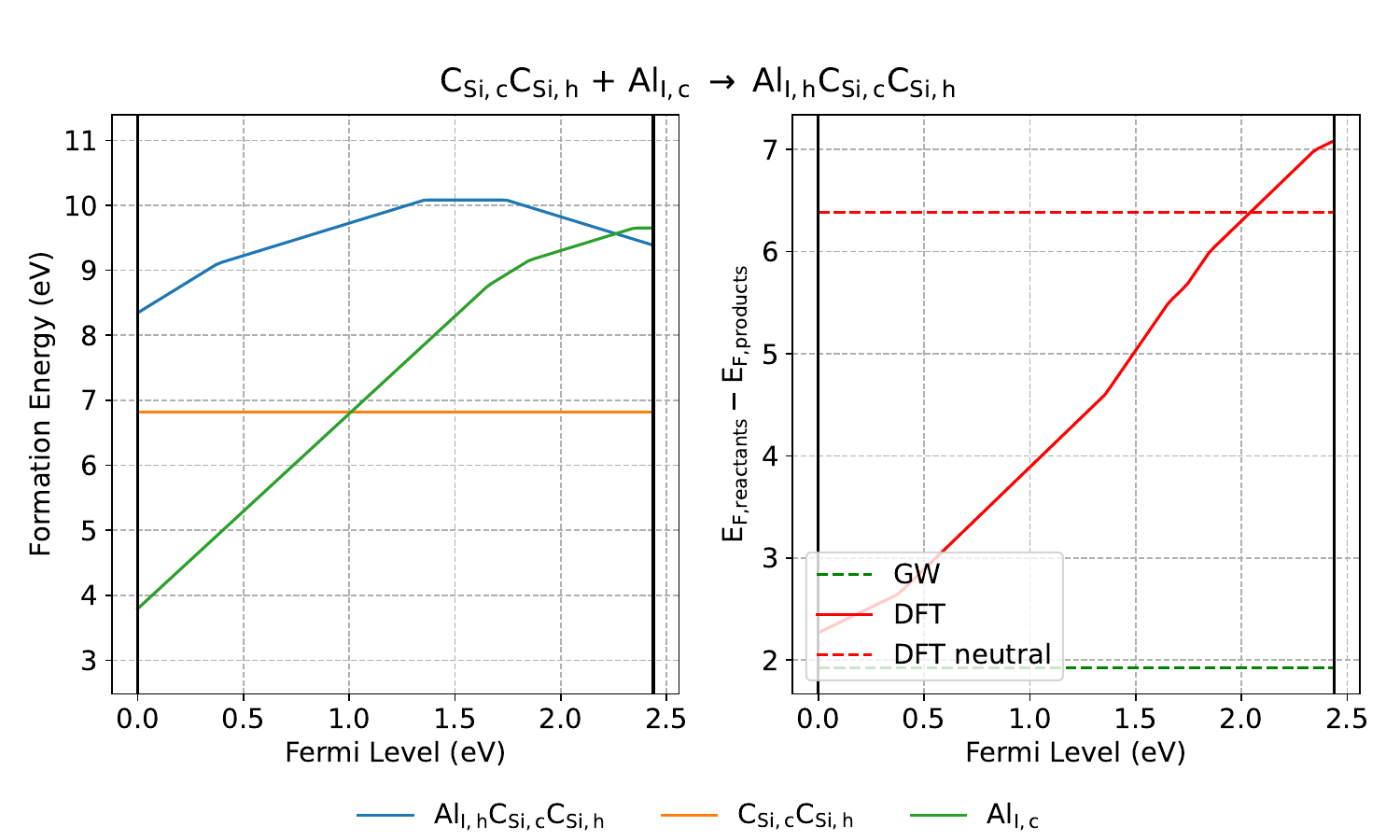}
\caption{Formation energy and binding energy of the $\mathrm{Al}_\mathrm{I,h}\mathrm{C}_\mathrm{Si,c}\mathrm{C}_\mathrm{Si,h}$ complex along the c-axis in the hexagonal plane. A positive energy means that the complex is more stable than its reactants.}
\label{fig:AlCSiCSi}
\end{figure}

After having discussed the defect levels and their compensation properties, we now investigate the thermodynamic stability of the complexes. In addition to the Fermi-level dependent binding energies, the neutral binding energies obtained from DFT and GW are shown as dashed lines in Figures~\ref{fig:AlAlSi} to ~\ref{fig:AlCSiCSi}.
A clear trend emerges: complexes containing substitutional Al ($\mathrm{Al_{Si}}$) generally become less stable as the Fermi level increases, whereas complexes containing Al interstitials ($\mathrm{Al_I}$) exhibit increasing binding energies towards n-type conditions. This behavior reflects the different charge states involved in the dissociation processes. Electron capture for complexes with Al$_{Si}$ when the carbon or silicon interstitial diffuses away and electron emission for complexes containing Al$_I$ when the Al interstitial is released. 

Among the investigated trimers, $\mathrm{Al_{I,h}C_{Si,c}C_{Si,h}}$ is the most stable configuration. Remarkable is that the out of plane configuration of this complex is considerably more stable than the in-plane configurations $\mathrm{Al}_\mathrm{I}2\mathrm{C}_\mathrm{Si,c}$ and $\mathrm{<Al-C>}_\mathrm{Si,c}\mathrm{C}_\mathrm{Si,c}$.
Moreover, the $\mathrm{Al}_\mathrm{Si,h}\mathrm{Al}_\mathrm{Si,c}\mathrm{C}_\mathrm{I,h}$ configuration is significantly more stable than $\mathrm{Al}_\mathrm{Si,h}\mathrm{Al}_\mathrm{Si,c}\mathrm{C}_\mathrm{I,c}$. Stability thus depends not only on the stoichiometry but also on the position of the interstitials within the defect and on the orientation of the defect within the crystal.
Overall, carbon-rich complexes containing Al interstitials are considerably more stable than the corresponding complexes formed by substitutional Al atoms. The GW potential overestimates the stability of trimers containing substitutional Al and underestimates the stability of complexes containing Al interstitials. 

The thermodynamic stability, dominant charge states, and compensation properties of all investigated complexes are summarized in Table~\ref{tab:clusters}.
\begin{table}
\caption{Bond lengths in (\AA) within the nearest neighbor shell of Al inside the trimers, where $I$ denotes a placeholder for C$_{Si}$, C$_I$ or Si$_I$. For comparison the $Si-C$ bond length in pristine SiC is $1.89$~\AA.}
\label{tab:bonds}
\begin{center}
\begin{tabular}{||l|lllll||}
\hline
   defect& Al & C & C & C & I\\
\hline
$\mathrm{Al}_\mathrm{Si,h}\mathrm{Al} _\mathrm{Si,c}\mathrm{C}_\mathrm{I,c}$ &$\mathrm{Al}_\mathrm{Si,h}$ & 1.9 & 2.0 & 1.9 & 1.9\\
&$\mathrm{Al}_\mathrm{Si,c}$ & 1.9 & 2.0 & 1.9 & 2.4\\
$\mathrm{Al}_\mathrm{Si,h}\mathrm{Al} _\mathrm{Si,c}\mathrm{C}_\mathrm{I,h}$ & $\mathrm{Al}_\mathrm{Si,c}$ & 2.1 & 2.0 & 2.1 & 2.0\\
&  $\mathrm{Al}_\mathrm{Si,h}$ & 2.3 & 2.2 & 2.1 & 2.1\\
$2\mathrm{Al}_\mathrm{Si,h}\mathrm{C}_\mathrm{I,h}$ & $\mathrm{Al}_\mathrm{Si,h}$ & 1.9 & 2.0 & 2.0 & 1.9\\
$\mathrm{Al}_\mathrm{I,c}2\mathrm{C}_\mathrm{Si,c}$ & $\mathrm{Al}_\mathrm{I,c}$ & 2.0 & 1.8 & - & 1.8\\
$\mathrm{Al}_\mathrm{I,h}\mathrm{C}_\mathrm{Si,c}\mathrm{C}_\mathrm{Si,h}$ & $\mathrm{Al}_\mathrm{I,h}$ & 2.1 & 2.0 & 2.0 & 1.9\\
$\mathrm{Al}_\mathrm{Si,h}\mathrm{Al}_\mathrm{Si,c}\mathrm{Si}_\mathrm{I,h}$ & $\mathrm{Al}_\mathrm{Si,h}$ & 1.9 & 2.1 & 1.8 & 2.2\\
& $\mathrm{Al}_\mathrm{Si,c}$ &  2.0 & 2.1 & 1.8 & 2.4\\
\hline
\end{tabular}
\end{center}
\end{table}

The calculated Al–C bond lengths in Table~\ref{tab:bonds} range from approximately $1.8$ to $2.3$~\AA, indicating moderate local lattice distortions around the defect complexes. Most Al–C distances are close to the Si–C bond length in pristine 4H-SiC ($1.89$~\AA), suggesting that substitutional Al preserves much of the tetrahedral bonding environment of the host lattice. Larger bond lengths, particularly those exceeding $2.1$\AA, indicate local strain induced by neighboring interstitials or antisite defects. These structural distortions modify the local electrostatic potential and orbital hybridization around the Al atom, which can influence the stability of different charge states and the position of defect-induced electronic levels within the band gap. The relatively small variation of the Al–C bond lengths ($1.8$–$2.3$~\AA) indicates that the Al impurity largely retains the tetrahedral coordination characteristic of the 4H-SiC lattice. Consequently, the electronic behavior is primarily governed by the defect charge state and the presence of neighboring intrinsic defects rather than by major structural reconstructions.
\pagestyle{empty}
\begin{sidewaystable}
\caption{Summary of binding energies, most stable charge states and compensation/trapping properties of all investigated defects in the p-type ($0.2$\,eV left column), intrinsic ($1.2$\,eV middle column) and n-type ($2.4$\,eV right column) regime, where negative U transitions are highlighted in bolt. At the end of the table the neutral binding energy E$_B^{GW}$ from the GW potential is added. 
Complexes marked with \textasciitilde are data from reference \cite{gali2007}, respectively.}
\resizebox{\linewidth}{!}{
\begin{tabular}{||l|lll|lll|lll|l||}
\hline
  & E$_B$ (eV) & q & comp/trap & E$_B$ (eV)&q & comp/trap & E$_B$ (eV)& q &  comp/trap & E$_B^{GW}$ (eV)\\
\hline
$\mathrm{Al}_\mathrm{Si,h}\mathrm{Al}_\mathrm{Si,c}\mathrm{C}_\mathrm{I,c}$ & 0.5   & 1+ &  & -1.0  & 1+ & & -4.0  & 2- & {\bf (+1/-2): 2} & 3.3\\
$\mathrm{Al}_\mathrm{Si,h}\mathrm{Al}_\mathrm{Si,c}\mathrm{C}_\mathrm{I,h}$ & 1.8   & 2+ & (+2/+1): 0.36 & 0.5  & 0 & (+1/0): 0.66, (0/-1): 1.6 & -1.0  & 1- & {\bf (-1/-2): 2.39} & 3.3\\
$2\mathrm{Al}_\mathrm{Si,h}\mathrm{C}_\mathrm{I,h}$ & 1.5  & 2+ & (+2/+1): 0.24  & 0.8 & 0 & {\bf (0/-2): 1.4}  & 0.0  & 2- &  & 3.5\\
$\mathrm{Al}_\mathrm{I}2\mathrm{C}_\mathrm{Si,c}$ & 0.5   & 1+ &  {\bf (+3/+1): 0.09} &  3  & 1+ &(+1/0): 1.73  & 5.5  & 2- & (0/1-):2.1, (-1/-2): 2.3 & 0.2\\
$\mathrm{<Al-C>}_\mathrm{Si,c}\mathrm{C}_\mathrm{Si,c}$ & 1.5    & 2+ &  (+2/+1): 0.2 & 3.0  & 0 &(+1/0): 1.29, (0/-1): 1.5 & 6.5  & 1- &  & 3\\
$\mathrm{Al}_\mathrm{I,h}\mathrm{C}_\mathrm{Si,c}\mathrm{C}_\mathrm{Si,h}$ & 2.5    & 2+ & (+2/+1): 0.38 & 4.5  & 1+ & (+1/0): 1.36, (0/-1): 1.7 & 7.0  & 1- &  & 2 \\
$\mathrm{Al}_\mathrm{Si,h}\mathrm{Al}_\mathrm{Si,c}\mathrm{Si}_\mathrm{I,h}$ & 4  & 2+ &  & 2.5 & 2+& {\bf (+2/-1): 1.54} & 1.1 & 1- &  & 5.8\\
\hline
$\mathrm{Al}_\mathrm{Si,h}\mathrm{Si}_\mathrm{I,h}$& 1.8 & 3+&     (+3/+2): 0.2 &  1.1 & 1+ & (+2/+1): 0.87, & -0.1 & 1- & (+1/0); 2.28, (0/-1): 2.34 & 1.3\\
$\mathrm{Al}_\mathrm{Si,h}\mathrm{Si}_\mathrm{I,c}$& 2.2 & 3+ &    &    1.8 & 3+ & (+3/+2): 1.58, {\bf (+2/-1): 1.7} & 0.3 & 1- &  & \\
$\mathrm{Al}_\mathrm{Si,h}\mathrm{C}_\mathrm{I,c}$& 1.3 & 1+ &    &   0.5 & 0 & (+1/0): 1.17, (0/-1): 1.6 & -0.3 & 2- &  (-1/-2): 2.36 & 1.4\\
$\mathrm{Al}_\mathrm{Si,h}\mathrm{C}_\mathrm{I,h}$& 1.1 & 1+ &    &   0.5 & 0 & (+1/0): 1.01, (0/-1): 1.83 & -0.6 & 2- &   (-1/-2): 2.38 &\\
$2\mathrm{Al}_\mathrm{Si,h}$ & 0.2  & 1- & (0/-1): 0.17  & -0.1 & 2- &  (-1/-2): 0.6 & -0.5  & 2- & &\\
$\mathrm{Al}_\mathrm{Si}\mathrm{V}_\mathrm{C}$\textasciitilde & 0.5 & 1+ &     &  0.5 &  1+ & (1/0): 1.5, (0/-1): 1.61 & -0.5 & 2- &  -1/-2: 2.33  & \\
\hline
\end{tabular}
}
\label{tab:clusters}
\end{sidewaystable}
\pagestyle{plain}
\newpage
\subsection{Al diffusion}
\begin{figure}[htbp]
\centering
    \subfloat[]{\includegraphics[width=0.7\columnwidth]{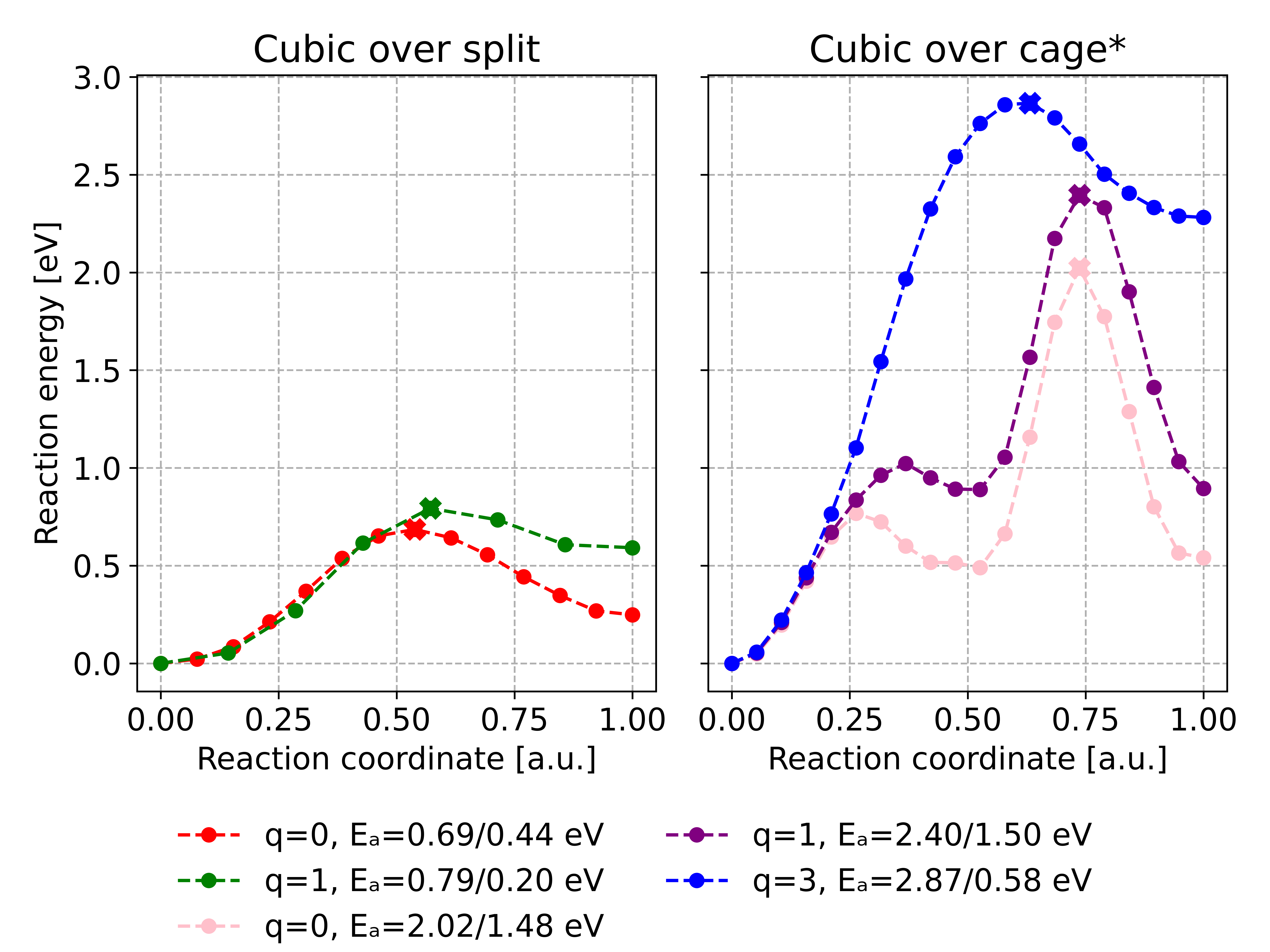}}\\
    \subfloat[]{\includegraphics[width=0.7\columnwidth]{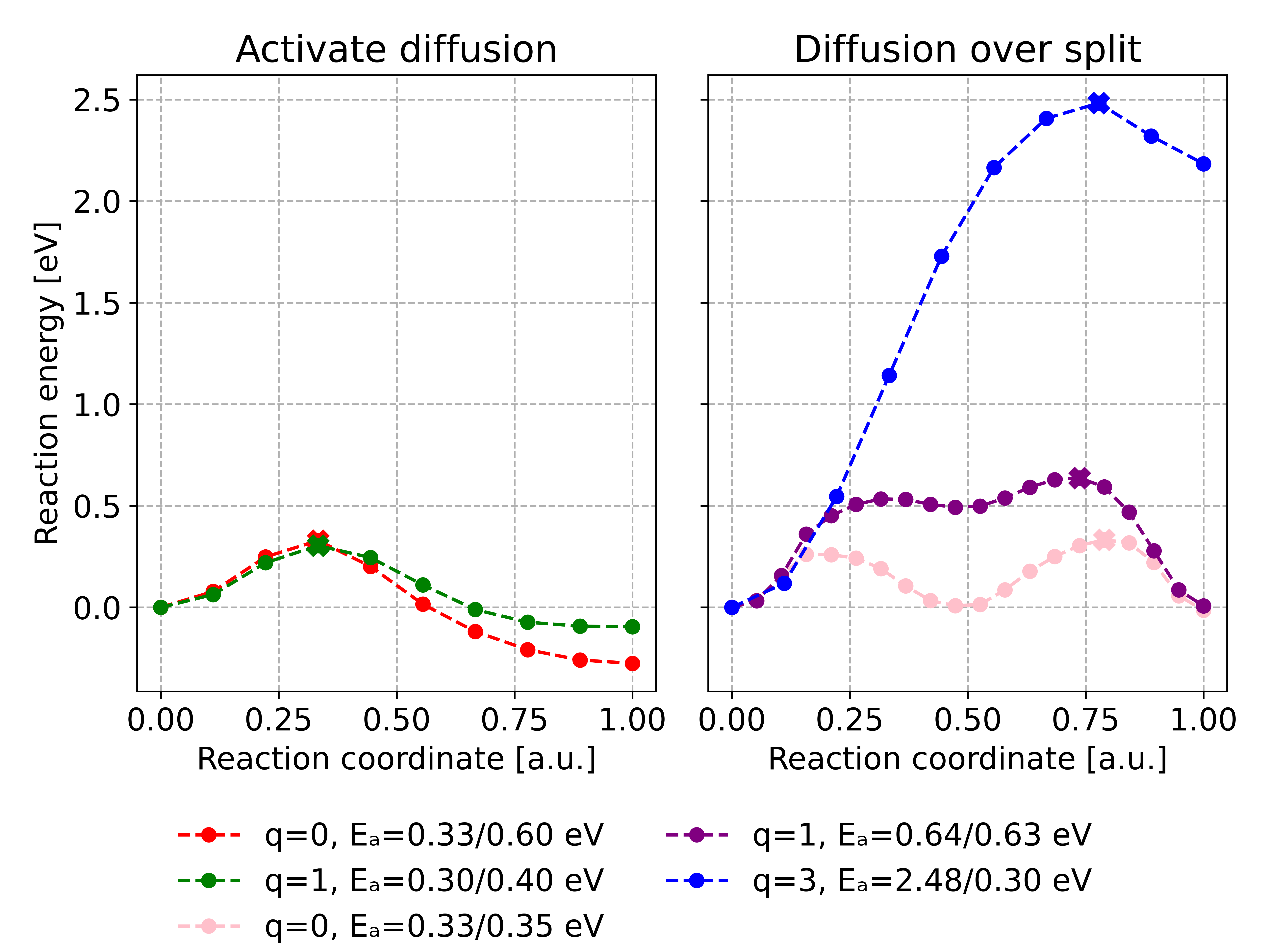}}
\caption{a) (left) Migration barrier and diffusion path for charge state $0$ and $1+$ in the intrinsic to n-type regime, where diffusion occurs by hops between Al$_{I,c}$ and $<110>(Al-Si)_h$ configurations. (right) Migration barrier and diffusion path for the $3+$ charge state in p-type, where the diffusion occurs between Al$_{T,c}$ and Al$_{T,Si}$ cage configurations in the cubic layer. (b) For charge $0$ and $1+$ the diffusion in the hexagonal layer proceeds via hops between $<110>(Al-Si)_h$ and $<110>(Al-Si)_c$ split configurations passing over the Al$_h$ as metastable state, while for charge state $3+$ in p-type the hops are directly between Al$_h$ and $<110>(Al-Si)_c$ configurations.}
\label{fig:diffusion}
\end{figure}

In the following reaction equations the energies in brackets correspond to forward and backward barriers. The configuration in the middle of the reaction equation is the unstable transition state. 
The diffusion paths in the cubic and hexagonal layer are shown in Figure~\ref{fig:diffusion}. The diffusion barriers and paths in hexagonal and cubic layer under p-type, intrinsic and n-type conditions are summarized in Table~\ref{tab:diffusion}. The calculation of the barriers is based on DFT CI-NEB calculations as described in the main document. 
\paragraph{{\bf p-type via cage in the cubic layer:}}
\begin{equation}
\boxed{\ce{(Al_{I,c})^3+ ->[(2.9/0.6) eV](Al_{TSi})^3+}}
\label{eqn:diff3+_cub}
\end{equation}
\paragraph{\bf p-type via split in the hexagonal layer:} 
The diffusion path in the hexagonal plane is different from the path in the cubic plane, which now proceeds over a split configuration. For constraint charge $3+$ the migration starts in the Al$_h$ position and proceeds via a transition state to the basal (Al-Si) split configuration. However, the split configuration in p-type is most stable in charge state $2+$ and not in $3+$.
The $3$+ charge stabilizes Al$_I$ at the beginning of the diffusion reaction and $2$+ stabilizes the split at the end of the reaction.  By applying the adiabatic approximation \cite{coutinho2021} it is assumed that the reaction proceeds on the potential energy surface (PES) $3+$, and at the end the (Al-Si) split configuration spontaneously captures an electron and jumps to the ground state on PES $2+$.
\begin{equation}
\boxed{\ce{(Al_{I,h})^3+ +e^- ->[(2.5/0.3)] 110(Al-Si)_{c}^2+}}
\label{eqn:diff3+_hex}
\end{equation}
\paragraph{\bf intrinsic cubic layer:}
In the intrinsic regime ($0.9$-$1.6$\,eV) the diffusion reaction is the same as in equation (\ref{eqn:diff3+_cub}) in cubic p-type.
\paragraph{\bf intrinsic hexagonal layer:}
For Fermi level between $1.0-1.3$\,eV the diffusion process proceeds as described in equation \ref{eqn:diff3+_hex} with a barrier of $2.5$\,eV. For higher Fermi levels it is energetically most favorable when Al$_{I,h}$$^{2+}$ which has a transition level ($+2/+1$) at $1.7$\,eV jumps to PES $1+$ at the beginning of the reaction by capturing an electron (at a cost of $1.7-E_F$). 
\begin{equation}
\boxed{\ce{(Al_{I,h})^2+ +e^- ->[(0.6 + (1.7 -E_F )/ 0.6)] 110(Al-Si)^1+}_{c}}\:\:\: (1.3\le E_F<1.7)
\label{eqn:diff2+_hex}
\end{equation}
\paragraph{\bf n-type cubic layer:}
\begin{equation}
\boxed{\ce{(Al_{I,c})^1+ ->[(0.8/0.2) eV] 110(Al-Si)_{h}^1+}}\:\:\:1.9\le E_F<2.3
\label{eqn:diff1+_cub}
\end{equation}
\begin{equation}
\boxed{\ce{(Al_{I,c}) ->[(0.7/0.4) eV] 110(Al-Si)_{h}}}\:\:\:E_F>2.3
\label{eqn:diff0_cub}
\end{equation}
{\bf n-type hexagonal layer:}
In the hexagonal layer the diffusion path changes, since the split configuration in the hexagonal layer is energetically lower than the Al$_h$ configuration
The diffusion starts with an activation hop from $<110>(Al-Si)_{h}$ split interstitial and proceeds on PES 1+ or PES 0, depending on the Fermi level:
\begin{equation}
\boxed{\ce{110(Al-Si)_{h}^1+ ->[(0.4/0.3)] Al_{I,h}^{1+} ->[(0.6/0.6)] 110(Al-Si)_{c}^1+}}\:\:\:1.7\le E_F<2.2
\label{eqn:diff1+_hex}
\end{equation}
\begin{equation}
\boxed{\ce{110(Al-Si)_{h} ->[(0.6/0.3)] Al_{I,h} ->[(0.3/0.4) eV ]110(Al-Si)_{c}}}\:\:\:E_F>2.2
\label{eqn:diff0_hex}
\end{equation}
\pagestyle{empty}
\begin{sidewaystable}
\caption{Summary of diffusion barriers, stable charge states and diffusion paths in the p-type ($0 - 0.5$\,eV, left column), intrinsic ($1.3 - 1.6$\,eV, middle column) and n-type ($1.9 - 2.4$\,eV, right column) regime.}
\begin{tabular}{||l|lll|lll|lll||}
\hline
  layer & mechanism & q & barrier (eV)& mechanism &q & barrier (eV) & mechanism & q & barrier (eV)\\
\hline
cubic & cage  & 3+ & (2.9/0.6) & cage & 3+ & (2.9/0.6) & split  & 1+  & (0.8/0.2) \\
&    & &  & & & & split & 0  & (0.7/0.4) \\
hexagonal & split   & 3+ &  (2.5/1.1) & split & 3+  &  (2.5/1.1) & split + activation & 1+  & (0.6/0.6) \\
&   &  &  & split & 2+ & (1.7-E$_F$+0.6/0.6) & split + activation  & 0 & (0.6/0.4) \\
\hline
\end{tabular}
\label{tab:diffusion}
\end{sidewaystable}
\pagestyle{plain}
Summary: In p-type the Al interstitial in the cubic layer Al$_c^{3+}$ diffuses via cage jumps from Al$_{TC}^{3+}$ on the tetragonal site to the metastable Al$_{TSi}^{3+}$ cage with a barrier of $2.9$\,eV. Al$_h$ diffuses via a barrier of $2.5$\,eV on PES 3+ by capturing an electron at the end of the reaction. In the intrinsic to n-type regime Al diffuses via the meta-stable $\langle$110$\rangle$(Al-Si) split configuration, with significantly lower barriers depending on the Fermi level in the hexagonal layer of around $0.6$-$1.0$\,eV. The diffusion paths and barriers in dependence of the Fermi level is once again summarized in Table~\ref{tab:diffusion}.
\newpage
\subsection{Al activation}
Al activation via the Si or carbon antisite is a two-step process that begins with the Al kick-in, which leads to the formation of the Al complex; in the second step, this complex decomposes into a substitutional Al and a free intrinsic interstitial.
\begin{figure}[htbp]
\includegraphics[width=0.9\columnwidth]{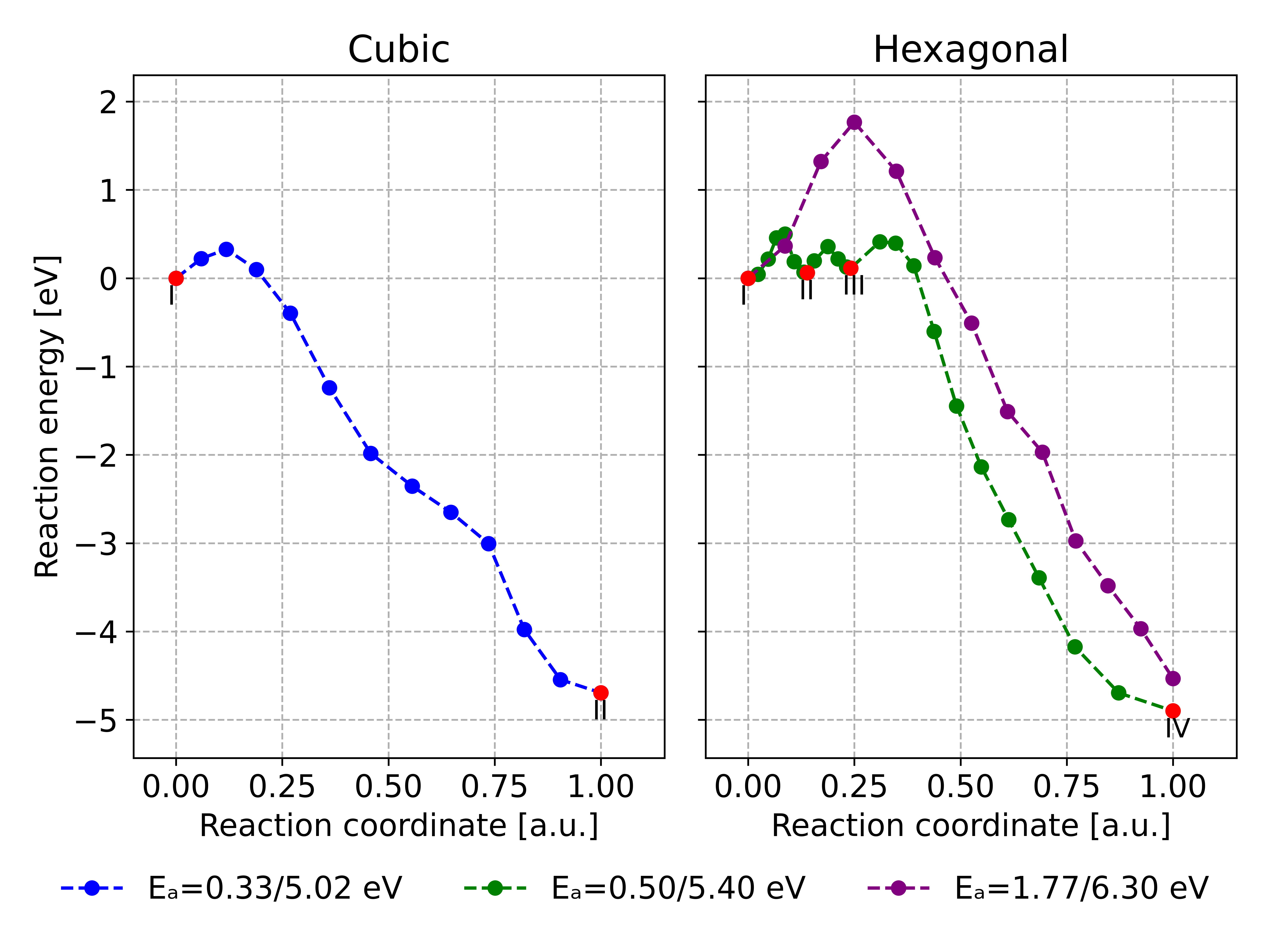}
\caption{Kick-in barrier for the neutral Al interstitial to the carbon antisite starting from the cubic or hexagonal layer. The formed neutral (Al-C)$_{Si}$ split-interstitial is considerably more stable than the neutral Al interstitial in cage configuration.}
\label{fig:kickin}
\end{figure}
\begin{figure}[htbp]
\centering
\vspace{0.0cm}
\subfloat[]{
    \centering
       \includegraphics[width=.45\columnwidth]{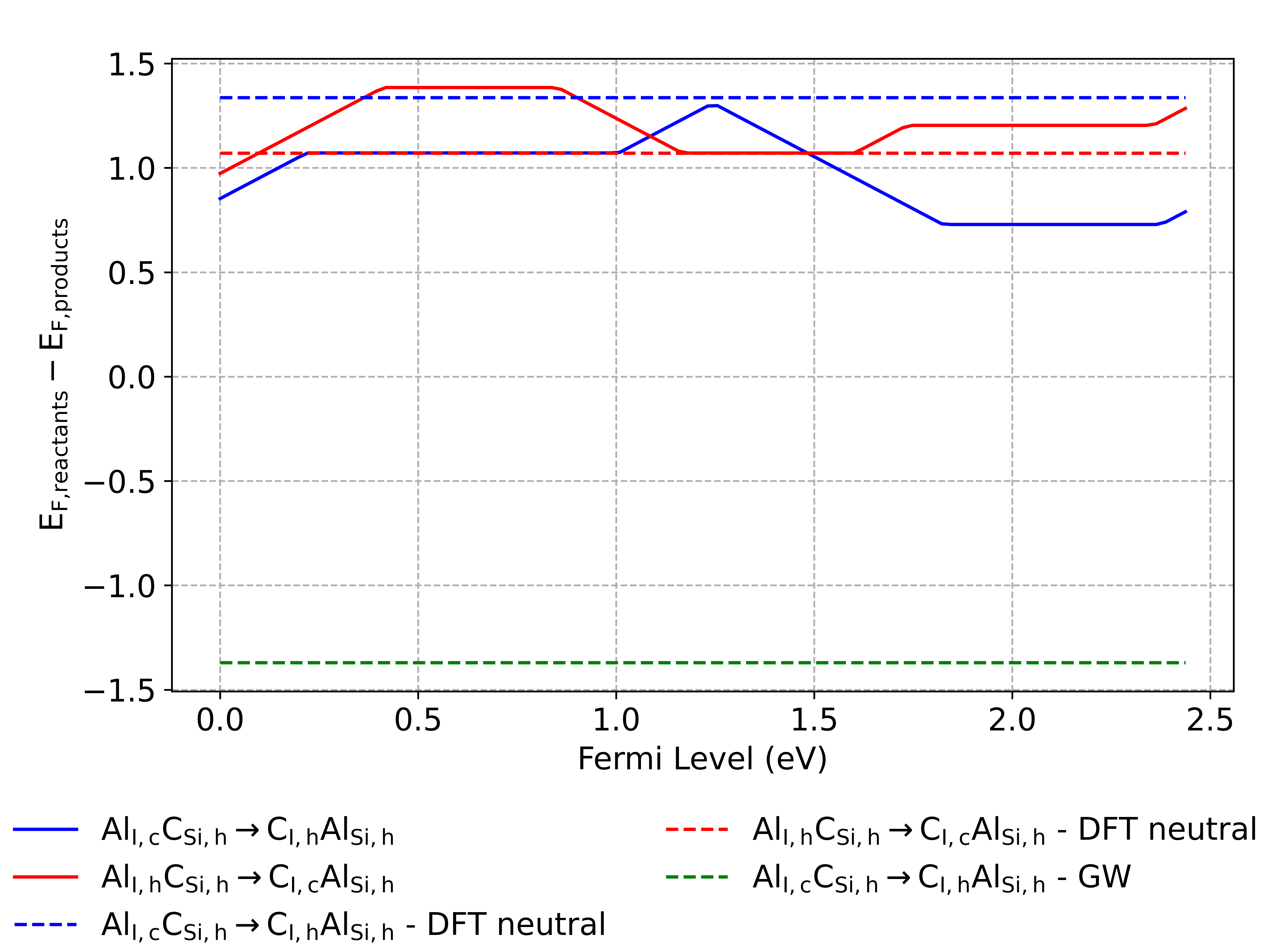}
}
\hfill
\subfloat[]{
    \centering
        \includegraphics[width=.45\columnwidth]{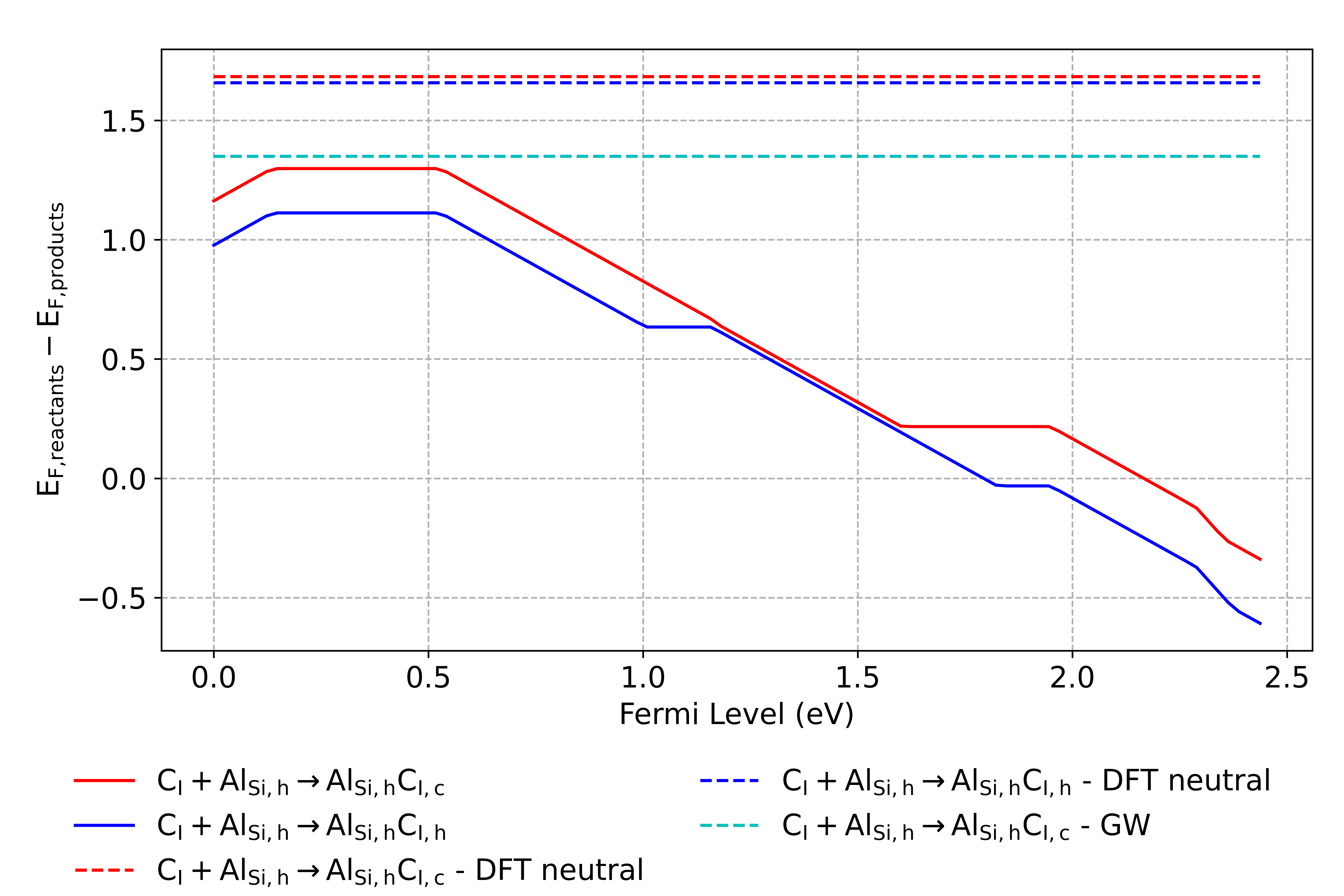}
}
\caption{Binding energy diagrams for reactions in the cubic (blue) and hexagonal (red) plane in dependence of Fermi level, where $\mathrm{Al}_\text{I}\mathrm{C}_\text{Si}$ is referred to as the (Al-C)$_{Si}$ split in the band gap. The dashed lines in (b) represent  binding energies for neutral charges, where the line in cyan corresponds to the energy deduced for the GW potential. Positive energies mean that the complex is thermodynamically more stable than the split configuration or the isolated defects.}
\label{fig:binding_energies_AlC_defects}
\end{figure}
\subsubsection{Activation via the carbon antisite}
The kick-in of the Al interstitial to the carbon antisite is always accompanied by electron capture which destabilizes the interstitial and lowers the kick-in barrier as shown as an example for the neutral Al interstitial in Figure~\ref{fig:kickin}. 
The further progress of the reaction from the (Al-C) split configuration to the complex and its subsequent decay into isolated defects is displayed in Figure ~\ref{fig:binding_energies_AlC_defects} (a) and (b) respectively. In the Figure one can see plateau regions where the reactions proceed on constant charge PES. If the binding energy increases or decreases with the Fermi level, electron absorption or emission occurs. The transition from the Al$_I$C$_{Si}$ split configuration to the more stable Al$_{Si}$C$_I$ complex in part (a) of the figure occurs on both the hexagonal and cubic planes across wide regions of the band gap without electron transfer on  PES $1+$, $0$ or $1-$. The dissolution of the complex in part (b), on the other hand, is always associated with the absorption of electrons, except in small plateau regions, which is accompanied by a weakening/drop of the complex binding energy. For selected Fermi levels the activation via the carbon antisite is further summarized in Table~\ref{tab:activation}.
Binding energies are taken from Figures~\ref{fig:point_defects}, ~\ref{fig:kickin} and ~\ref{fig:binding_energies_AlC_defects}, where positive energies indicate that the product on the right side of the equation is more stable than the reactants, thus judging the thermodynamic stability of Al-complexes. Whenever full kinetics is available for selected charge states barriers from Figure~10 of the main document are included in brackets.

In summary: the Al$_{Si}$C$_I$ complex is a trapping state in the entire band gap. With rising Fermi level the probability to separate into Al$_{Si}$ and C$_I$ increases. In n-type the complex becomes more and more metastable and decays more or less spontaneously in dependence of kinetic barriers into its components. In p-type there is a high recombination barrier for the formation of the split configuration of around $3$\,eV, which lowers with rising Fermi level. Thus, the activation process happens preferentially in the intrinsic to n-type regime.
\subsubsection{Activation via Silicon site}
\begin{figure}[htbp]
\includegraphics[width=0.9\columnwidth]{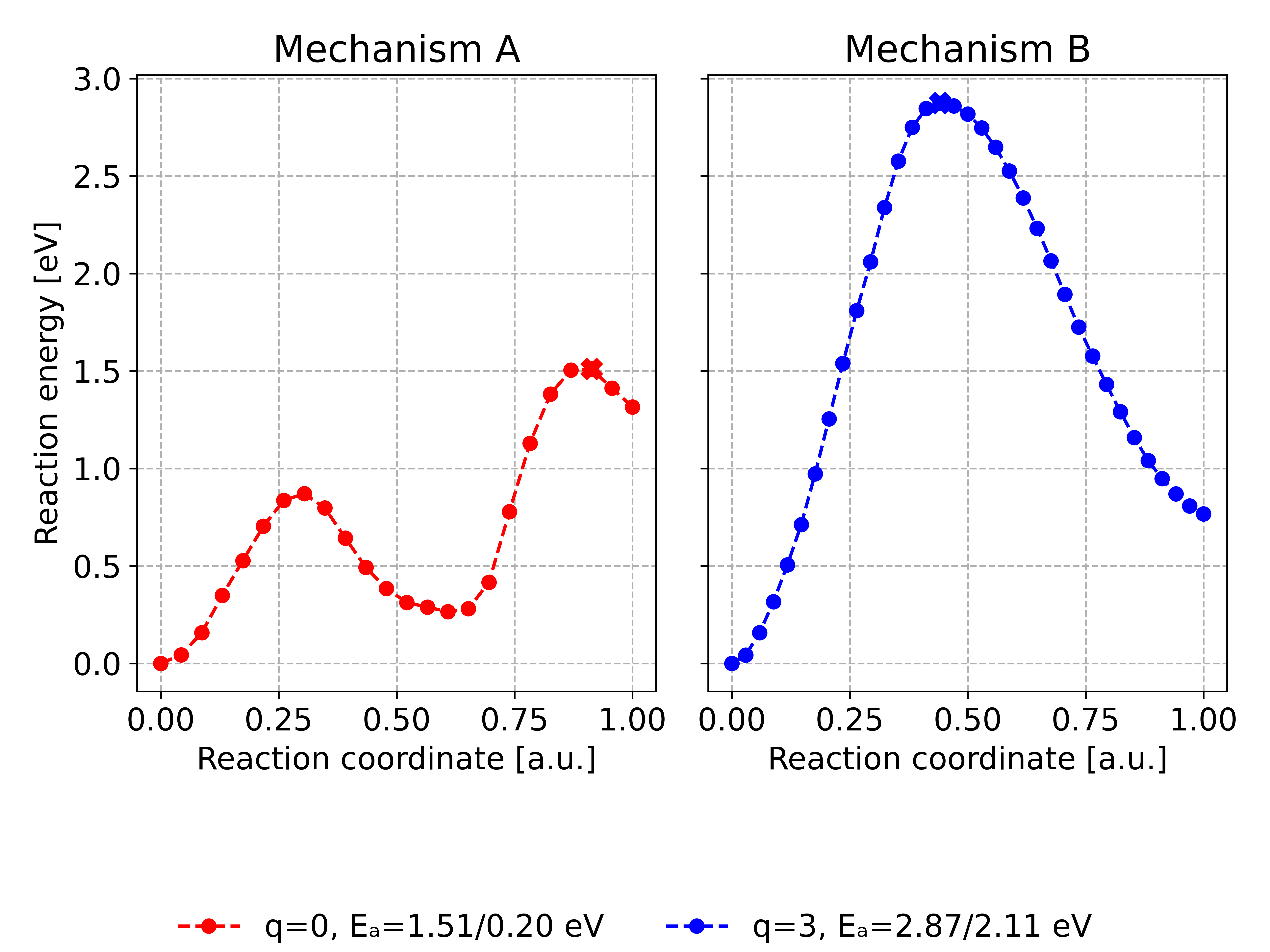}
\caption{Al$_{h}$ kick-in to the Silicon on the cubic site. Two different path are shown: Mechanism A for charges $0$ and $1+$ proceeds via the $(\overline{1}2\overline{1}0)<Al-Si>$ split configuration as metastable state and mechanism B for the $3+$ charge state which advances directly to the transition state without passing over the basal split.}
\label{fig:kickinSi}
\end{figure}
\begin{figure}[htbp]
\centering
\vspace{0.0cm}
    \subfloat[]{
       \includegraphics[width=.45\columnwidth]{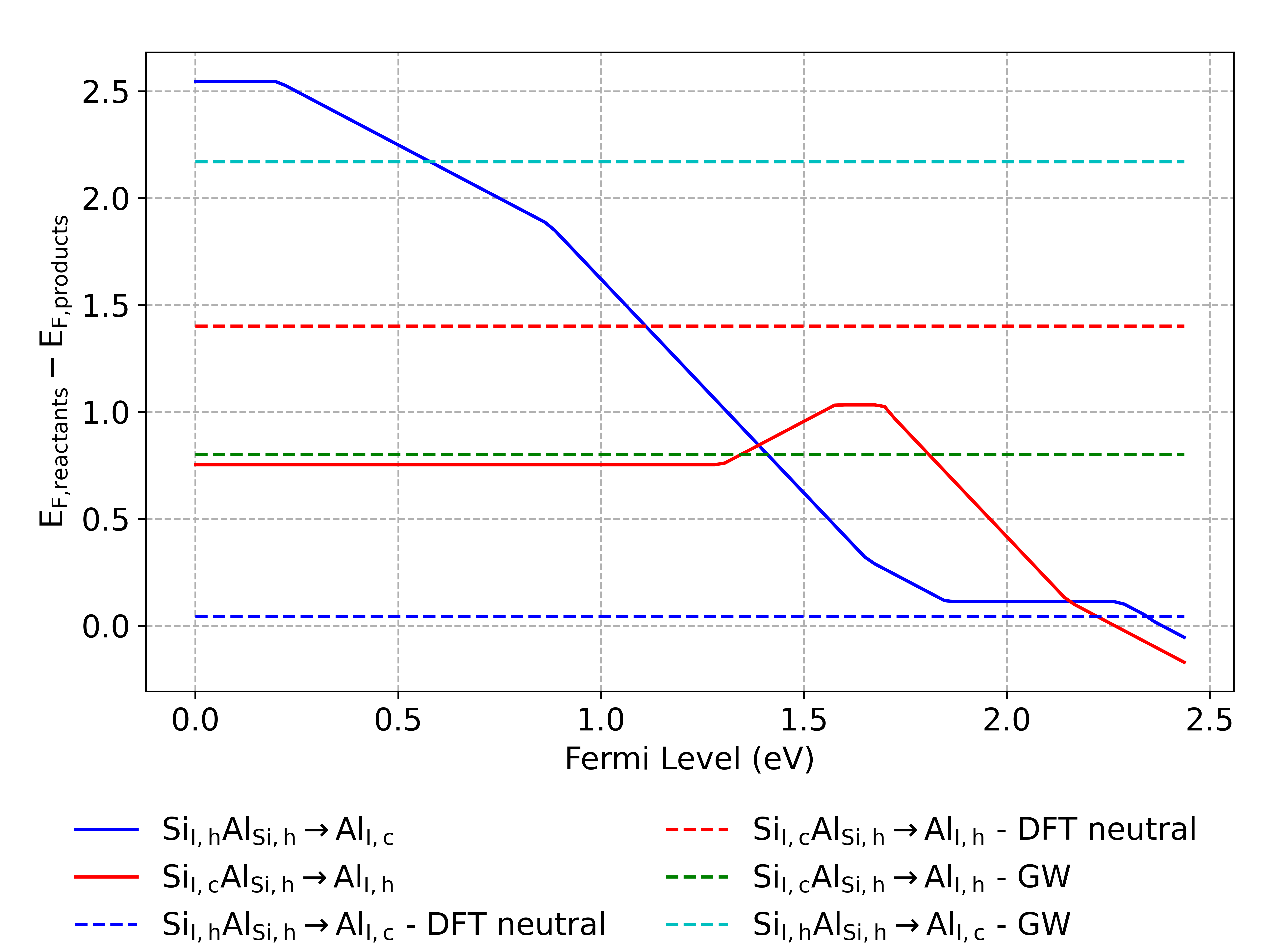}}       
    \subfloat[]{
       \includegraphics[width=.45\columnwidth]{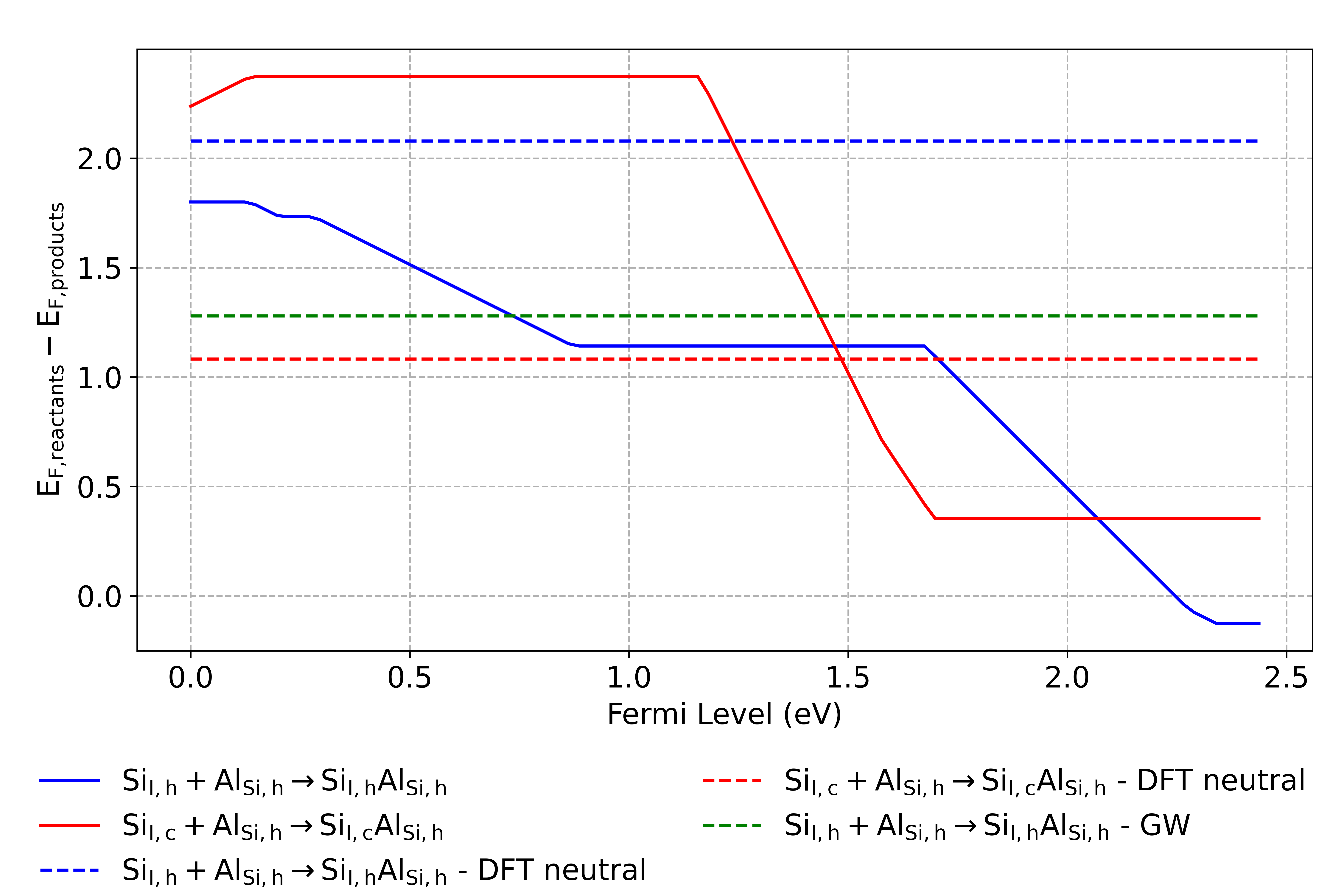}}
\caption{Binding energy diagrams for reactions in the cubic (blue) and hexagonal (red) plane in dependence of Fermi level, where $\mathrm{Al}_\text{Si}\mathrm{Si}_\text{I}$ is referred to as the (Al-Si) complex. The dashed lines  represent  binding energies for neutral charges, where the line in green and cyan corresponds to the energy deduced for the GW potential. Negative energies in (a) mean that the complex is thermodynamically more stable than Al$_I$, positive energies in (b) mean that it is more stable than the isolated defects.}
\label{fig:binding_energies_AlSi_defects}
\end{figure}

When the (Al-Si) complex forms via the kick-in of the Al interstitial, the complex remains thermodynamically unstable against Al$_I$ throughout most of the band gap; it is more stable only in the n-type regime as can be seen in Figure~\ref{fig:binding_energies_AlSi_defects}(a). In the cubic plane, complex formation is accompanied by electron capture in the  intrinsic regime whereas in the hexagonal plane, electrons are captured only under n-type conditions. When the complex decays in part (b) of the figure, the process is reversed: in the hexagonal plane, the reaction proceeds without electron capture in the intrinsic regime, while in the cubic plane, the complex dissolves at constant charge in the n-type regime. In both cases, the complex is more stable throughout the band gap than the isolated defects, only in the n-type region Al$_{Si,h}$Si$_h$ becomes thermodynamically unstable, allowing activation. The binding energies are taken from Figure~\ref{fig:binding_energies_AlSi_defects}, the barriers from Figure~\ref{fig:kickinSi}. 

The activation reactions in dependence of Fermi level including barriers are once more summarized in Table~\ref{tab:activation}.
The table shows the stability of the complex; a negative energy indicates that the defect is unstable, while a positive energy indicates that it is stable along the reaction pathway under consideration. The energy on the left-hand side describes the difference in formation energies between the interstitial Al and the complex; the energy on the right-hand side describes the difference between the complex and the substitutional Al. 

For the hexagonal layer, the kinetic barriers for the two decay processes are also given.  Only on the hexagonal plane the process of complex formation and decay proceeds entirely on PES 3+ up to mid gap. 
Very striking is the absorption of $4$ electrons taking place during decay of the complex for Fermi levels between $1.3$ and $1.7$\,eV visible as jump in the red curve in \ref{fig:binding_energies_AlSi_defects}(b), which is caused by the negative U-behavior of the Al$_{Si,h}$Si$_c$ complex. In the n-type regime, the complex on the hexagonal plane forms on PES 0; at the end, the complex spontaneously transitions to the $1-$ ground state and decays. 

On the cubic plane, both formation and decay occur entirely on the PES 1- for a Fermi level close to the CBM. At mid-gap, Al$^{3}_{I,c}$ has to captures $2$ electrons at a cost of $0.7$\,eV, before the reaction takes place on PES $1+$. 

In addition the GW barrier for the kick-in of (Al$_{I,c}$)$^0$  together with the binding energies are shown in comparison to the kick-in of Al$_{I,c}^{2+}$ at Fermi level $1.7$\,eV accompanied by the capture of an electron.
Summary: The complex Al$_{Si,h}$Si$_{i,c}$ is thermodynamically very stable in p-type up to the intrinsic regime where it shows constant binding energies. In n-type its stability diminishes due to the the capture of electrons as can be seen in Figure~\ref{fig:binding_energies_AlSi_defects}. However, because of the high migration barrier of the silicon interstitial the complex represents a trapping state, which with higher probability returns to the more stable Al$_{I,h}$ configuration than to dissolve into separate defects. The stability of the (Al-Si) complex is depending on whether the silicon interstitial can be found in the hexagonal or cubic layer. The complex Al$_{Si,h}$Si$_{i,h}$ is considerably less stable under p-type conditions, under n-type conditions it becomes thermodynamically unstable and just persists because of high kinetic barriers.  In contrast to the complex in the cubic layer it has a moderate constant binding energy over the entire intrinsic regime. 
The (Al-Si) complex thus does not contribute to Al activation, but supports transient enhanced diffusion. 
\pagestyle{empty}
\begin{sidewaystable}
\caption{Summary of stable charge states and binding energies with respect to the decay towards separation into Al$_I$ on the left and towards Al$_{Si}$ on the right side (if available including barriers) for (Al-C) and (Al-Si) complexes in the p-type ($0.2$\,eV left column), intrinsic ($1.2$\,eV, middle column) and n-type ($2.4$\,eV, right column) regime. In the last line the kick-in from the cubic plane at Fermi level $1.7$\,eV is added, which is closest to the GW barriers in the neutral state.}
\resizebox{\linewidth}{!}{
\begin{tabular}{||l|ll|l||ll|l||ll|l||}
\hline
  defect & q & binding energy & (barrier) (eV) & q & binding energy & (barrier) (eV) & q & binding energy & (barrier) (eV)\\
\hline
 $\mathrm{Al}_\mathrm{Si,h}\mathrm{C}_\mathrm{I,h}$ & 1+ & 1.1/1.1 & & 0 & 1.2/0.5 & & 1-  & 0.7/-0.5 &\\
$\mathrm{Al}_\mathrm{Si,h}\mathrm{C}_\mathrm{I,c}$ & 1+ &  1.2/1.3  & & 0  & 1.1/0.5 & (0.25/1.3) (1.6/1.1) & 1-  & 1.1/-0.2 &\\
&   &  &  &  &  &  &  & & \\

$\mathrm{Al}_\mathrm{Si,h}\mathrm{Si}_\mathrm{I,h}$ & 3+ & -2.5/1.8 &  & 1+ & -1/1.1 &  & 1-  & 0.1/-0.1 & \\
$\mathrm{Al}_\mathrm{Si,h}\mathrm{Si}_\mathrm{I,c}$ & 3+ &  -0.8/2.2  & (2.9/2.1), (5.7/3.5)& 3+  & -0.8/2.2 & (2.9/2.1), (5.7/3.5) & 1-  & 0.2/0.4 & (1.5/1.7), (1.9/1.5)\\
\hline
$\mathrm{Al}_\mathrm{Si,h}\mathrm{Si}_\mathrm{I,h}$ $E_F=1.7$\,eV &  & &  &  & & & 1+ & -0.2/0.7 & (1.8/1.6), (2.2/1.5)\\
$\mathrm{Al}_\mathrm{Si,h}\mathrm{Si}_\mathrm{I,h}$ GW &  & &  &  &  &  & 0 & -0.7/1.2 & (1.9/1.2), (2.7/1.5)\\

\hline
\end{tabular}
}
\label{tab:activation}
\end{sidewaystable}
\pagestyle{plain}
\subsection{Binding energies of neutral clusters}
In addition to the binding energies as a function of the Fermi level, the neutral binding energies are shown as dashed lines in the Figures~\ref{fig:binding_energies_AlC_defects} and ~\ref{fig:binding_energies_AlSi_defects}. The neutral binding energies are those energies obtained by subtracting the formation energies of the neutral reactants from the  formation energy of the neutral product without taking into account the position in the band gap. These values are of interest because they are typically used in kinetic Monte Carlo simulations. It is common practice \cite{Ko2017} to calculate the neutral formation energies for a sequence of cluster sizes using DFT and then to determine the binding energies of smaller sub-clusters by subtracting their formation energy from that of the larger initial cluster
\[
E_b = E_{f_{n-1}} + E_{f_{1}}- E_{f_{n}}
\]
gives the binding energy of a single nucleon that is detached from a cluster of size $n$. Where E$_b$ is different from $\Delta$E defined before, since it does not account for charge states. 
While it is justified to treat pure carbon clusters as charge neutral in most of the band gap \cite{Li2023} as can be seen in table~\ref{tab:levels}, this no longer holds for Si-C clusters \cite{hornos2008} nor for (Al-Si) and (Al-C) complexes. Moreover, when electrons are captured or emitted during dissolution the binding energy of the complex depends on the position of the Fermi level. As can be seen in  Figures~\ref{fig:binding_energies_AlC_defects} and ~\ref{fig:binding_energies_AlSi_defects}, the neutral binding energy can deviate significantly from the actual binding energy, especially in the case of Al$_{Si}C_I$, where the stability of the complex is strongly overestimated in the neutral case. In addition to the DFT energy, the binding energy from the GW potential is drawn in green. It can be seen that the GW binding energies are feasible approximations for (Al-Si) complexes in the intrinsic regime while for Al$_{Si}$C$_I$ the binding energy is overestimated by about $1$\,eV like in neutral DFT. While for dimers DFT and GW neutral binding energies agree well, for trimers the stability statements from DFT and GW partially contradict each other especially for the Al$_I$ related complexes as can be seen in Figures~\ref{fig:AlAlSi} to \ref{fig:AlCSiCSi}. Neutral DFT binding energies are fair approximations for acceptor related trimers for Fermi levels close to VBM, while for Al$_I$ related trimers they are feasible approximations under n-type conditions, in the intrinsic regime the binding energies are either strongly over- or under estimated by the neutral energies.